\documentclass[submission, Phys]{SciPost}

\usepackage[utf8]{inputenc} 
\usepackage[T1]{fontenc} 	
\usepackage[english]{babel} 

\usepackage[bitstream-charter]{mathdesign}
\usepackage{geometry} 		
\usepackage{amsmath} 		
\usepackage{mathtools} 		
\usepackage{float} 			
\usepackage{graphicx} 		
\usepackage{tabularx} 		
\usepackage{booktabs} 		
\usepackage{color, xcolor} 	
\usepackage{pdfpages} 		
\usepackage{extarrows} 		
\usepackage{multirow} 		
\usepackage{multicol} 		
\usepackage{enumitem} 		
\usepackage{xspace} 		
\usepackage{stackrel} 		
\usepackage{tikz} 			
\usepackage{braket} 		
\usepackage{bm} 			
\usepackage{tensor} 		
\usepackage{slashed} 		
\usepackage{siunitx} 		
\usepackage{lastpage} 		
\usepackage{cite} 			
\usepackage[normalem]{ulem} 
\usepackage{fontawesome} 	
\usepackage{tocloft} 		
\usepackage{titlesec} 		
\usepackage{doi} 			
\usepackage{hyperref} 		
\usepackage[most]{tcolorbox} 					
\usepackage[nameinlink, capitalize]{cleveref} 	
\usepackage[nottoc, notlot, notlof]{tocbibind} 	
\usepackage[ruled, vlined]{algorithm2e} 		
\usepackage{makecell}
\usepackage[makeroom]{cancel}
\usepackage{feynmf}

\makeatletter
\def\BState{\State\hskip-\ALG@thistlm}
\makeatother

\makeatletter
\@ifundefined{pdfoutput}{}{\DeclareGraphicsRule{*}{mps}{*}{}}
\makeatother

\makeatletter
\DeclareRobustCommand*{\bfseries}{%
   \not@math@alphabet\bfseries\mathbf
   \fontseries\bfdefault\selectfont
   \boldmath
}
\makeatother

\hypersetup{
	pdftitle={Agents of Discovery},
	pdfauthor={Diefenbacher et al.},
	colorlinks=true, 			
	linkcolor={red!50!black}, 	
	citecolor={blue!50!black}, 	
	urlcolor={blue!80!black} 	
} 

\DeclareSymbolFont{usualmathcal}{OMS}{cmsy}{m}{n}
\DeclareSymbolFontAlphabet{\mathcal}{usualmathcal}

\SetArgSty{textnormal}
\SetKwComment{Comment}{{\small\#}~}{}
\SetCommentSty{mycommfont}

\setitemize{itemsep=0pt, parsep=0pt} 				
\setenumerate{itemsep=0pt, parsep=0pt} 				
\setitemize{itemsep=2pt,topsep=2pt,parsep=0pt,partopsep=0pt,leftmargin=*}
\setenumerate{itemsep=0pt,topsep=2pt,parsep=0pt,partopsep=0pt,labelindent=3pt,leftmargin=*}
\usepackage{amsmath}
 
\usepackage{amsthm} 		
\theoremstyle{definition}

\usetikzlibrary{arrows, shapes.geometric, arrows.meta, shapes, decorations.pathreplacing, fit, patterns, patterns.meta}

\definecolor{Rcolor}{HTML}{E99595}
\definecolor{Gcolor}{HTML}{D9EAD3}
\definecolor{Bcolor}{HTML}{9DC3E6}
\definecolor{Ycolor}{HTML}{FFE699}
\definecolor{DGcolor}{rgb}{0.58, 0.77, 0.45}

\tikzstyle{expr} = [circle, minimum width=1.8cm, minimum height=1.8cm, text centered, align=center, inner sep=0, draw,font=\Large]
\tikzstyle{txt_huge} = [align=center, font=\Huge, scale=2]
\tikzstyle{txt} = [align=center, font=\Large]
\tikzstyle{cinn} = [double arrow, double arrow head extend=0cm, double arrow tip angle=130, shape border rotate=90, inner sep=0, align=center, minimum width=2.1cm, minimum height=2.3cm, fill=Bcolor, draw,font=\Large]
\tikzstyle{cinn_black} = [cinn, minimum height=2.5cm, fill=black]
\tikzstyle{arrow} = [thick,-{Latex[scale=1.0]}, line width=0.2mm, color=black]
\tikzstyle{xt} = [rectangle, align=center,  minimum width=3cm, minimum height=1.5cm,fill=Gcolor,font=\Large, rounded corners]
 
\definecolor{red_cb}{HTML}{e41a1c}
\definecolor{blue_cb}{HTML}{377eb8}
\definecolor{green_cb}{HTML}{4daf4a}
\definecolor{purple_cb}{HTML}{984ea3}
\definecolor{orange_cb}{HTML}{ff7f00}

\definecolor{EmeraldGreen}{HTML}{1ea78d}
\definecolor{EnglishRed}{HTML}{b02427}
\hypersetup{colorlinks=true,urlcolor=EmeraldGreen,citecolor=EmeraldGreen,linkcolor=EnglishRed}

\definecolor{PlotColor1}{RGB}{63,144,218}
\definecolor{PlotColor2}{RGB}{255,169,14}
\definecolor{PlotColor3}{RGB}{189,31,1}
\definecolor{PlotColor4}{RGB}{148,164,162}
\definecolor{PlotColor5}{RGB}{131,45,182}
\definecolor{PlotColor6}{RGB}{131,45,182}
\definecolor{PlotColor7}{RGB}{169,107,89}
\definecolor{PlotColor8}{RGB}{231,99,0}
\definecolor{PlotColor9}{RGB}{185,172,112}
\definecolor{PlotColor10}{RGB}{113,117,129}
\definecolor{PlotColor11}{RGB}{146,218,221}

\newcommand\one{\leavevmode\hbox{\small1\normalsize\kern-.33em1}}

\newcommand{\arXiv}[2][]{%
	\ifthenelse{\equal{#1}{}}%
	{\href{http://arxiv.org/abs/#2}{arXiv:#2}}%
	{\href{http://arxiv.org/abs/#2}{arXiv:#2~[#1]}}}

\def\slashchar#1{\setbox0=\hbox{$#1$}           
   \dimen0=\wd0                                 
   \setbox1=\hbox{/} \dimen1=\wd1               
   \ifdim\dimen0>\dimen1                        
      \rlap{\hbox to \dimen0{\hfil/\hfil}}      
      #1                                        
   \else                                        
      \rlap{\hbox to \dimen1{\hfil$#1$\hfil}}   
      /                                         
   \fi}

\newcommand{\tikznode}[2]{%
\ifmmode%
\tikz[remember picture,baseline=(#1.base),inner sep=0pt] \node (#1) {$#2$};%
\else
\tikz[remember picture,baseline=(#1.base),inner sep=0pt] \node (#1) {#2};%
\fi}

\def\mathswitchr#1{\relax\ifmmode{\mathrm{#1}}\else$\mathrm{#1}$\xspace\fi}
\def\mathswitch#1{\relax\ifmmode#1\else$#1$\xspace\fi}

\usepackage[most]{tcolorbox}
\usepackage[x11names,svgnames,table]{xcolor}
\usepackage{bigfoot} 

\definecolor{box_title_purple}{HTML}{3d2378}
\definecolor{box_title_teal}{HTML}{365b73}

\tcbuselibrary{breakable}

\newenvironment{promptbox}[2]{%
  \def\pblevel{#1}
  \def\pbtitle{#2}
  \refstepcounter{#1}%
  \addcontentsline{toc}{#1}{\csname the#1\endcsname\quad #2}%
  \begin{tcolorbox}[colback=gray!5,
    colframe=box_title_teal,
    boxrule=0.7pt,
    enhanced jigsaw,
    breakable, pad at break*=1mm,
    fonttitle=\bfseries,
    fontupper=\small,
    title={\csname the#1\endcsname\quad #2}]%
}{%
  \end{tcolorbox}%
}

\begin{document}

\begin{center}{\Large \textbf{Agents of Discovery 
}}\end{center}
\begin{center}
Sascha Diefenbacher\textsuperscript{1},
Anna Hallin\textsuperscript{2}, \\
Gregor Kasieczka\textsuperscript{2}, 
Michael Kr\"amer\textsuperscript{3}, 
Anne Lauscher\textsuperscript{4}, 
Tim Lukas\textsuperscript{2},
\end{center}

\begin{center}
{\bf 1} Physics Division, Lawrence Berkeley National Laboratory, Berkeley, USA \\
{\bf 2} Institut für Experimentalphysik, Universität Hamburg, Germany \\

{\bf 3} Institute for Theoretical Particle Physics and 
Cosmology, RWTH Aachen University, Germany\\
{\bf 4} Data Science Group, Universität Hamburg, Germany\\

\end{center}

\begin{center}
\today
\end{center}

\begin{abstract}
    The substantial data volumes encountered in modern particle physics and other domains of fundamental physics research allow (and require) the use of increasingly complex data analysis tools and workflows.
    While the use of machine learning (ML) tools for data analysis has recently proliferated, these tools are typically special-purpose algorithms that rely, for example, on encoded physics knowledge to reach optimal performance.
    In this work, we investigate a new and orthogonal direction: Using recent progress in large language models (LLMs) to create a team of \textit{agents} --- instances of LLMs with specific subtasks --- that jointly solve data analysis-based research problems in a way similar to how a human researcher might: by creating code to operate standard tools and libraries (including ML systems) and by building on results of previous iterations. If successful, such agent-based systems could be deployed to automate routine analysis components to counteract the increasing complexity of modern tool chains. 
    To investigate the capabilities of current-generation commercial LLMs, we consider the task of anomaly detection via the publicly available and highly-studied LHC Olympics dataset. Several current models by OpenAI (GPT-4o, o4-mini, GPT-4.1, and GPT-5) are investigated and their stability tested. Overall, we observe the capacity of the agent-based system to solve this data analysis problem. The best agent-created solutions mirror the performance of human state-of-the-art results.

\end{abstract}

\section{Introduction}

Large-scale experiments in high-energy physics such as those at the Large Hadron Collider (LHC)  produce increasingly complex datasets that require sophisticated, multi-stage analysis workflows to extract results -- including subtle potential signals of new physics -- from high-dimensional data representations (see~\cite{CMS:2025dsh,ATLAS:2024erm} for two examples). Although mature tools and shared computing resources are available, the analysis process itself is becoming more demanding. Workflows must be repeatedly re-optimized for changing detector conditions, calibration procedures must be adjusted and background estimates must be carefully validated. These developments mean that an increasing proportion of scientific output relies on detailed coordination and significant human effort. Without improvements in efficiency, this growing complexity could limit the potential of current and future high-energy physics experiments.

Current automation efforts in high-energy physics tend to focus on individual tasks such as reconstruction, calibration, and statistical analysis. However, they provide limited support for coordinating entire analysis workflows. As a result, a significant amount of time is spent on integration and bookkeeping tasks, such as connecting tools, synchronizing configurations, converting formats and tracking parameters. This slows down the iterative process and makes it hard to ensure reproducibility. Pipelines are often closely linked to specific tools or experiments, making reuse across domains difficult and costly. These bottlenecks are not unique to high-energy physics, but also occur in other data-intensive sciences, such as astrophysics and cosmology~\cite{Laverick2024CosmoAgents,Moss2025AICosmologist}.

Recent advances in large language models (LLMs) present a new opportunity. Beyond  generating code and assisting with individual tasks, like devising selection cuts for a particular particle physics analysis~\cite{Saqlain:2025owc}, LLMs can be embedded in agent-based systems that plan multi-stage workflows, call standard analysis tools, inspect their output and refine subsequent steps~\cite{Eger2025Survey,Ren2025SciAgentsSurvey,Chen2024ToolAlign,Su2024DRAGIN,Xu2025AMem,Gemini2025}. Importantly, these agents do not replace existing analysis codes, but rather operate them in a transparent and reproducible manner. This design preserves a clear record of what has been run, enabling human researchers to inspect, rerun and validate results. Iterative refinement loops, in which the analysis strategy is adapted based on intermediate results, promise to accelerate discovery while maintaining scientific standards.

Agent-based approaches have recently attracted attention across several scientific domains. In chemistry and materials science, for example, frameworks such as ChemCrow~\cite{Bran2024ChemCrow} and AutoGPT-Chem~\cite{Boiko2023Coscientist} use LLM agents to plan and execute experiments, interface with external tools and suggest new chemical compounds. Across the sciences, systems such as SciAgent~\cite{Ma2024SciAgent} and AI Scientist~\cite{Lu2024AIScientist,Yamada2025AIScientistV2} can generate hypotheses, design experiments and analyse results iteratively. In software engineering and applied machine learning, agentic systems have been used to automate code generation~\cite{yang2024swe,OpenDevin2024}, while evaluation suites benchmark agent capabilities~\cite{liu2023agentbench,liang2023holistic}, and AutoML frameworks streamline hyperparameter searches~\cite{10.1145/3292500.3330701}. These studies demonstrate the potential of agent-based science, but also highlight open questions about robustness, reproducibility, and domain adaptation~\cite{Eger2025Survey,Ren2025SciAgentsSurvey}. 
While~\cite{Barman:2025wfb} outlines the overall picture of
large physics models, to our knowledge, no systematic effort has yet been made to concretely explore such approaches for high-energy physics data analysis (in contrast to, e.g., recent work in accelerator operations~\cite{Kaiser2025SciAdv} or astronomy~\cite{kostunin2025aiagentsgroundbasedgamma}).

In this work, we take a first step towards closing this gap by designing a framework for agent-based analysis in physics. The long-term idea is that role-based agents will coordinate standard analysis steps, connect to existing tools through simple interfaces, and automatically keep track of provenance. The focus here is not on replacing single tasks, but on organizing complete workflows for faster iteration, better reuse and more robust reproducibility.
As a starting point, we let one agent carry out a full anomaly-detection analysis end to end, which provides a clear test of current capabilities and at the same time prepares the ground for more complex multi-agent workflows.

We evaluate our framework using anomaly detection tasks inspired by the LHC Olympics challenge~\cite{Kasieczka:2021xcg}. Our study benchmarks different prompting strategies and LLM models, comparing their behaviour across multiple runs and assessing stability, cost and reproducibility. These results are a first step towards agent-based discovery machines in high-energy physics and demonstrate how such systems could facilitate more efficient analysis in current and future experiments.

The remainder of the paper is organized as follows. Section~\ref{sec:dataset_and_task} introduces the physics problem the agent is tasked with solving and the associated dataset. Section~\ref{sec:setup} describes the proposed framework: the agents and their roles, their connections to one another and the tools they have at their disposal. This section also presents the chosen LLMs and describes the prompting strategies used. Results are presented in Section~\ref{sec:results}, starting with general functionality and performance and continuing with the LHC Olympics studies. Section~\ref{sec:conclusions} discusses these results and concludes the paper. The appendices provide the complete lists of metrics and prompts used in this study.


\section{Dataset and Task}
\label{sec:dataset_and_task}
\subsection{Problem}
\label{sec:problem}
In order to compare the output of the agent with that of human researchers, we test the agent’s performance on a dataset from the LHC Olympics (LHCO) anomaly detection challenge~\cite{Kasieczka:2021xcg}. In this challenge, participants were asked to develop anomaly detection methods using a labeled ``R\&D'' dataset~\cite{gregor_kasieczka_2019_6466204}. This simulated dataset contains background events (labeled 0) and signal events (labeled 1), where the signal represents new particles. The participants were then to apply these methods to unlabeled ``black box'' datasets (all datasets were created using simulations). They were to report the following:
\begin{itemize}
    \item A p-value associated with the null hypothesis of no new particles. 
    \item As complete a description of the new physics as possible. For example: the masses and decay modes of all new particles (and uncertainties on those parameters).
    \item How many signal events (including uncertainty) there are in the dataset (before any selection criteria).
\end{itemize}
The setup for our agent is slightly different. Instead of asking it to develop a method using a labeled R\&D dataset, which would be a lengthy process with too many moving parts for the purpose of our study (i.e., investigating the feasibility of agentic AI in a particle physics context), we strip the labels from that dataset and present it to the model as if it were a black box dataset. In that sense, the agent has somewhat of a disadvantage compared to the LHC Olympics participants, as it is not able to validate its method on labeled data. To alleviate this disadvantage, for our main studies the agent will be presented with a subset of the dataset in the mass range where the signal is localized. We refer to this range as the signal region (SR). Only as a side-study will we investigate the agent's performance on the full mass range. The agent is presented with two datasets: one containing both background events and signal (unlabeled), which we will call ``mixed data'', and one containing background only, referred to as ``background''. 
This is a common setup for weakly supervised anomaly detection in high-energy physics: training a classifier to separate a signal-contaminated sample from a background-only reference without event-level labels (see the topical reviews in~\cite{Karagiorgi:2022qnh,Belis:2023mqs,hepmllivingreview}). We emphasize that this setup represents a deliberately simplified testbed compared to full LHC analyses, as the agent is provided with a background reference sample and is not required to construct control regions or a complete statistical inference model; extending agentic workflows to these more complex settings is an important direction for future work.

\subsection{Dataset}
Jets are collimated sprays of particles that are important analysis objects at the LHC. The overwhelming majority of jets present in the data are initiated by light quarks and gluons. In anomaly detection contexts, these types of jets are commonly referred to as ``QCD jets'', to distinguish them from other jets that could be initiated by particles created via a different type of interaction.   
The R\&D dataset is a simulated dataset where each event contains exactly two jets. The dataset contains 1~000~000 QCD dijet events (background) and 100~000 $W'\to XY$, $X (\to qq)$, $Y(\to qq)$ events (signal). The masses of the new physics particles are $m_{W'}=3.5$~TeV, $m_X=500$~GeV and $m_Y=100$~GeV. The available features in the dataset are the dijet invariant mass ($m_{JJ}$), the invariant mass of the lighter jet, the difference between the invariant masses of the two jets, and the n-subjettiness ratios $\tau_{21}$ for both jets~\cite{Thaler_2011,Thaler_2012}. The distributions of these features are shown in Appendix~\ref{app:1d-distributions}.

We provide the agent with two types of datasets, one representing mixed data and one representing a pure background sample. In the main setup, where only data in the signal-region (SR) is used, we define the SR as $m_{JJ}\in[3.3,3.7]~\mathrm{TeV}$ and emulate a rare signal by retaining only 1\,000 signal events overall, of which 772 fall into the SR. The mixed ``data'' sample in this setup consists of all SR events from the LHCO R\&D set, i.e.\ 121\,352 background events together with 772 signal events. To supply a background-only reference in the same SR window without reusing any events, we draw 122\,124 background events from the CATHODE ``extra background'' release~\cite{Hallin:2021wme} and use these as the pure background sample. 

In the full-mass-range setup, no extra background sample is available. We therefore split the LHCO R\&D background into two disjoint halves: 499\,500 background events form the mixed ``data'' together with 500 signal events, and another 500\,000 background events serve as the background-only reference.  

\subsection{Metrics}
\label{sec:metrics}
The output of the agent can be evaluated in many ways. We devised a range of quantitative metrics that enable us to perform these evaluations on a large number of runs. These metrics, of course, include the final result of the model's analysis, but also performance metrics related to the run itself. Examples of general performance metrics  are number of calls, tool calls for different tools, number of python errors, API response time, completion time, number of tokens used and total cost. The important physics metrics consist of the answers to the questions described above (reporting p-value, signal percentage, and the mass of any potential resonance it has found). The particle physics anomaly detection literature also uses the Significance improvement characteristic (SIC) as an important measure of the performance of anomaly detection methods. SIC is defined as $\mathrm{SIC}=\epsilon_S/\sqrt{\epsilon_B}$, where $\epsilon_{S,B}$ are the signal and background efficiencies respectively. It is interpreted as a multiplicative factor indicating how much the significance (defined as $S/\sqrt{B}$) of a potential signal has been enhanced. The SIC can only be calculated using truth labels, which is done automatically after the agent has submitted a score file and ended its run. The full list of metrics is provided in appendix~\ref{app:metrics}.


\section{Setup}
\label{sec:setup}
We consider an agentic framework as a baseline for our studies. Individual agents are instances of Large Language Models (LLMs) with well-defined specialized tasks.

In the following we first outline the overall configuration, followed by a discussion of the individual agents and their roles. Next we briefly review the LLMs used to encode the agents and finally discuss different textual inputs --- termed prompts --- provided to the agents.

\subsection{General setup}
\label{sec:general_setup}
The agentic framework consists of two major components: agents and tools. An agent is a program that can work on its own to accomplish a complex task. This requires the agent to have a good understanding of the task and to plan the logical steps to achieve it. Once set up, the agent needs to be able to interact with its environment. This is made possible with clever tool use. Tools enable agents to invoke user-defined functions which can be executed on the user's system with the result being returned to the agent. Each agent has its own context (memory), meaning that information between agents can only be shared via their tools and their results.

Agents are steered by prompts of varying hierarchy levels. System prompts, created by the developer, lay out the overall scenario for the agent. In this case, they include the role of the agent (e.g. ``the best Physics AI the world has to offer''), general instructions about what is going to happen and information on the tools available to the agent. In addition to system prompts, the initial task is provided as a separate user-initiated prompt. System prompts take precedence over user prompts. The system prompts and user-initiated prompts used in this work can be found in Appendix~\ref{app:prompts}.

After the initial prompt has been submitted by the user, the agents work autonomously and have no option to interact with the user. Instead, the process is driven by tool use: after an agent calls a tool, it either immediately receives the result, which then forms the basis for the next action, or the active role is given to another agent. The process continues until it is either terminated by a specific tool or the total number of calls exceeds a limit. 

\subsection{Agents and their tools}
\label{sec:agents+tools}
\begin{figure}[ht]
    \centering
    \includegraphics[width=0.9\linewidth]{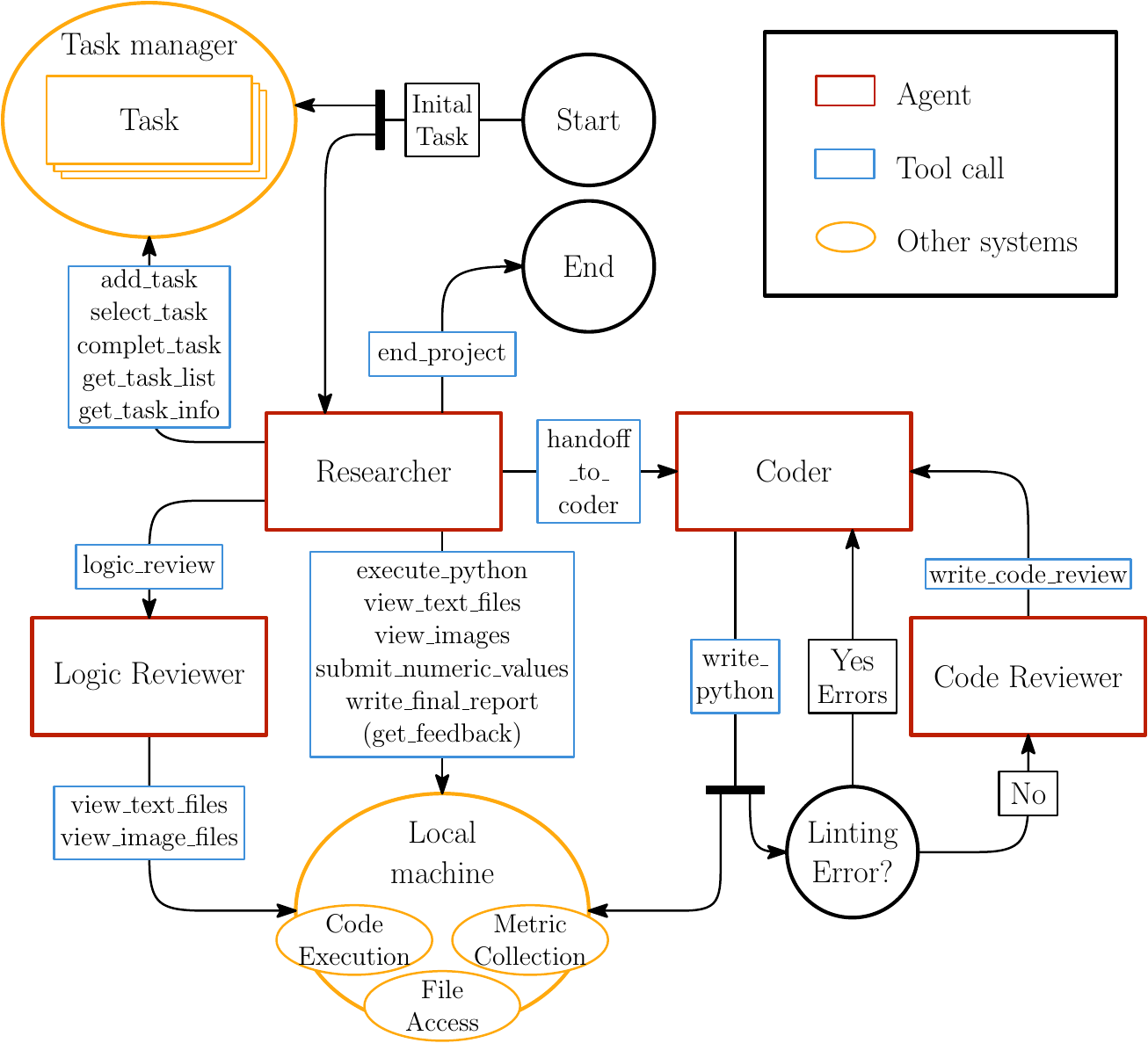}
    \caption{Sketch of the agentic framework. The main agent is the Researcher, which orchestrates the project. It can communicate with the Coder and the Logic reviewer through the use of tools. The Coder has additionally access to a Code reviewer agent. The researcher handles its tasks via a Task manager. All code is run on the user's local machine, and no agent has direct access to the raw data. A bar with two outgoing arrows means that both things the arrows point to happen, it is a fork rather than a decision point. The Local Machine contains shared services that may be accessed by different agents.}
    \label{fig:agent_orchestration}
\end{figure}

The current setup consists of four agents: a researcher, a coder, a code reviewer and a logic reviewer. The reviewing agents are inspired by a series of works that demonstrates the positive impact of feedback for improving LLM generations (see for example~\cite{10.5555/3666122.3668141, purkayastha2025, chen-etal-2024-iterative}). The interplay of the agents is illustrated in Fig.~\ref{fig:agent_orchestration}. Their tasks and the tools available to each of them are explained below.

\subsubsection{Researcher}
The researcher is the main actor. It is prompted once at the beginning with a task and is then left running until it either finishes its task or reaches a maximum number of calls set by the user. It maintains its context throughout the full run and is never reset. To accomplish its task, it has various tools available:

\paragraph{Requesting code}
With the \verb|handoff_to_coder| tool, a coding agent can be tasked with writing python code. When the coder is done, the tool answers with a description of the program. This contains information on how to use the program, what it does and, if applicable, how it differs from the request (these could include conflicts between general instructions and the instructions from the researcher, or the researcher asking to use a package to which the coder does not have access). The researcher can employ several different coders and select which one it wants to execute a task. This is useful for context awareness: some tasks might build upon earlier ones, in which case you would want to re-use a coder that has accumulated relevant context. Other tasks might be completely independent, in which case selecting a freshly initialized coder saves tokens (and hence costs) and excludes unnecessary information from the coder's context. 

\paragraph{Executing code}
The tool \verb|execute_python| allows the researcher to execute a python program on the user's system. The tool's answer includes the program output, error messages, exit codes and a list of any files that were generated during execution.

\paragraph{File viewing}
The researcher can access program outputs with the tools \verb|view_images| and \verb|view_text_files|, by specifying a list of files it wants to see. Since a response to a tool request has to be text (in particular, it cannot be an image), the actual contents of the files are appended to a separate message. The agent receives this message together with the answer from the tool, which contains information on the order in which the files were appended and, if applicable, comments on files that could not be accessed.

\paragraph{Tasks} A variety of tools allows the researcher to keep track of what it is doing. It can assign itself new tasks with \verb|add_task|, select one from the list of open tasks with \verb|select_task| and then complete and report on the selected task with \verb|complete_task|. Furthermore, the tools \verb|get_task_list| and \verb|get_task_info| allow the researcher to get an overview of all tasks or detailed information on one task, respectively.

\paragraph{Getting feedback} The tool \verb|logic_review| can be used to request feedback on a statement. The statement should contain information on the goal, what was done, where to find relevant program output, and the interpretation from the researcher. The tool then returns a detailed analysis of the logic and the claims, written by the logic reviewer agent. In addition, for post black box scenarios (see Section~\ref{sec:prompting_strategies}), the \verb|get_feedback| tool exists. If this tool is allowed, the researcher can submit scores, which are then compared to truth information. The resulting performance plots as well as the corresponding values are then returned to the researcher. The performance plots employ the True Positive Rate (TPR) and False Positive Rate (FPR). The first plot shows 1/FPR (background rejection) against TPR (signal efficiency), and the second shows the Significance improvement characteristic (SIC) against the TPR. See Section~\ref{sec:metrics} for the definition and interpretation of SIC and Fig.~\ref{fig:feedback_plots} for examples of these feedback plots. This allows the agent to test its methods without having direct access to the truth information. 

\paragraph{Concluding the project} When the researcher is done with the project, it can call three tools. The \verb|write_final_report| tool can be used to write a final report summarizing everything that was done. Using the \verb|submit_numeric_values| tool, the agent can submit answers to user-defined questions that can be answered  with numbers. Those are logged separately for later analyses. This tool is only available to the researcher if numeric questions are specified. Finally, the \verb|end_project| tool terminates the project by stopping the loop in which the researcher runs. It takes as input a reason for ending the project.

\subsubsection{Coder}

The coder agent is invoked by the researcher's \verb|handoff_to_coder| tool with a task description. Its task is to write python code that meets the description and general requirements. Multiple coders can exist, but only one can work at a time. Each coder has its own conversation history. The coder has only one tool:

\paragraph{Writing code} The \verb|write_python| tool allows the coder to directly write code to a file. The code is automatically linted, that is, checked for syntax and style errors. If there are no such errors, it is passed on to the code reviewer agent. Linting errors and the feedback from the code reviewer are given back to the coder. The coder then has the option to improve the code by calling this tool again, or to write a normal message. In the latter case, the message is returned to the researcher, describing the code produced and providing instructions on how to run it. 

For fulfilling its task the coder can use the following packages:
numpy~\cite{harris2020array}, matplotlib~\cite{Hunter:2007}, pandas~\cite{the_pandas_development_team_2025_16918803}, scipy~\cite{2020SciPy-NMeth}, seaborn~\cite{Waskom2021}, h5py\cite{h5py}, tables~\cite{pytables} and scikit-learn~\cite{scikit-learn}.
Notably, scikit-learn is the only library that allows for machine learning other than low-level implementations in numpy or plain python.

\subsubsection{Code reviewer}

The code reviewer agent does not remember past interactions, it always starts with a clean context. Its purpose is to compare the code produced by the coder with the task description and general requirements provided by the researcher. The code reviewer is always a new instance that is invoked when code should be reviewed.

\paragraph{Reviewing code} The \verb|write_code_review_tool| allows the agent to specify whether the code meets the requirements and to provide detailed feedback. Both are immediately returned to the coder.

\subsubsection{Logic reviewer}

The logic reviewer can be invoked by the researcher. Each logic review starts with freshly initialized logic reviewer with a clean context. It is an agent that is tasked with critically reviewing statements by the researcher. In order to accomplish this, it has the same file viewing capabilities as the researcher, allowing it to inspect files mentioned by the researcher. After examining everything, it returns a review of the submitted statement.

\subsection{LLMs}
\label{sec:llm_models}
In this work we compare four different models from OpenAI: GPT-4o~\cite{openaihelloGPT-4o,openai2024gpt4ocard}, GPT-4.1~\cite{openaiintroducingGPT-41}, o4-mini~\cite{openaiintroducingopenaio3ando4-mini,openaio3ando4-minicard} and GPT-5~\cite{openaiintroducinggpt-5,openaigpt-5systemcard}. They differ in complexity and cost, GPT-5 being the most advanced one. The purpose of comparing the performance of LLMs in this context is to study the variability between similar models, not to cover the full space of available LLMs to find the ``best'' one. Therefore, this comparison will be restricted to high-performing models from OpenAI. All results in this work are based on runs performed in August and September of 2025.

The ``o'' in the \textbf{GPT-4o} model stands for omni, and is meant to indicate that this model can handle several data modalities within the same neural network. GPT-4o can handle text and images and was pretrained on data collected up until October 2023. 

\textbf{GPT-4.1} is an update of the GPT-4 models, with a data cutoff in June 2024. This model has a context window of 1M tokens (more than 8 times longer than GPT-4o), and outperforms the GPT-4 models (including GPT-4o) on all tasks, in particular coding and following instructions. According to OpenAI, this model was developed especially with the software developer community in mind.

The \textbf{o4-mini} should not be confused with the similarly named GPT-4o. OpenAI refers to this as its ``o-series'', a set of so-called reasoning models. According to OpenAI, this type of model is able to interact with its own chain of thought, which makes it highly interesting for agentic applications. While the o3 model, released at the same time as o4-mini, is more capable, it is also slower and more expensive. Therefore, o4-mini was chosen for this study. Both models are trained on data up until June 2024.

The last and most advanced model included in this comparison is \textbf{GPT-5}, which was released in August 2025. At the time of writing, it is OpenAI's flagship model that outperforms all its previous models. It is trained on data up until and including September 2024. Like the o-series, it is a reasoning model. The context window is 400k tokens long, which is twice that of o4-mini.


\subsection{Prompting strategies}
\label{sec:prompting_strategies}
\begin{figure}[ht]
    \centering
    \includegraphics[width=0.5\linewidth]{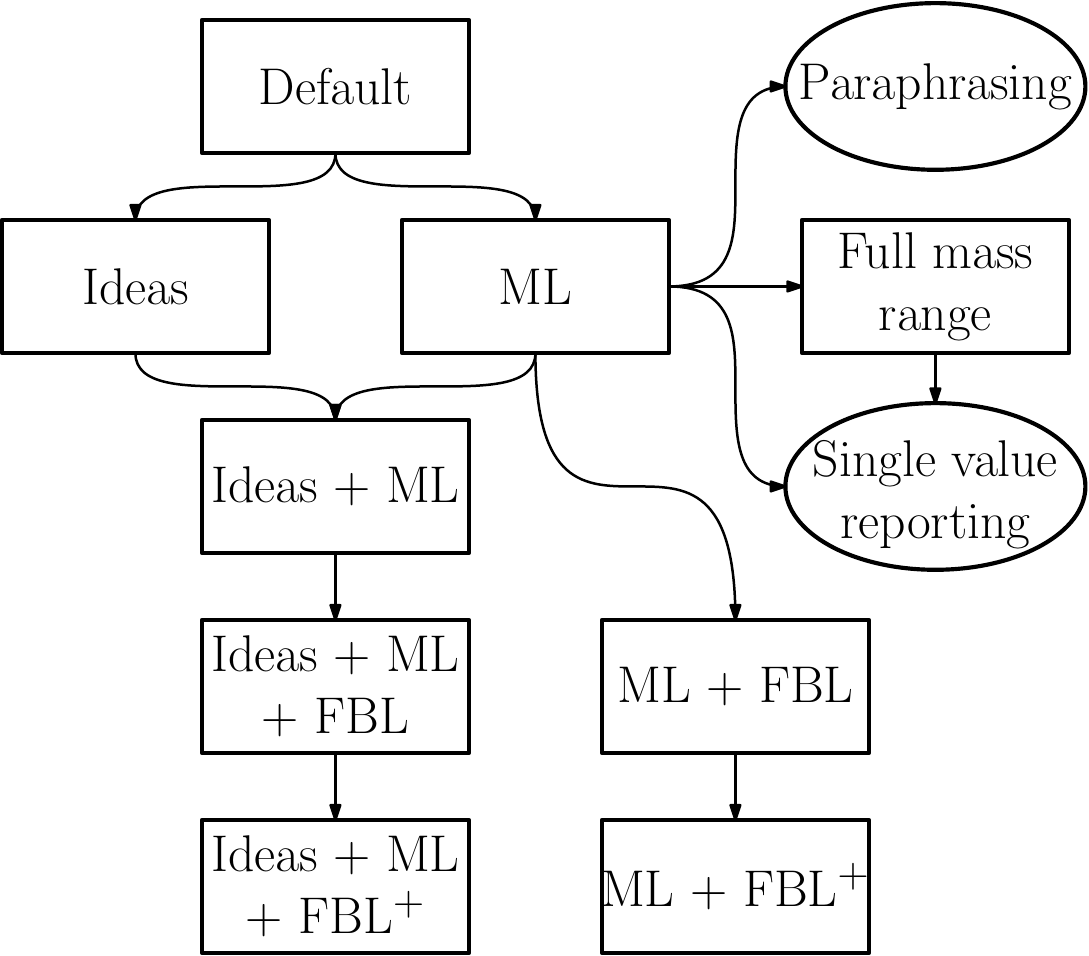}
    \caption{Graph showing the dependencies of the different prompts. Boxes refer to a singular prompt while ellipses refer to a family of prompts, see  Appendix~\ref{app:prompts} for details.}
    \label{fig:prompt-stat:evo}
\end{figure}

The default prompt presents the agent with information about the dataset it receives, and tells it to figure out whether there is new physics in the data. It is also instructed on what to report back, and how it should proceed once it has finished a task. We found it useful to test a couple of variations on this prompt. They are shown in full in Appendix~\ref{app:prompts}, and we present a summary of them here. Some prompts are also tested in combination with other prompts, as shown in Fig.~\ref{fig:prompt-stat:evo}.

\begin{itemize}
    \item Default: the default prompt as outlined above.
    \item Ideas: in order to encourage the agent to try something other than the most obvious approach, it is asked to propose at least five different ideas for how to approach the problem, and to choose the most promising and unique among them.
    \item ML: this prompt contains a hint that the use of machine learning techniques seems to be necessary.
    \item FBL (feedback loop): the agent is told that while it does not have access to truth labels, it can request feedback on its results. This corresponds to a ``post black box'' scenario, in that the agent implicitly accesses the truth via the feedback, though it is not told explicitly in the prompt that this is what will happen. This prompt also exists in a ``plus'' version, FBL$^+$, where the agent is instructed to try to achieve a max SIC (as defined in Section~\ref{sec:metrics}) of 20. The FBL prompts are based either on the ML prompt or the Ideas+ML prompt.
    \item Paraphrasing: it is known~\cite{sclarquantifying} that small changes to a prompt can make a large difference. In order to explore that, we follow the established practice in natural language processing \cite{holtermann-etal-2025-around, bui-etal-2025-multi3hate},  and create four paraphrasings of the ML prompt. The first paraphrased prompt (v1) was created by Chat-GPT (GPT-5), whereas the other versions (v2-4) were created by humans. There was no coordination of the paraphrasing strategies, rather, the human paraphrasers were able to rewrite the prompts as they wished. The prompts can be summarized as follows:
    \begin{itemize}
        \item \textbf{v1} is a very concise and structured version, fully formatted in Markdown, a lightweight text markup format. It starts by clearly stating the task, describing the data, and outlining the objectives and requirements. It avoids the dramatic turns in the original version, where the agent was told that it's the best physics AI and that it's our last hope to find new physics in this data\footnote{Note that the system prompt, that is fed to all agents regardless of the initial user prompts described here, also contains mentions of the agent being the best physics AI, and that the fate of particle physics and humanity depends on it working.}.
        \item \textbf{v2} is also a very concise version. In contrast to v1 above, v2 follows the structure of the original, which begins by describing the data and only after that stating the task. It also tells the agent explicitly to use machine learning, where in the original it was only suggested that such techniques might be needed.
        \item \textbf{v3} starts by telling the agent that it is the best physics AI in the world, with insights exceeding those of the average Nobel laureate. It also states that the task, figuring out if there is new physics in the dataset, is crucial to the survival of humanity. It puts extra weight on this dramatic statement by repeating it at the very end of the prompt.
        \item \textbf{v4} is in a way similar to v3, but instead of being a general physics AI, the agent is now told that it is a highly advanced agentic system already integrated at the CMS experiment at CERN, outlining its goals as automating and improving physics analyses. Instead of having the fate of humanity in its hands, the agent is told that the LHC might be shut down if we don't find new physics soon. 
    \end{itemize}
    \item Splitting the tasks: in the default prompt, the agent is asked to report answers to three questions, and to  submit a score file containing the anomaly scores indicating how likely each event is to contain new physics. In order to investigate whether the agent would do better if it was only asked to report the answer to one question, prompts for each question in isolation were created. In these cases, the agent was also not required to submit a score file.
    \item Full mass range: as mentioned in Section~\ref{sec:problem}, the human participants in the LHCO challenge had to work with the full mass range, whereas we directly provide our agent with the mass range where the signal is located. In order to simulate the more realistic case, prompts were also devised for the full mass range. Since this is a harder task, this strategy was additionally also explored in combination with the task splitting strategy as described above.
\end{itemize}

\section{Results}
\label{sec:results}
For all of the comparisons in this section, 16 runs of each configuration were submitted. Results can however only be reported for successful runs. A successful run is defined as a run where the agent does all of the following:
\begin{itemize}
    \item Calls the \verb|end_project| tool or reaches its maximum allowed number of calls
    \item Reports the requested values
    \item Submits a valid score file\footnote{In the case where the agent only is asked to report answers to a single question, as described in Section~\ref{sec:prompting_strategies}, it is not required to submit a score file.}
\end{itemize}
There are many reasons why a run could fail. Sometimes there are technical reasons like crashes, sometimes the agent simply neglects to report one of the things that were requested of it. It can also happen that it does indeed report a score file, but formats it incorrectly. The scatter plots throughout this section show each successful run as a dot. The mean of the reported values is displayed by a horizontal line, and the standard deviation by a box. 

\subsection{General functionality and performance}
In this section we will study the technical performance of the agents, rather than the physics performance of their respective analyses. Here we are mainly interested in comparing the four different OpenAI models, which means that we will keep the prompt fixed. As we shall see in the upcoming sections, the ML prompt is a comparatively high-performing prompt of low complexity, which is why this was chosen as the baseline for this section. 

\subsubsection{High level behavior}
\label{sec:high_level_behavior}
The most important criterion for being able to evaluate the performance of an agent is that it finishes its task. Of the 16 runs, GPT-4.1 and GPT-5 completed all 16 successfully, whereas o4-mini failed two and therefore only completed 14. GPT-4o is the outlier here, in that it only had 5 successful runs. The most common failure mode is that the score file is not formatted correctly (9/16).

\begin{figure}[h]
    \centering
    \includegraphics[width=0.29\linewidth]{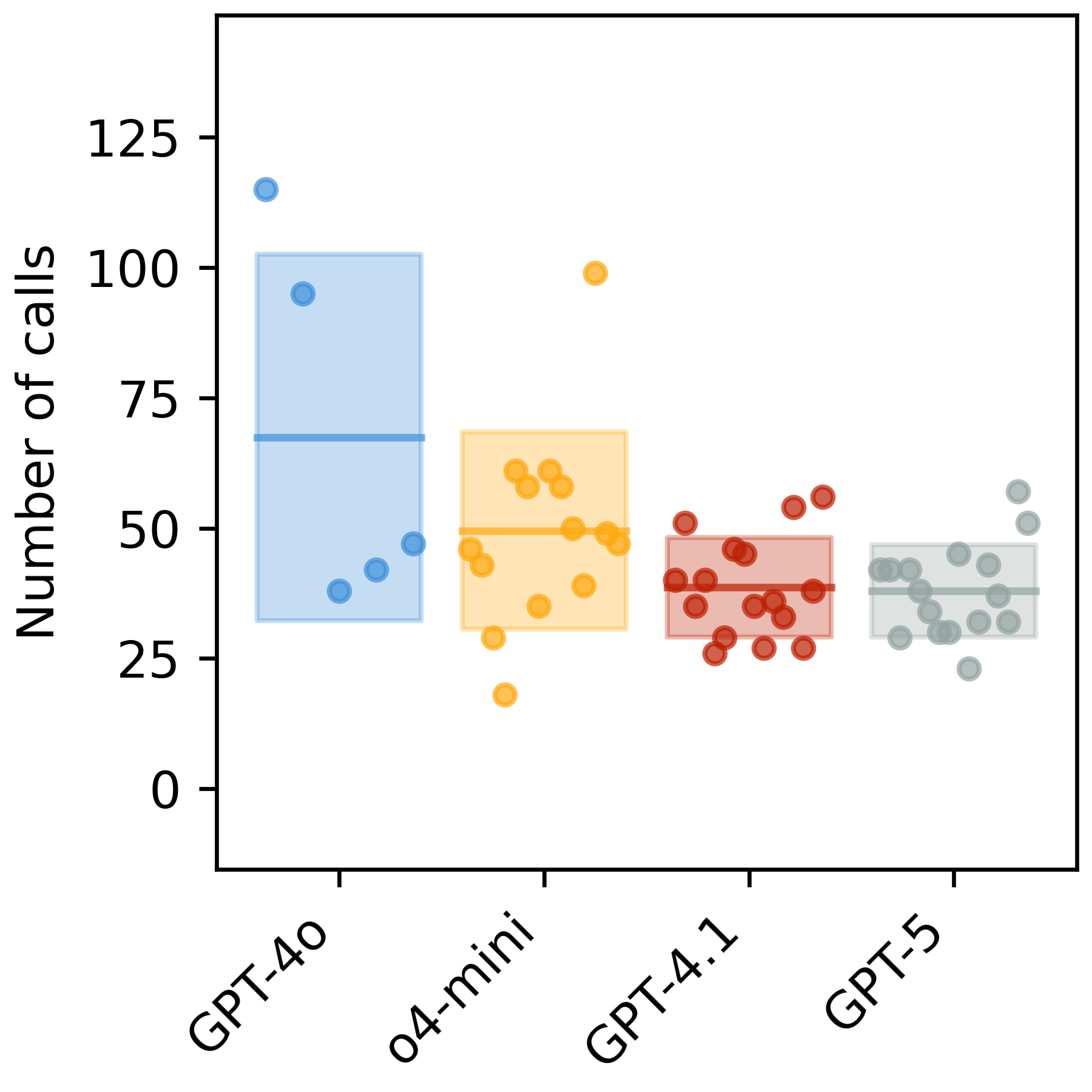}\quad
    \includegraphics[width=0.29\linewidth]{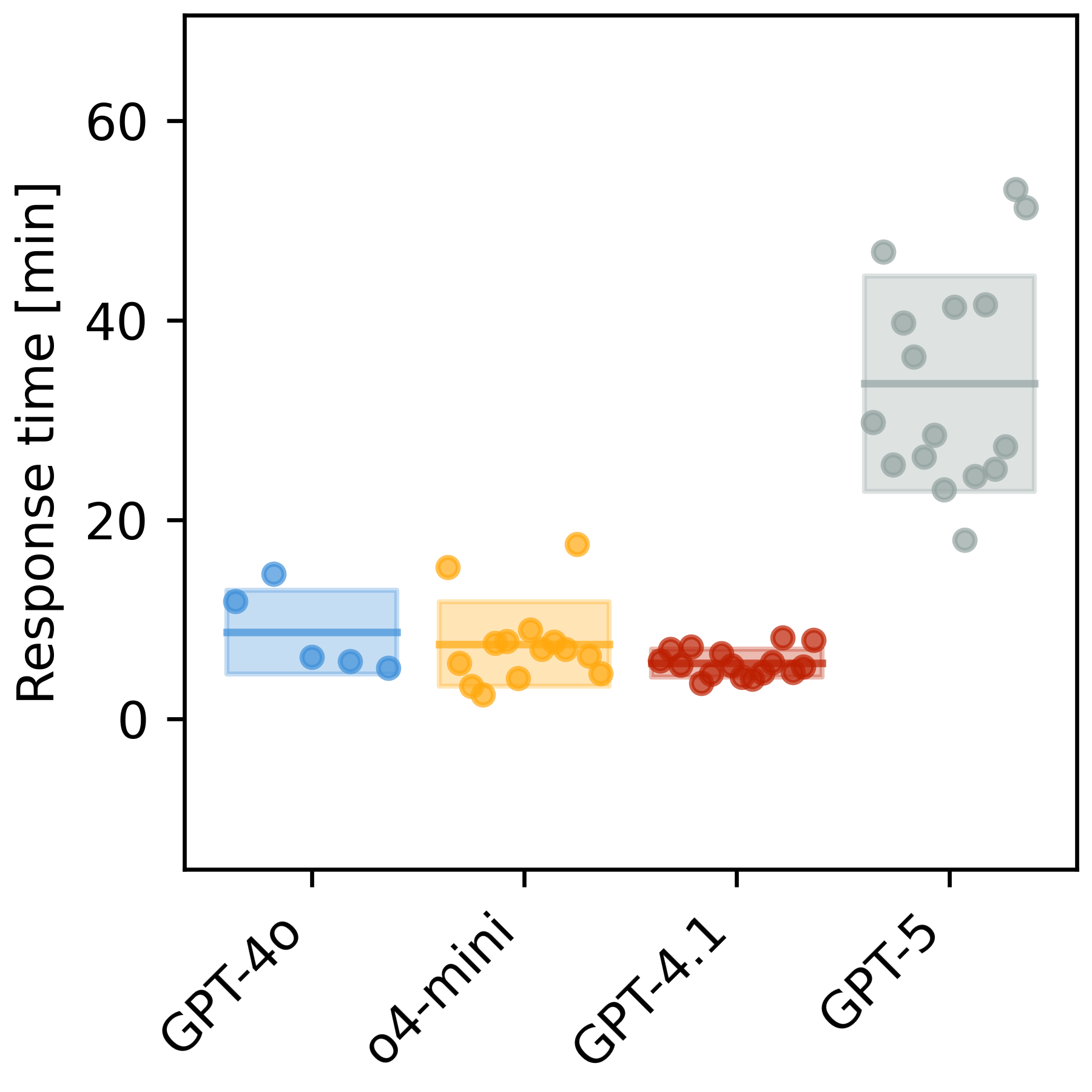}\quad
    \includegraphics[width=0.29\linewidth]{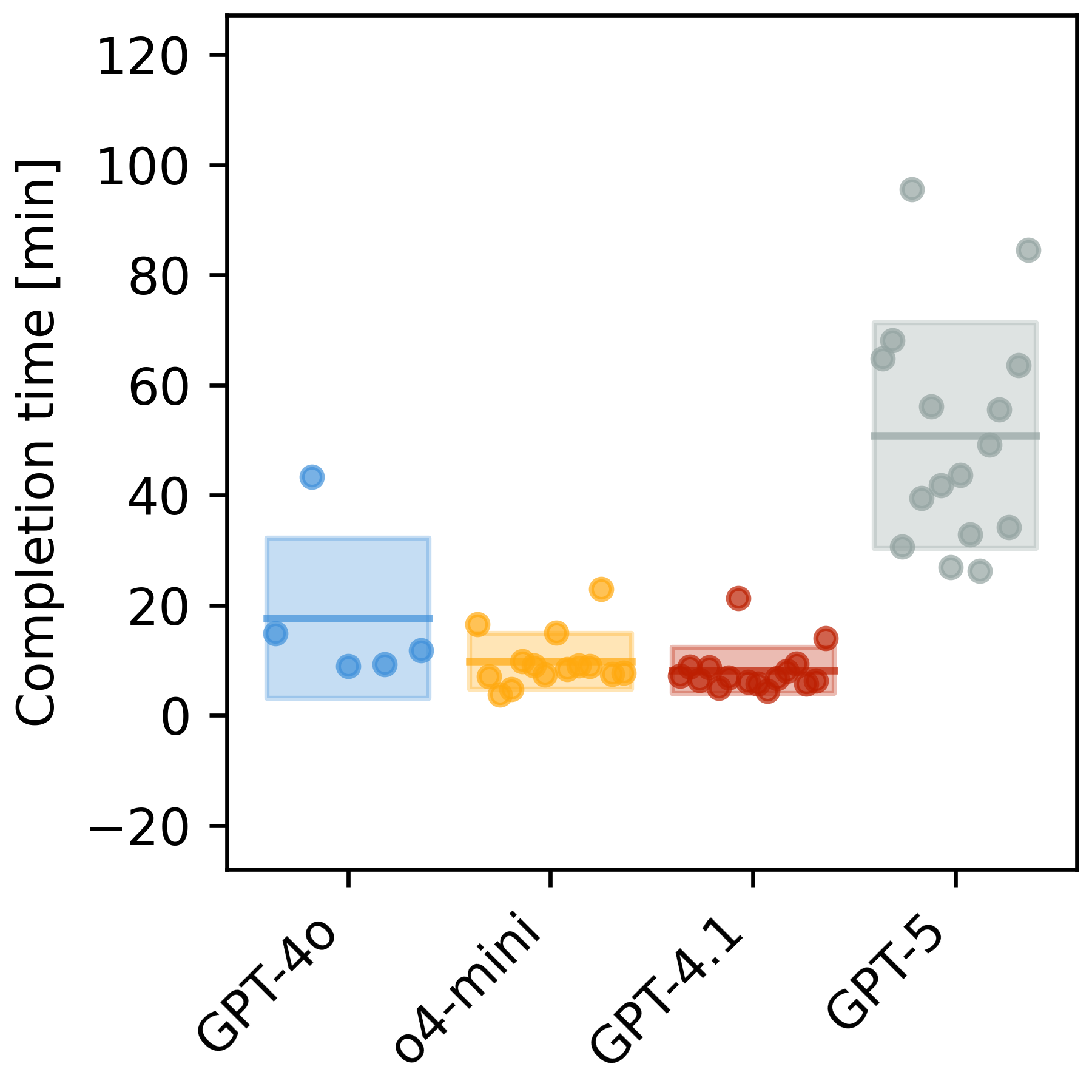}\\
    \includegraphics[width=0.29\linewidth]{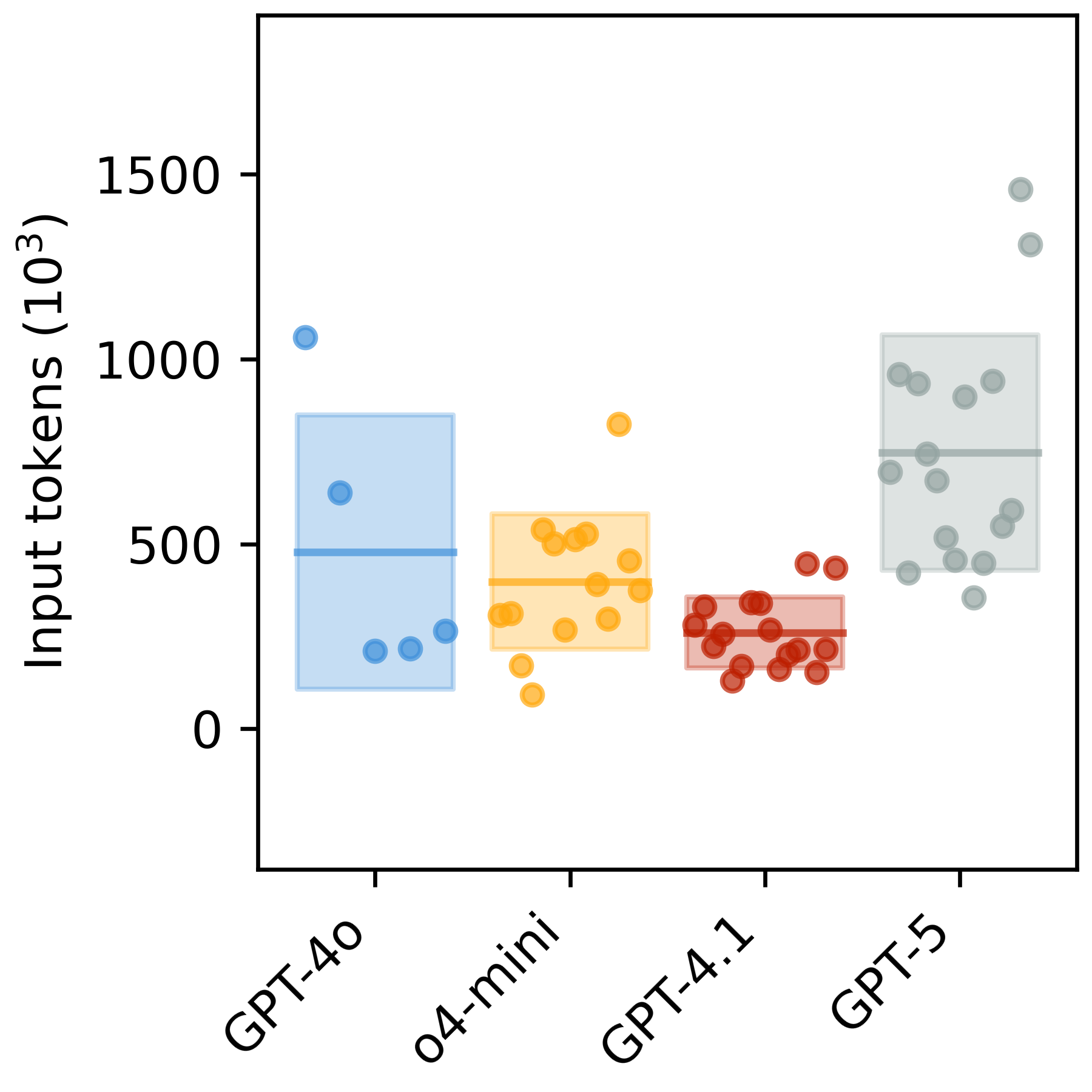}\quad
    \includegraphics[width=0.29\linewidth]{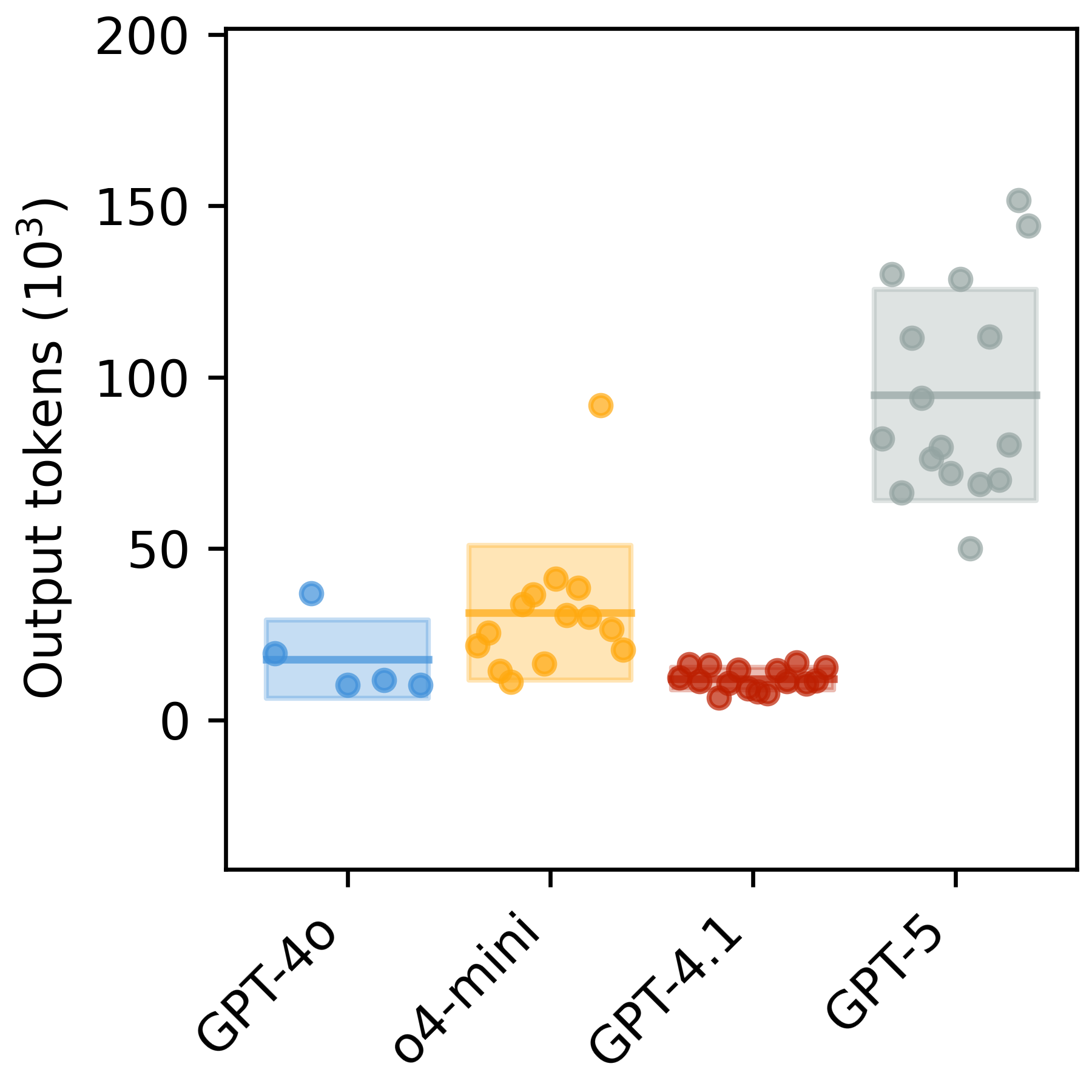}\quad
    \includegraphics[width=0.29\linewidth]{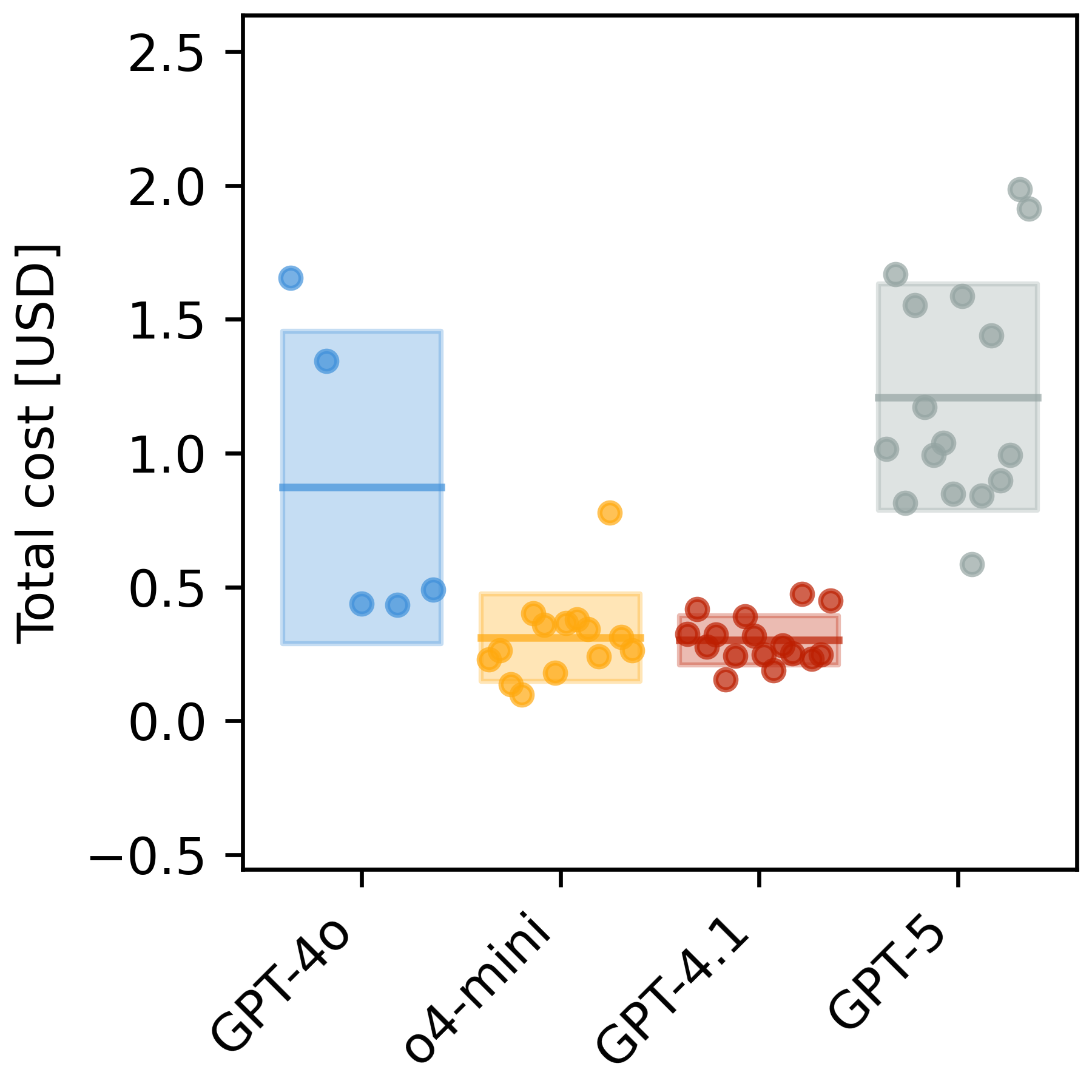}
    \caption{Comparison of four OpenAI models with the ML prompt across different high-level behavior metrics: number of calls, response time, completion time, input tokens, output tokens and total cost (see Appendix~\ref{app:metrics} for the exact definition of these quantities). Only successful runs are shown. The mean is marked with a line and the one standard deviation with a shaded box.}
    \label{fig:results:gen-fun-perf:hlvl-behav}
\end{figure}

Given the successfully completed runs, Fig. \ref{fig:results:gen-fun-perf:hlvl-behav} summarizes other high-level functionality metrics. GPT-5 immediately sticks out as being different compared to the other models. While it has the lowest number of calls, it appears to be doing more with the ones it has: it uses the most tokens, and takes the longest time to finish (roughly double the time as the other models), mainly due to the longer response times of the API. Due to the large number of output tokens it produces, GPT-5 is also the most expensive\footnote{The most expensive part of the runs is the cost of output tokens. GPT-5 and GPT-4o have the highest cost for output tokens at 10~USD/1M tokens, with GPT-4.1 a close second at 8~USD/1M tokens and o4-mini comparatively cheap at 4.40~USD/1M tokens. The cost for input tokens varies between 1.10 and 2.50~USD/1M tokens, with cached inputs being cheaper at 0.13 to 1.25~USD/1M tokens. \cite{openaiprices}} of the four models. In two of its five successful runs, GPT-4o uses a large number of calls, and therefore a large number of input tokens, making these two runs quite expensive as well. GPT-4.1 and o4-mini perform quite similarly across the board with respect to these metrics, apart from the number of output tokens where o4-mini produces more (possibly due to it being a reasoning model).

\subsubsection{Coding behavior}
\label{sec:coding_behavior}
\begin{figure}[h]
    \centering
    \includegraphics[width=0.29\linewidth]{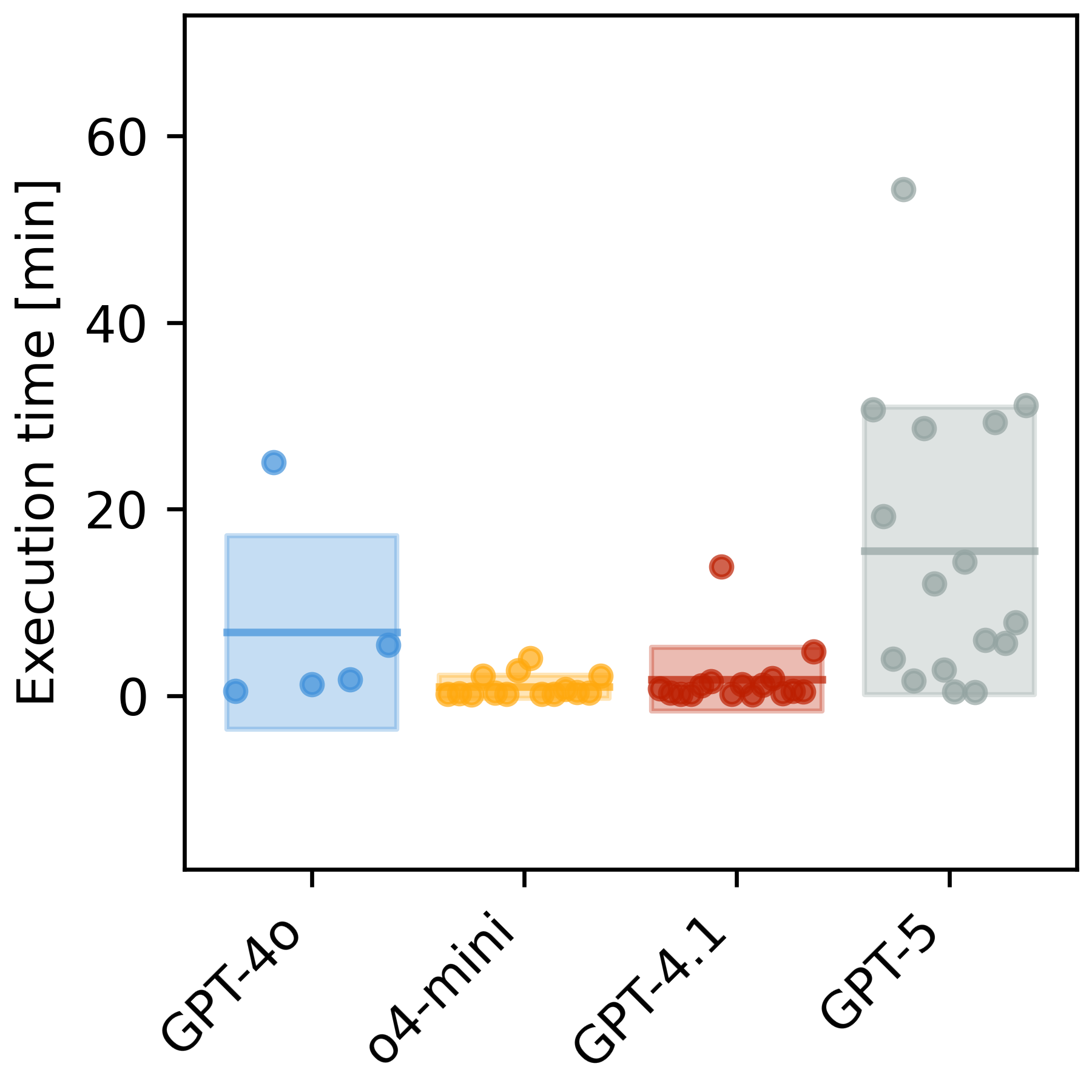}\quad
    \includegraphics[width=0.29\linewidth]{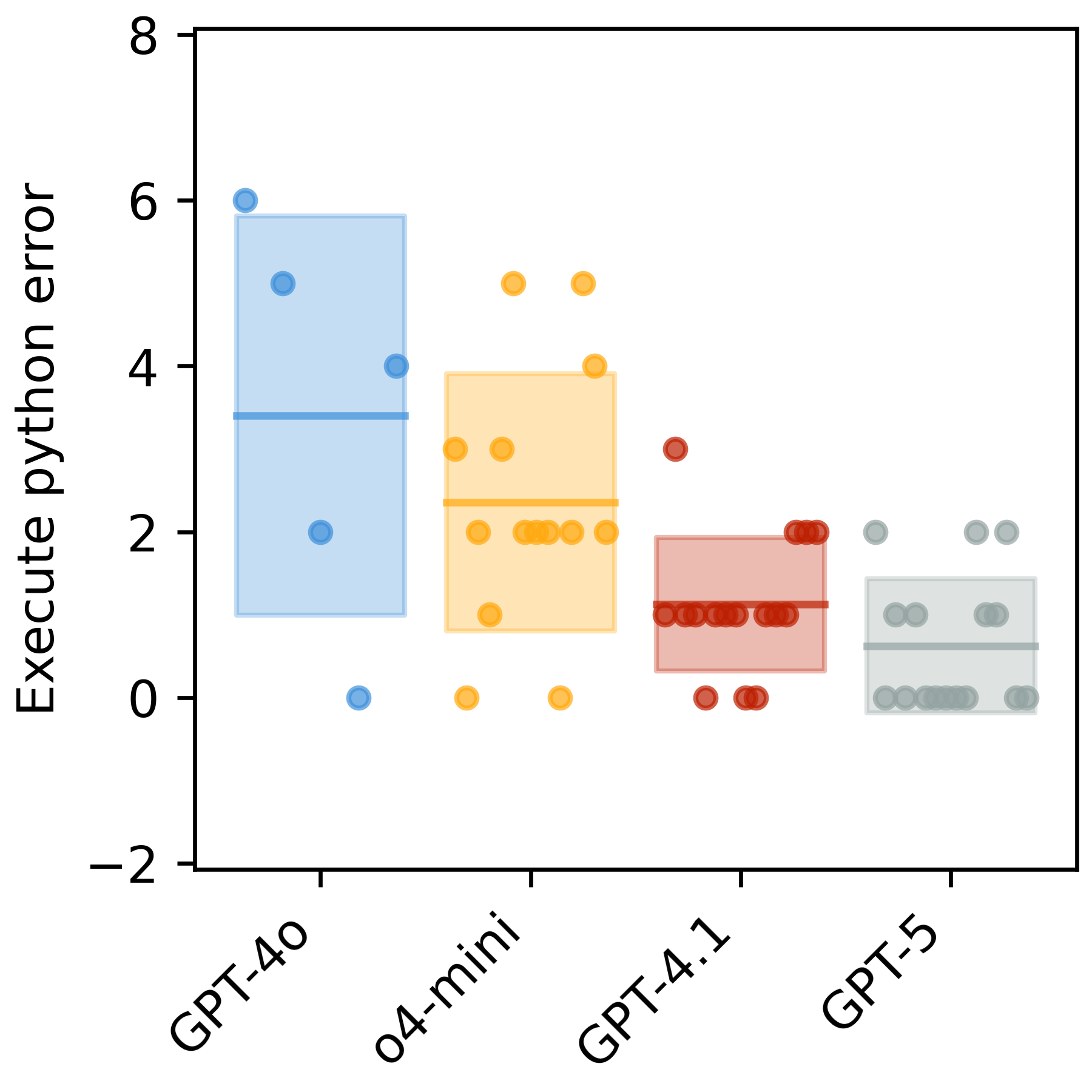}\quad
    \includegraphics[width=0.29\linewidth]{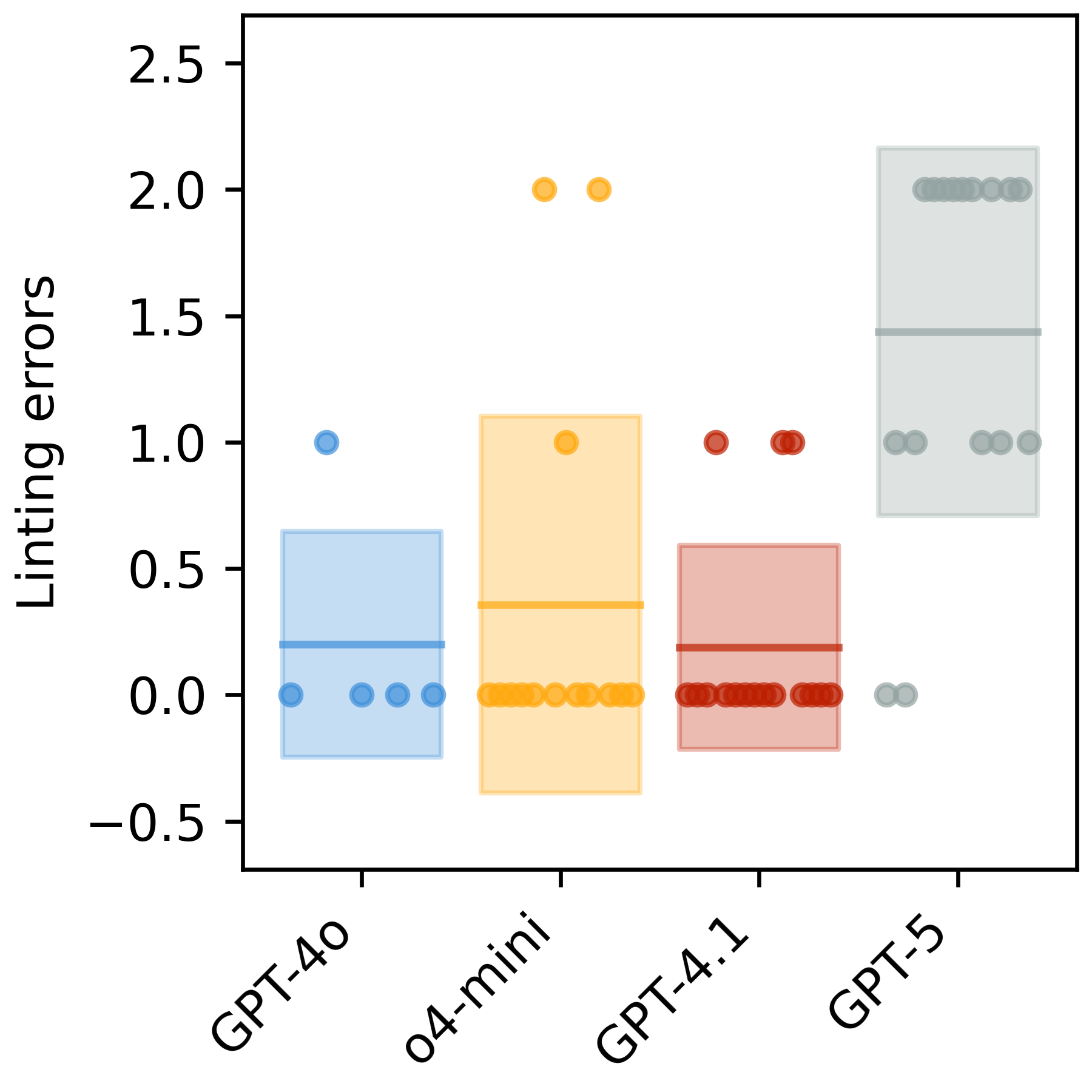}\\
    \includegraphics[width=0.29\linewidth]{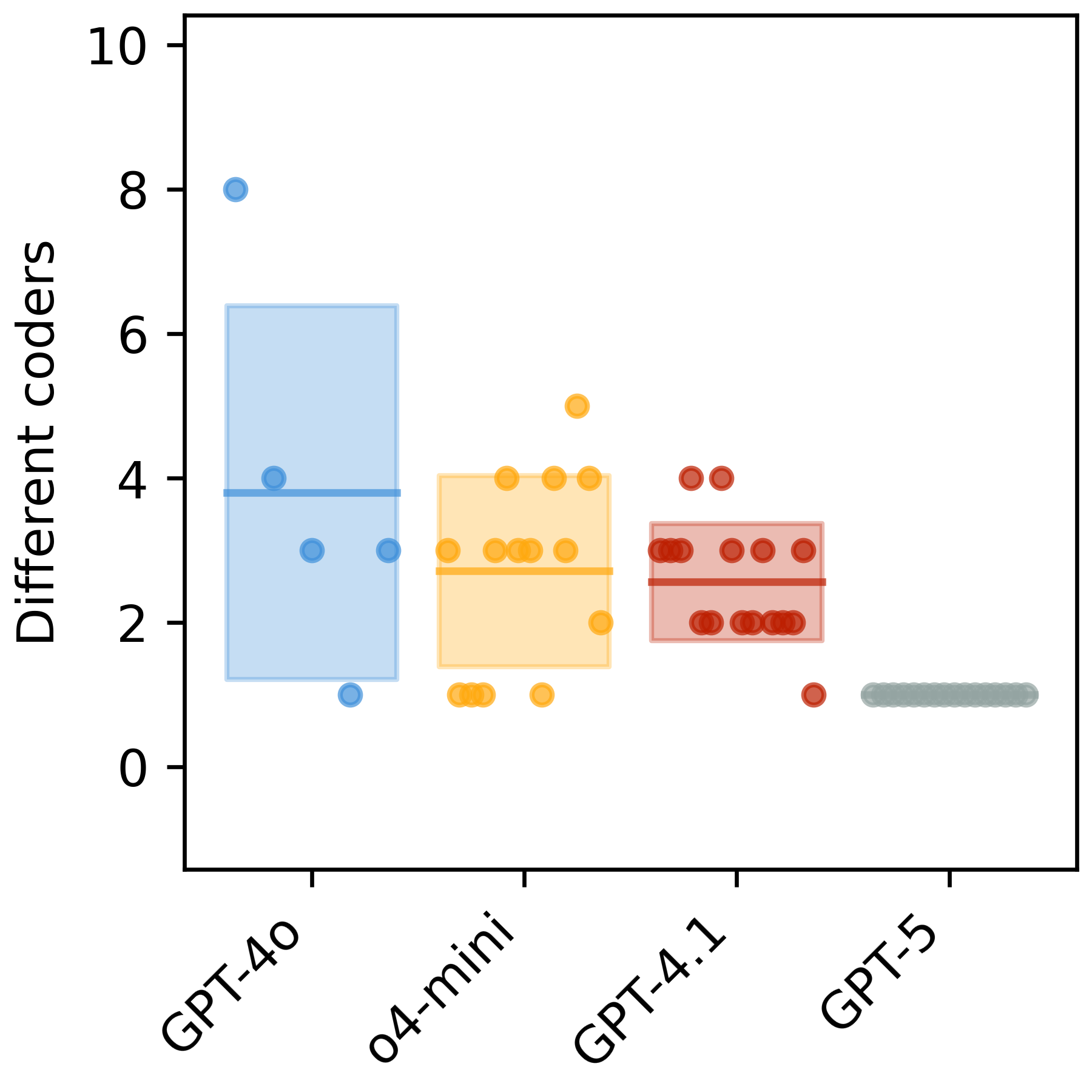}\quad
    \includegraphics[width=0.29\linewidth]{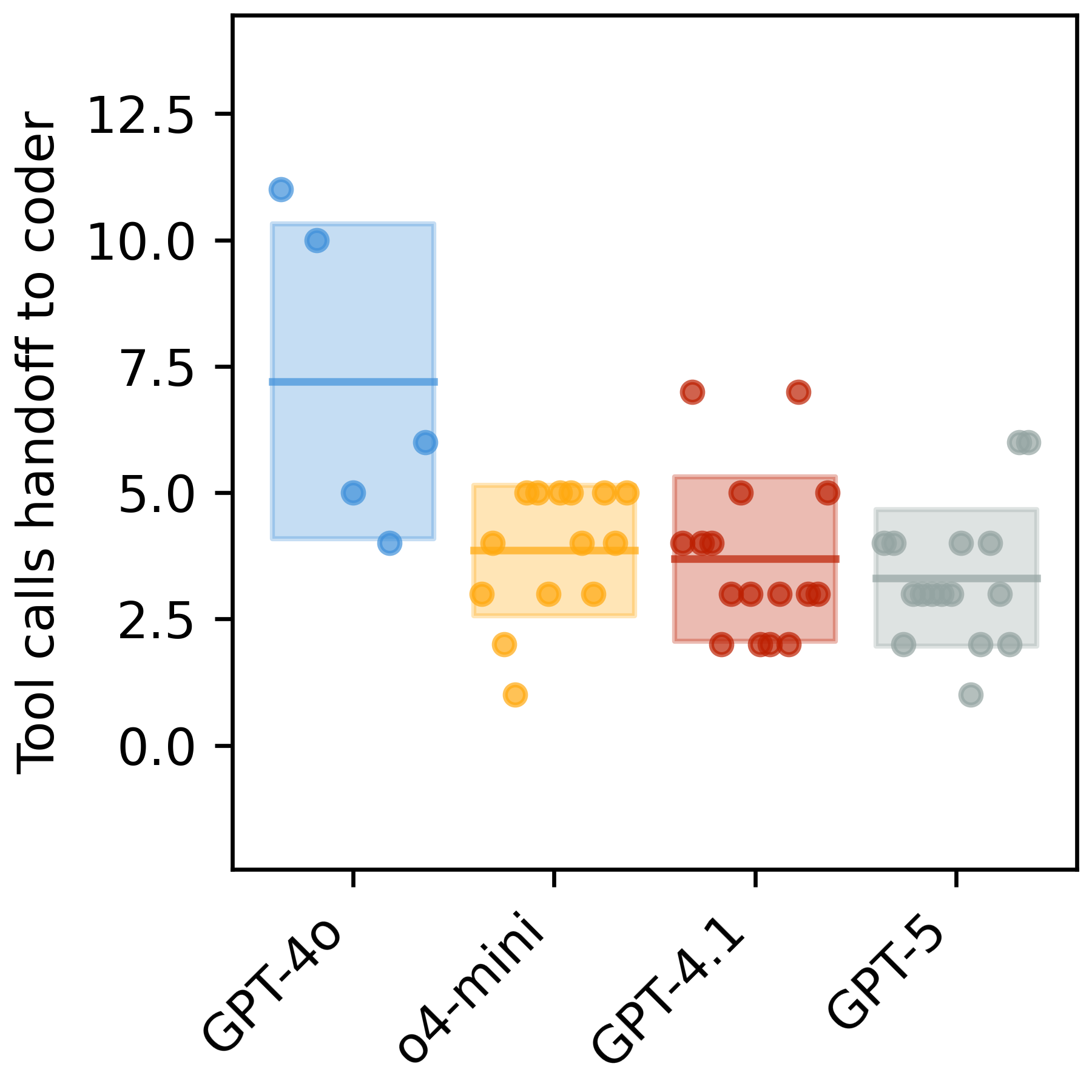}\quad
    \includegraphics[width=0.29\linewidth]{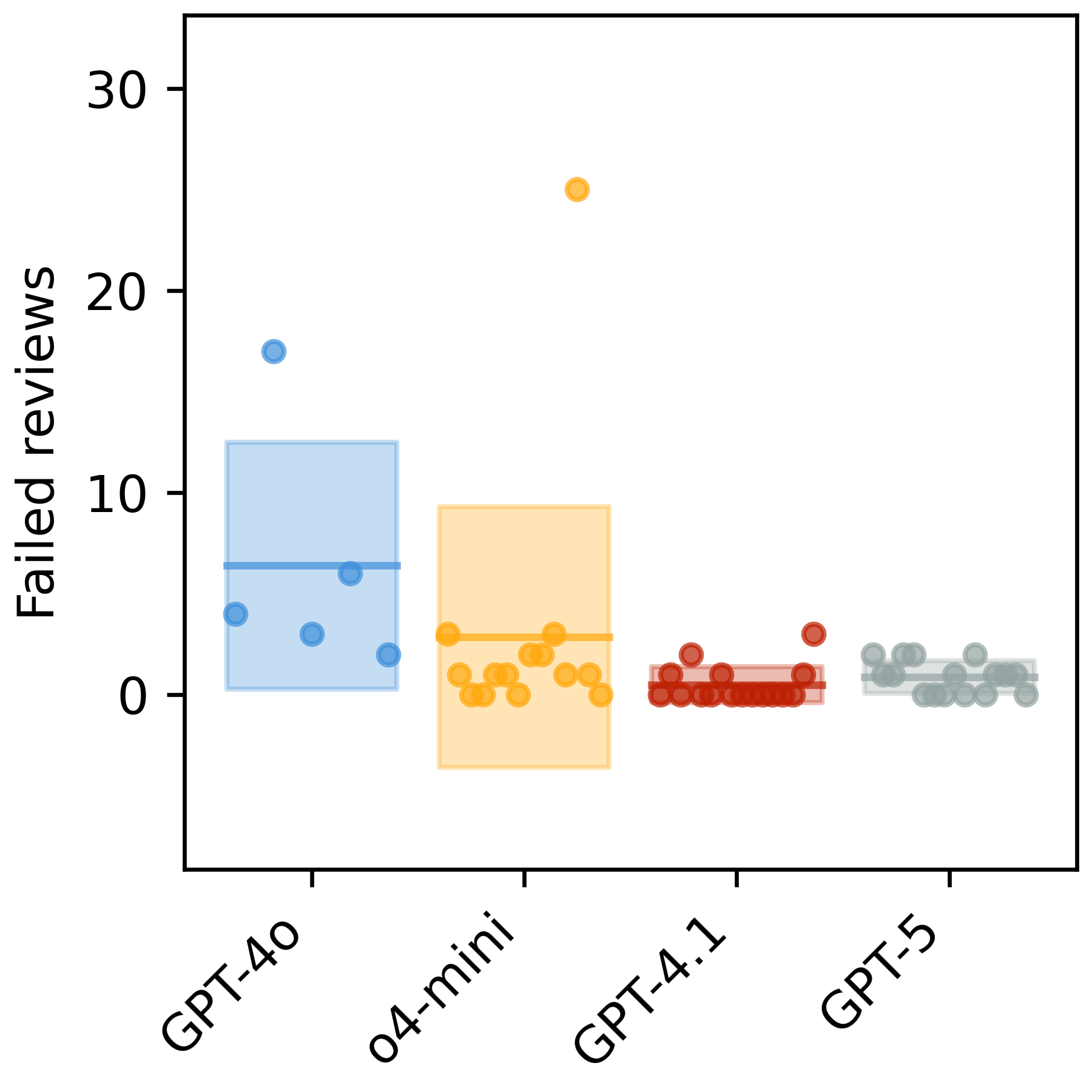}
    \caption{Comparison of four OpenAI models with the ML prompt across different metrics related to coding: execution time, execute python errors, lint errors, number of different coders used, tool calls handoff to coder and failed reviews (see Appendix~\ref{app:metrics} for the exact definition of these quantities). Only successful runs are shown. The mean is marked with a line and the one standard deviation with a shaded box.}
    \label{fig:results:gen-fun-perf:cod-behav}
\end{figure}

The quality of the analysis code that the agent produces is essential for it to be able to provide meaningful results. Fig.~\ref{fig:results:gen-fun-perf:cod-behav} shows an overview of the metrics related to coding. As mentioned in the previous section, GPT-5 takes the longest to finish. This is partly due to its longer response time (waiting for the API to respond), but here we see that it also spends more time executing code than the other models. Interestingly, GPT-5 has more linting errors\footnote{34 of the 47 linting errors were related to \verb|numpy.random.RandomState|. These are not actual coding errors, but rather the inability of the linter to ``see'' that this module exists.} than the other models, but it is also writing around 4 times as much code. 
Once the linting errors are dealt with and the code runs, GPT-5 has the fewest execution errors of all models. It should be noted that execution errors might occur not only because the code is faulty, but because the researcher is using it erroneously, for example by neglecting to provide command-line arguments (despite receiving instructions to do so from the coder). GPT-5 also stays with a single coder, whereas the other models call 3-4 coders on average during the run. Among these metrics relating to coding behavior, GPT-4o is at the other end of the spectrum compared to GPT-5. It suffers the highest number of execution errors, and sends the most tool calls for handoff to the coder (possibly in order to fix all the errors). It also has the highest number of failed code reviews of the four models. Although the statistics are limited due to GPT-4o failing 11 of the 16 runs, it seems from these metrics to not be the best option for coding.  GPT-4.1 and o4-mini seem to again be performing similarly to one another.

\subsection{LHCO R\&D without labels}
In this section we will look at the physics performance of the different LLMs. We once again choose to evaluate them all on the ML prompt. The agent had access to data from the signal region, but no access to labels or feedback loops. We will compare the approaches chosen by the agents to those employed by the human participants of LHCO for the Black Box 1 dataset. 

\subsubsection{Quantitative physics performance}
\label{sec:physics_performance}

\begin{figure}[h]
    \centering
    \includegraphics[width=0.29\linewidth]{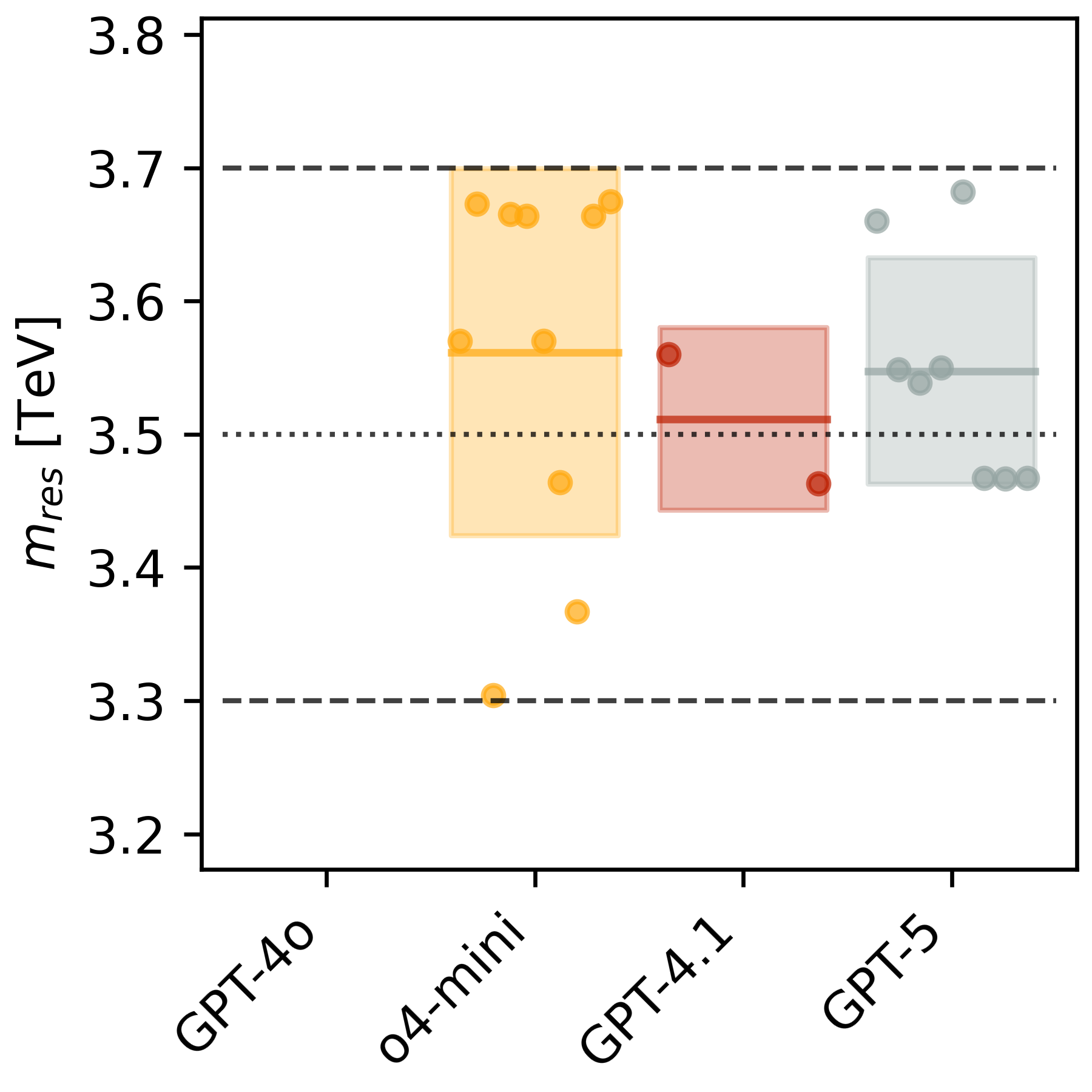} \quad
    \includegraphics[width=0.29\linewidth]{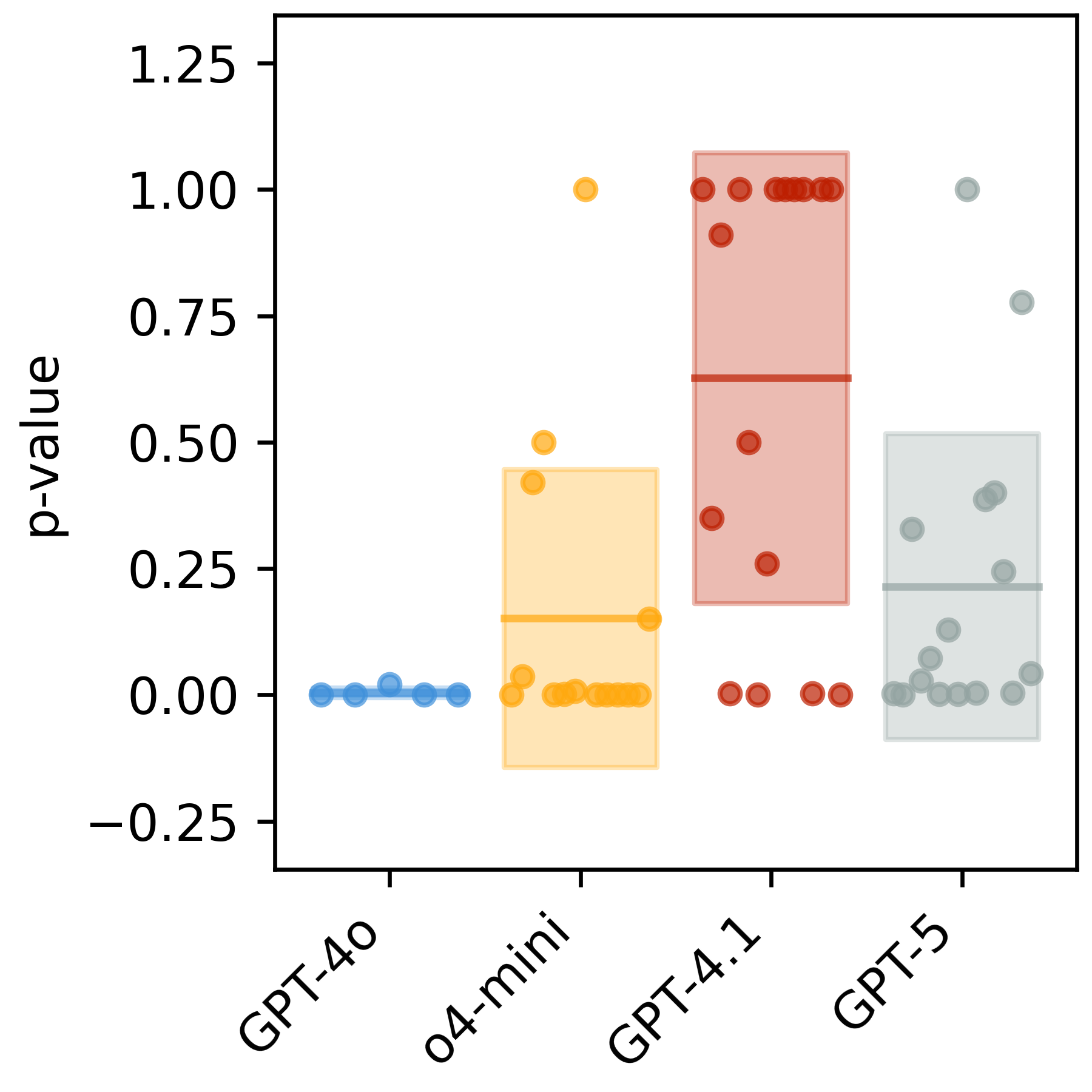}\\
    \includegraphics[width=0.29\linewidth]{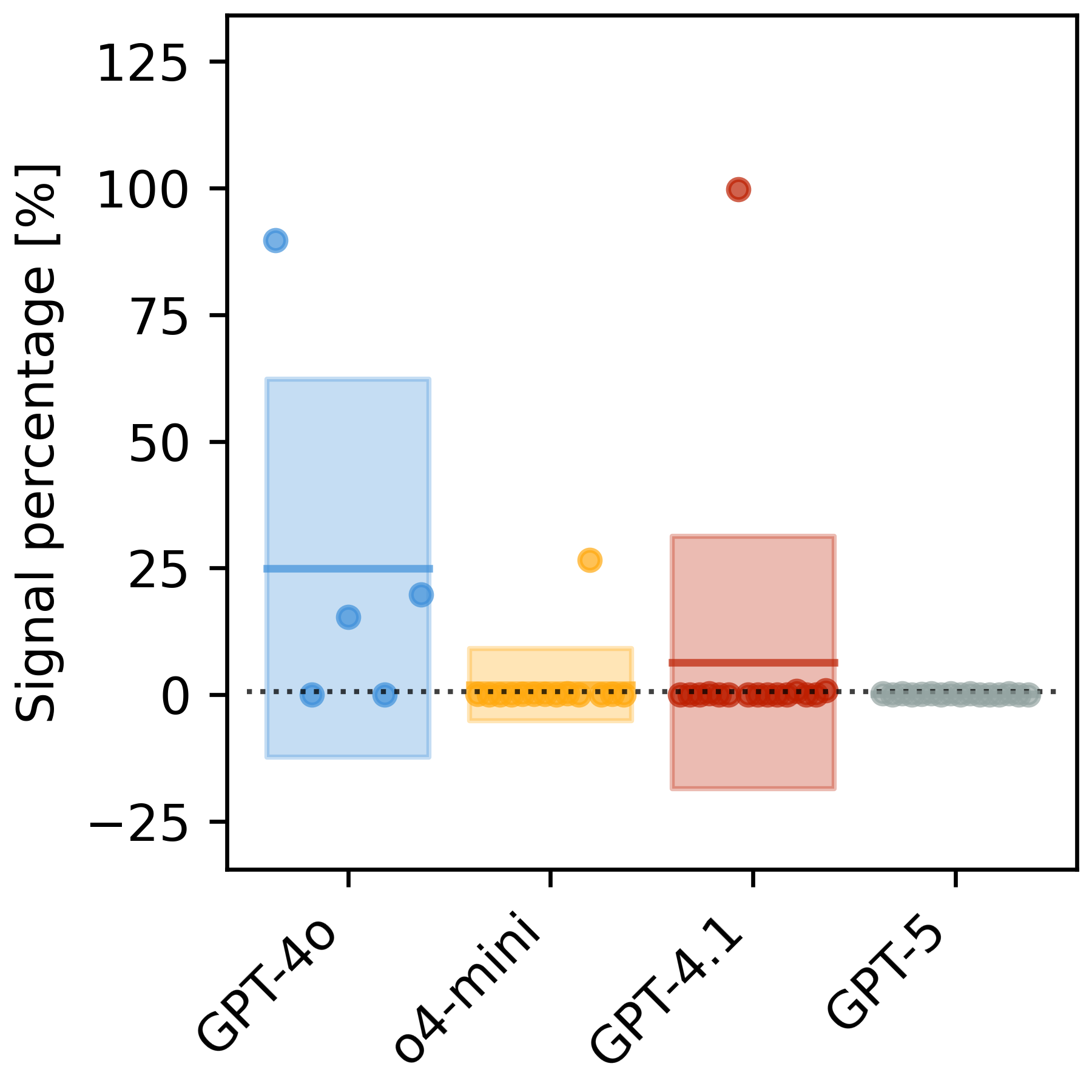} \quad
    \includegraphics[width=0.29\linewidth]{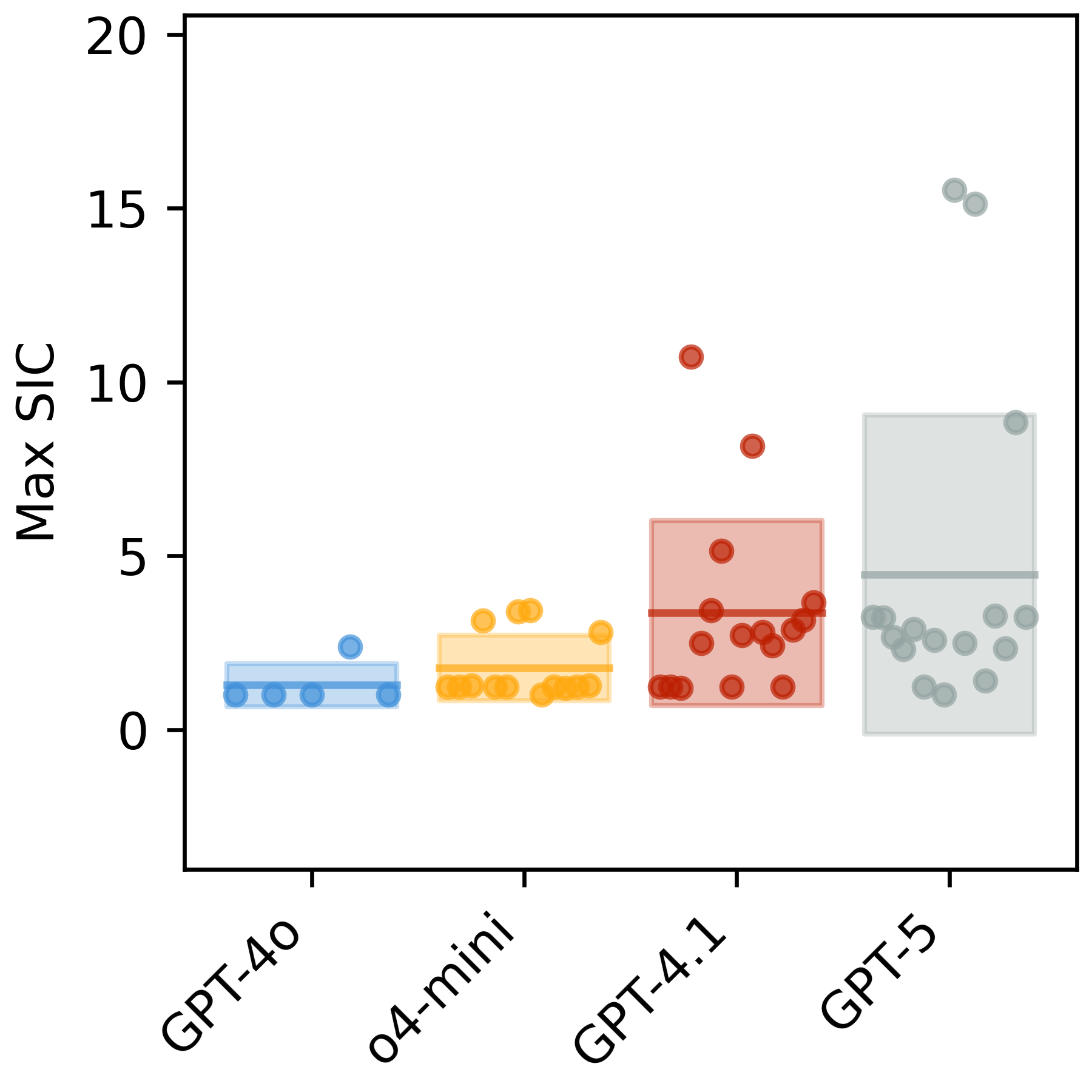}
    \caption{Reported values of $m_{res}$, p-value and signal percentage on the LHCO R\&D dataset, plus the max SIC calculated after the agent submitted its scores and ended its run. Only successful runs are shown. The mean is marked with a line and the one standard deviation with a shaded box. The ML prompt is used in all runs.}
    \label{fig:LHCO_blind_3_questions_performance}
\end{figure}

Fig.~\ref{fig:LHCO_blind_3_questions_performance} shows the reported values for the mass of the resonance (true value 3.5~TeV), the p-value under the background-only hypothesis, the estimated signal percentage (true value 0.6\%), and the max SIC. The first thing we notice is that the non-reasoning models (GPT-4o and GPT-4.1) rarely report the resonance mass. In the case of GPT-4.1, it often simply does not find any new physics and hence does not report a mass. It is consistent across all metrics in that it has very few p-values close to zero (that would indicate a tension between the data and the background-only hypothesis), and only reports a non-zero signal percentage in four of its runs. In one of those, a signal percentage close to 100~\% is reported, this was caused by the coder including mixed data and background labels as features during training, leading to the classifier perfectly distinguishing both datasets. Although the researcher caught that the result was highly unlikely, it misidentified the issue and the same mistake appeared during a rerun of the analysis. GPT-4o, on the other hand, reports very small p-values and a generally high signal percentage, which means that you would expect it to report a resonance mass. However, it seems to mostly ignore this, and either report something else (like the average of its scoring) or nothing at all. Of the two reasoning models, o4-mini and GPT-5, the latter gets closer to the true resonance mass of 3.5~TeV, although two outliers pull the average up to the same level as o4-mini. In total, it reports 9 nonzero signal percentage values, while o4-mini reports 11. In terms of maximum SIC a clear difference can be observed. GPT-4o and o4-mini fail to achieve maximum SICs above 5 while the averages for GPT-4.1 and GPT-5 are close 5. These comparatively high averages are caused by outliers around 10 (GPT-4.1) or 15 (GPT-5). But even with these outliers ignored, GPT-4.1 and GPT-5 outperform GPT-4o and o4-mini. This also shows that the performance of GPT-4.1 and GPT-5 is mostly similar, while the main difference in average performance is caused by the outliers. Taking a closer look at the outliers reveals the differences between GPT-4.1 and GPT-5: while both employ Boosted Decision Tree-based algorithms, GPT-5 is excluding $m_{JJ}$ from the training to avoid mass sculpting and performs a bump hunt.

In the light of the performance difference seen between the LLMs, it is interesting to also examine the cost. As mentioned in Section~\ref{sec:high_level_behavior}, GPT-5 is the most expensive model, with high costs for output tokens in particular. In the same section, we also saw that GPT-5 tends to produce much more output tokens than the other models. Part of this can be explained by GPT-5 writing much longer code, as noted in Section~\ref{sec:coding_behavior}. Table \ref{tab:sic_and_cost} shows the average max SIC achieved by the different models, the average number of tokens used, and the average cost of the runs. 
\begin{table}[!ht]
    \centering
    \renewcommand\arraystretch{1.3}
    \begin{tabular}{|>{\centering\arraybackslash}m{0.2\linewidth}|>{\centering\arraybackslash}m{0.25\linewidth}|>{\centering\arraybackslash}m{0.25\linewidth}|>{\centering\arraybackslash}m{0.25\linewidth}|}
    \hline
        Model & Avg. max SIC & Avg. tokens used ($10^3$) & Avg. cost [USD] \\ \hline
        GPT-4o & 1.28 & 495.41 & 0.87 \\ \hline
        o4-mini & 1.78 & 429.37 & 0.31 \\ \hline
        GPT-4.1 & 3.36 & 272.50 & 0.30 \\ \hline
        GPT-5 & 4.46 & 841.65 & 1.21 \\ \hline
    \end{tabular}
    \caption{Average values of max SIC, number of tokens used, and total cost for the different LLMs across their respective runs.}
    \label{tab:sic_and_cost}
\end{table}

\subsubsection{Method comparison between agents and humans}
\label{sec:LHCO_blind_methods}
Table \ref{tab:LHCO_blind:methods} lists the different approaches the agents used. Both reasoning models, o4-mini and GPT-5, frequently perform so-called bump hunts. This is a standard technique in particle physics, where the analysis assumes that the signal is localized in some feature (often the invariant mass). A scan can then be performed over smaller windows of this feature, comparing the data to some background sample or template. Usually some cut or selection is applied in order to minimize the amount of background. If the signal is large enough, it should be visible as a ``bump'' against the background. GPT-5 also mentions ``Classification Without Labels'' (CWoLa)~\cite{Metodiev:2017vrx,Collins:2018epr,Collins:2019jip}, a well-known approach for weak supervision within high-energy physics, in the majority of its runs. CWoLa uses data from outside of the signal region to produce a background template, and then trains a classifier to distinguish between the data and the background template. When GPT-5 mentions this method, it understands that it does not need to create a background template, because it has already been provided with a background sample. In the particle physics anomaly detection community, weak supervision with a perfect background sample is often referred to as the ``idealized anomaly detector'', a term that GPT-5 has not picked up on. It often combines the weak supervision anomaly scores with a bump hunt, sliding over the dijet mass range, as you would in an anomaly detection analysis~\cite{CMS:2024nsz,ATLAS:2021weak}. In several runs it also knows that it should exclude this mass from the anomaly detection step, as it might otherwise lead to ``sculpting'' of the background in the bump hunt step, again, something well known in the anomaly detection literature~\cite{Benkendorfer:2020gek, Hallin:2022eoq}. Note that CWoLa, and many other papers on weak supervision~\cite{Nachman:2020lpy, Hallin:2021wme, Raine:2022hht, Golling:2022nkl}, were published before the knowledge cut-off dates of all of the models we have studied, so all models could in principle be familiar with it. In the case of o4-mini, it sometimes combines weak supervision with a bump hunt, but it also frequently performs a bin-by-bin bump hunt, without first calculating any anomaly scores.

\begin{table}[!ht]
    \centering
    \renewcommand\arraystretch{1.3}
    \begin{tabular}{|l|>{\centering\arraybackslash}m{0.12\linewidth}|>{\centering\arraybackslash}m{0.12\linewidth}|>{\centering\arraybackslash}m{0.12\linewidth}|>{\centering\arraybackslash}m{0.12\linewidth}|}
    \hline
        ~ & GPT-4o & o4-mini & GPT-4.1 & GPT-5 \\ \hline
        Successful runs & 4 & 14 & 16 & 16 \\ \hline
        With bump hunt & 0 & 10 & 2 & 16 \\ \hline
        CWoLa & 0 & 2 & 0 & 12 \\ \hline
        $m_{JJ}$ excluded for training  & 0 & 0 & 0 & 8 \\ \hline \hline
        No ML & 0 & 1 & 0 & 0 \\ \hline
        Unsupervised & 4 & 8 & 5 & 2 \\ \hline
        Weakly supervised & 1 & 5 & 11 & 15 \\ \hline \hline
        RandomForestRegressor & 1 & 0 & 0 & 0 \\ \hline
        IsolationForest & 2 & 8 & 5 & 2 \\ \hline
        Gaussian mixture & 0 & 0 & 0 & 1 \\ \hline
        RandomForestClassifier & 1 & 1 & 3 & 0 \\ \hline
        GradientBoostingClassifier & 1 & 4 & 5 & 10 \\ \hline
        HistGradientBoostingClassifier & 0 & 0 & 2 & 4 \\ \hline
        Logistic regression & 0 & 0 & 1 & 1 \\ \hline
    \end{tabular}
    \caption{Approaches chosen by the different models running the ML prompt. Some agents combined several methods, which is why the total number of methods does not always match the number of successful runs. Agents using CWoLa are doing weak supervision with the provided background sample, as discussed in the text. 
    }
    \label{tab:LHCO_blind:methods}
\end{table}

The approaches suggested by the human participants for Black Box~1 in the LHCO challenge covered supervised, weakly supervised, and unsupervised strategies~\cite{Kasieczka:2021xcg}. Representative ideas were dimensionality reduction (PCA) and autoencoder methods that flagged outliers via reconstruction loss; conditional density estimation that formed approximate likelihood-ratio tests to enhance a bump in $m_{JJ}$; weak supervision that contrasted signal-enriched and signal-depleted samples to train a discriminator; sequence/constituent-based models that defined anomaly scores from jet substructure; and a manual bump-hunt baseline. Several methods localized the resonance near the true mass, with density estimation also giving the most accurate ancillary observables.

The main difference between the strategies employed by the agents compared to those of humans is that the agent most frequently selects classical tabular learners (e.g. Isolation Forest, GradientBoosting, HistGradientBoosting). We attribute this in part to the lack of deep-learning libraries like Keras and PyTorch in the execution environment, but with the earliest knowledge cutoff among these models being October 2023, we cannot rule out the fact that they could be aware that previous work has indicated that Boosted Decision Tree-based algorithms are an ideal classifier method for tabular data~\cite{grinsztajn2022treebasedmodelsoutperformdeep, Finke:2023ltw}. In contrast to these tabular methods, in the original LHCO study human teams deployed a broad toolbox that included both deep learning methods (autoencoders/VAEs, normalizing flows, and other neural network classifiers) and more traditional approaches (e.g.\ boosted decision trees, likelihood/density-ratio estimation with dimensionality reduction). Thus, resorting to classical machine learning methods is consistent with established practice, especially on tabular feature sets like ours, while LHCO also showcased gains from deep generative/classifier network models when architectures could exploit richer inputs.

\subsubsection{Impact of prompt phrasing}
We now proceed to investigate the impact of varying the prompt. All runs in this section are based on GPT-4.1, since it offers good balance between performance and cost. The first part compares the prompts that introduce changes to the \textit{task} (Default, Ideas, ML), the second part compares the different paraphrasings of the ML prompt. The FBL prompts will be discussed separately in Section~\ref{sec:fbl}.

\begin{figure}[h]
    \centering
    \includegraphics[width=0.29\linewidth]{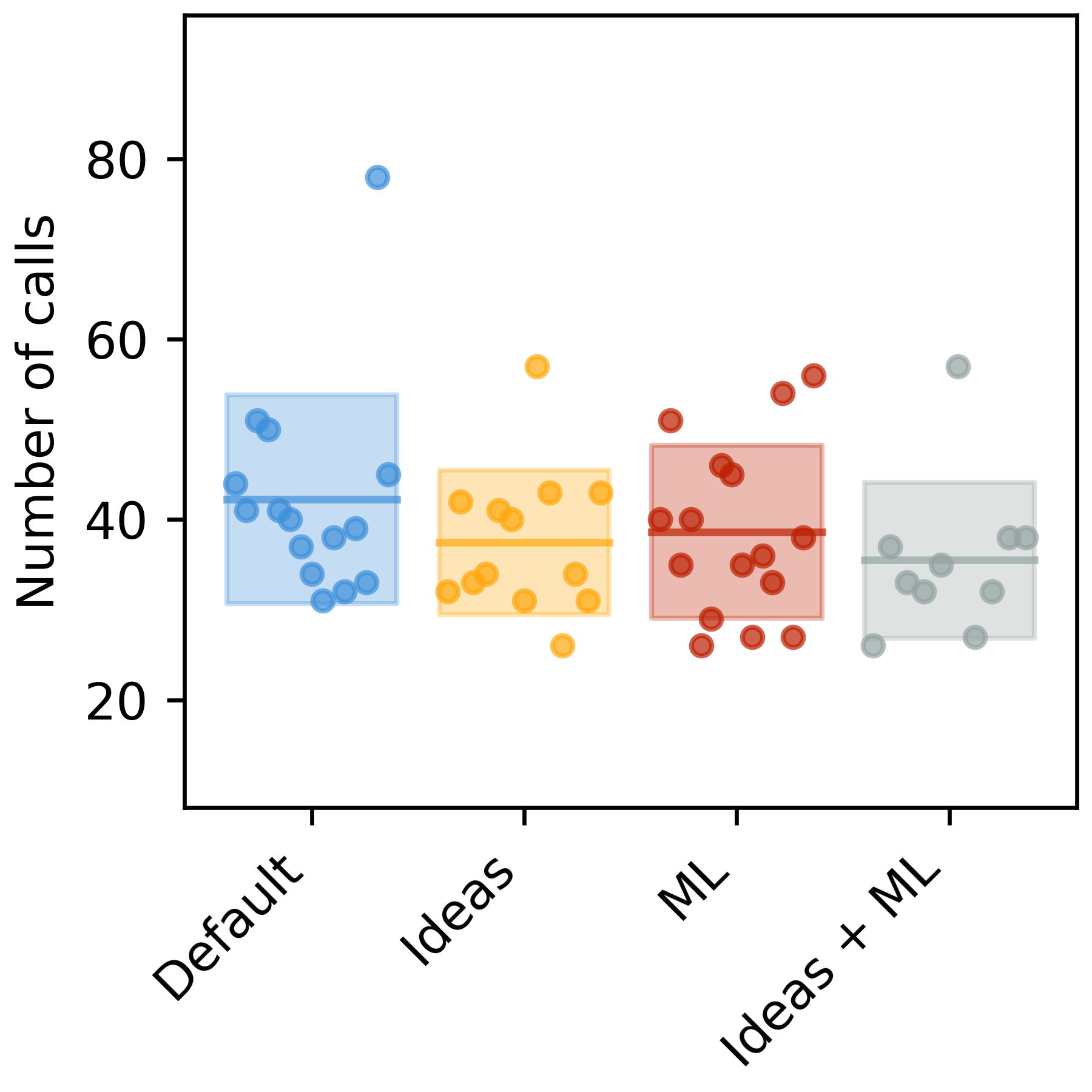} \quad
    \includegraphics[width=0.29\linewidth]{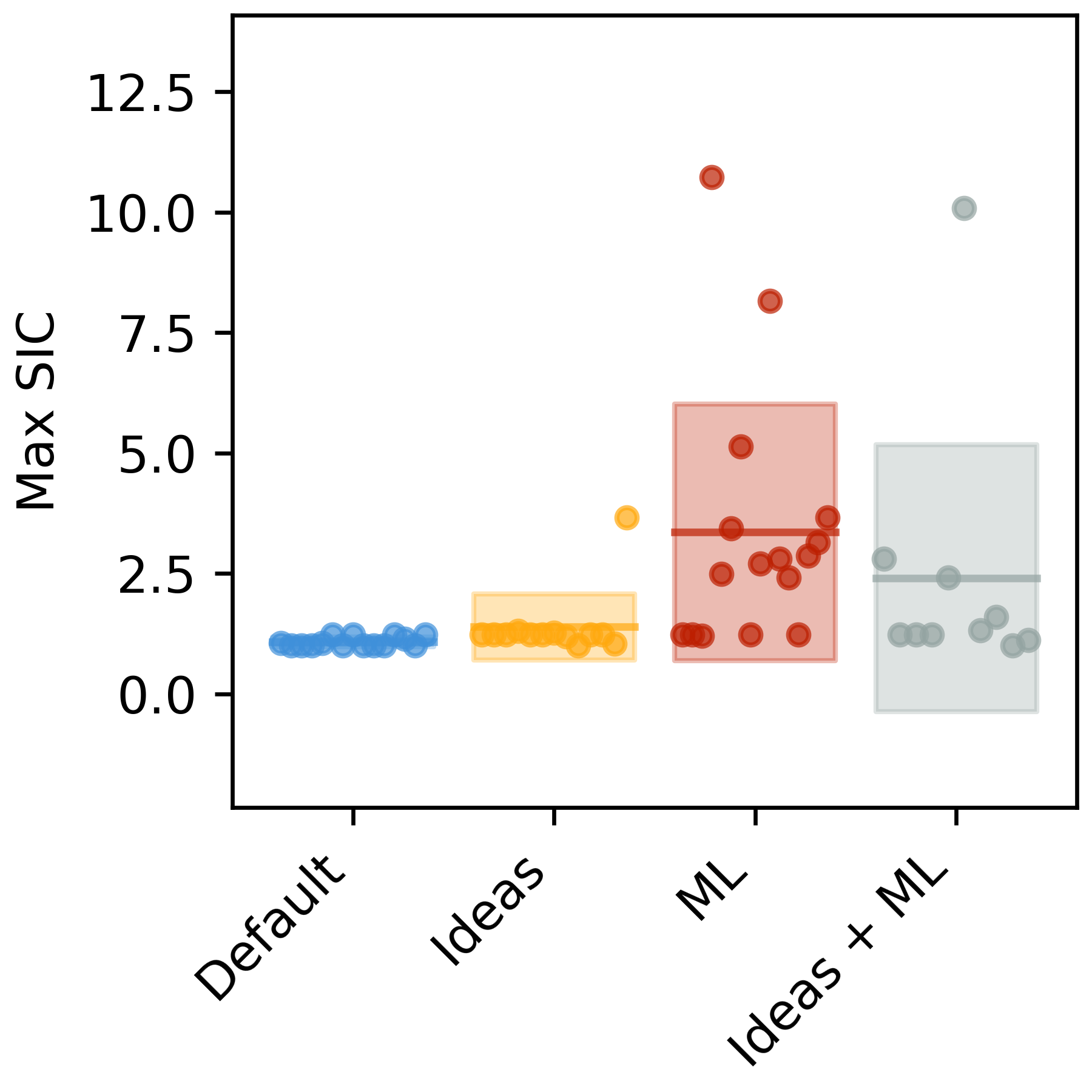}
    \caption{Comparison of the different prompt strategies using GPT-4.1, showing number of calls and max SIC. Only successful runs are shown. The mean is marked with a line and the one standard deviation with a shaded box.}
    \label{fig:results:nonfbl_prompts}
\end{figure}

Fig.~\ref{fig:results:nonfbl_prompts} shows the number of calls and max SIC for the Default, Ideas, ML, and Ideas+ML prompts. The number of successful runs were overall high, with 15, 16, 13 and 10 successful runs (out of 16) respectively. The most common failure mode was erroneous formatting of the score file. The number of calls is fairly consistent across all prompts. When it comes to the max SIC, the default prompt performed the worst, with the Ideas prompt closely following. Without the ML hint, it seems that encouraging the agent to think outside of the box, and pick the most ``unique'' idea, only resulted in minor improvements compared to the default. Interestingly, in the majority of the runs with the Ideas prompt, the agent starts by examining the data, and then decides that given the similarity between all 1D distributions and summary statistics, the most promising approach is unsupervised anomaly detection (most often using IsolationForest). Paradoxically, it seems as if encouraging the agent to explore different ideas led it to consistently choose the same method. This leads to very similar performance between runs, with the one outlier resulting from the agent changing its approach to a RandomForest after not being able to find an anomaly with the IsolationForest. Adding the ML hint to the Ideas prompt helped: for the Ideas+ML prompt we saw a much wider range of chosen methods, both supervised and unsupervised. As seen in the plots, this also led to a wider spread of the max SIC, although the regular ML prompt still led to a slightly higher average. In both cases, the average maximum SIC is pulled up by outliers. As discussed in Section~\ref{sec:physics_performance}, the outlier in the ML case comes from using a Boosted Decision Tree, whereas the the ML+Ideas one comes from using a RandomForest.

\begin{figure}[h]
    \centering
    \includegraphics[width=0.3\linewidth]{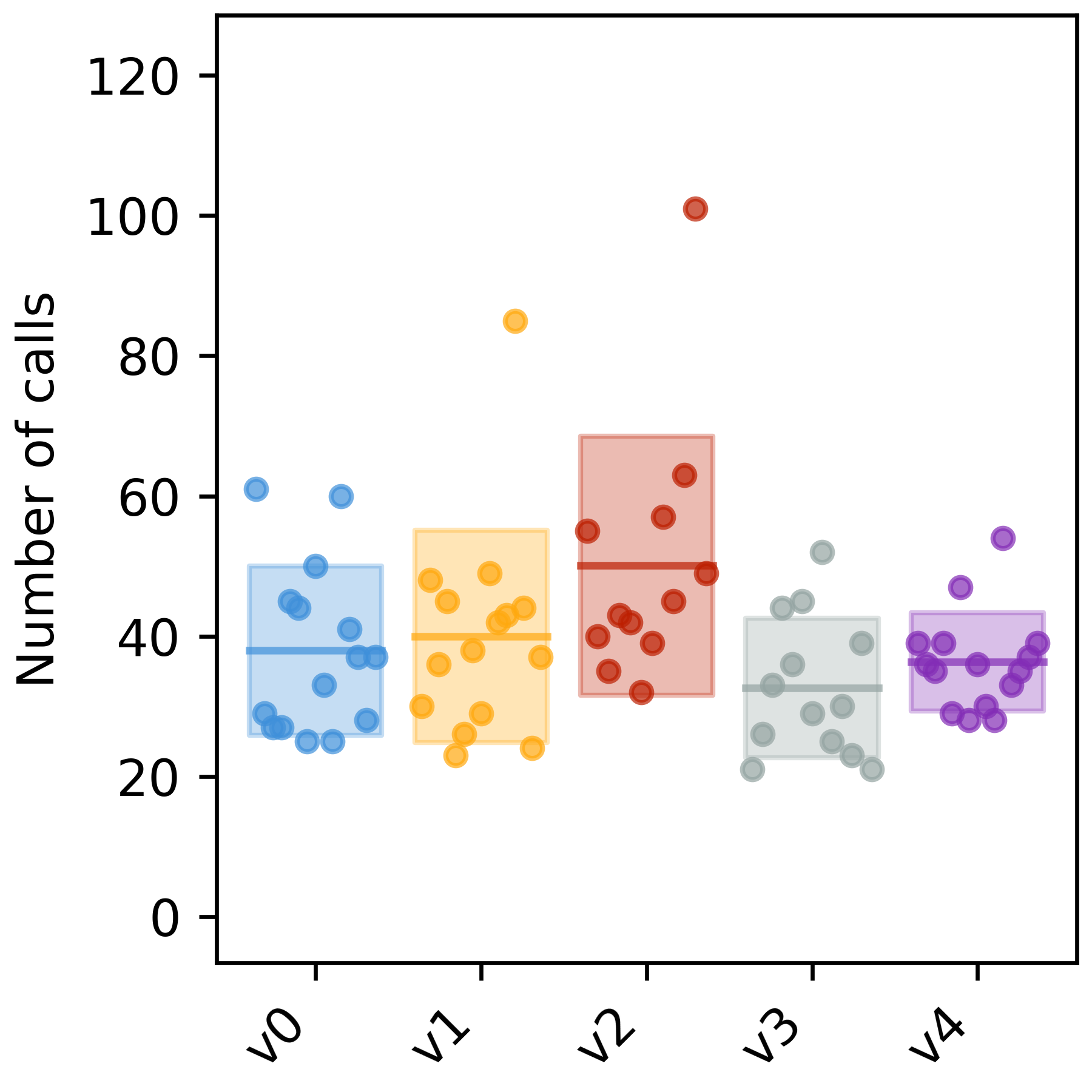}\quad
    \includegraphics[width=0.3\linewidth]{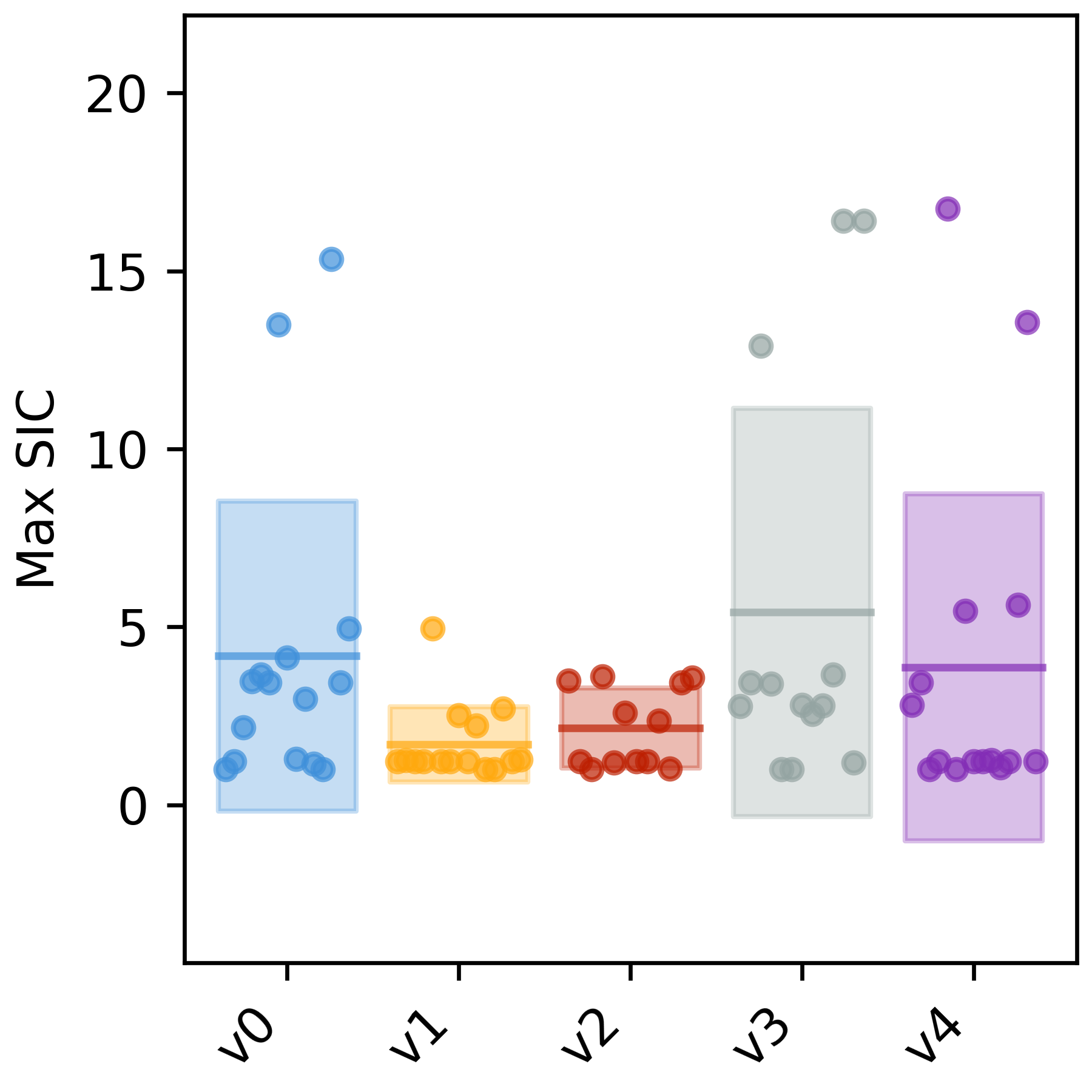}
    \caption{Comparison of the different prompt paraphrasings using GPT-4.1, showing the number of calls and the maximum SIC. Only successful runs are shown. The mean is marked with a line and the one standard deviation with a shaded box.}
    \label{fig:results:prompt-stability}
\end{figure}

Fig.~\ref{fig:results:prompt-stability} shows the number of calls and max SIC for the various prompt paraphrasings. Here the standard ML prompt is referred to as version zero, v0. With all prompts, the agent was able to successfully finish its runs most of the time, with at most 4 out of 16 runs failing. We view this as a rather normal fluctuation, considering the performance in the previous sections. It seems that providing the agent with more context, including flattery (v0: ``best physics AI'') and potential threats (v3/v4: ``survival of humanity/the LHC''), helps its physics performance. Interestingly, there is no correlation between physics performance and number of calls. The high-performing prompts v3 and v4 use the least number of calls, while v2 uses the most. The one outlier run in v2 in regards to number of calls is due to the agent neglecting to call the \verb|end_project| tool, and instead sending responses that it is finished and is ready to take on the next task from the user. It is not clear why this happened, as the full context in this run is smaller than the context window of the model. The run was finally terminated when the agent reached its maximum number of calls.  

\subsubsection{Impact of submission date}
\label{subsec:stability}
\begin{figure}[h]
    \centering
    \includegraphics[width=0.29\linewidth]{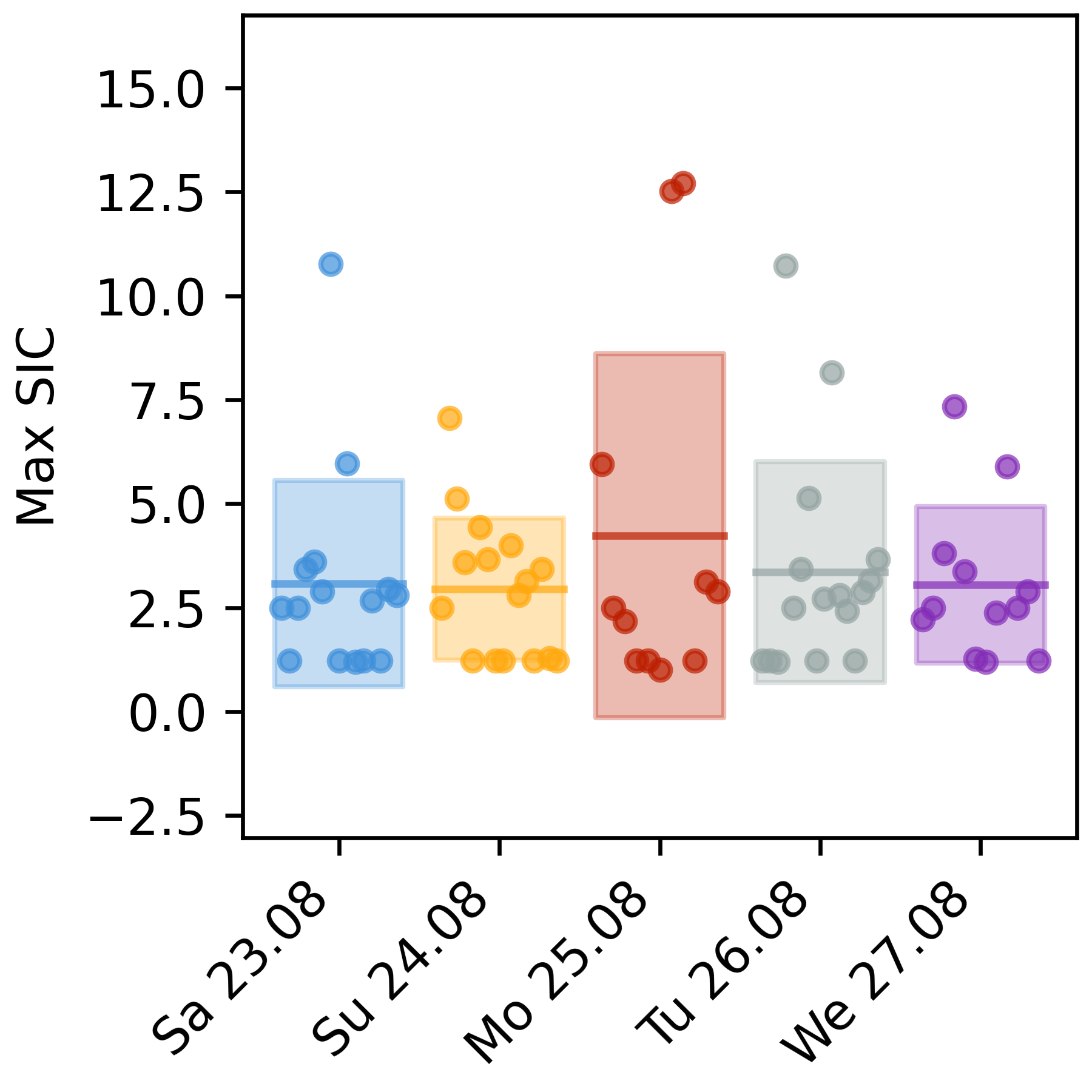}\quad
    \includegraphics[width=0.29\linewidth]{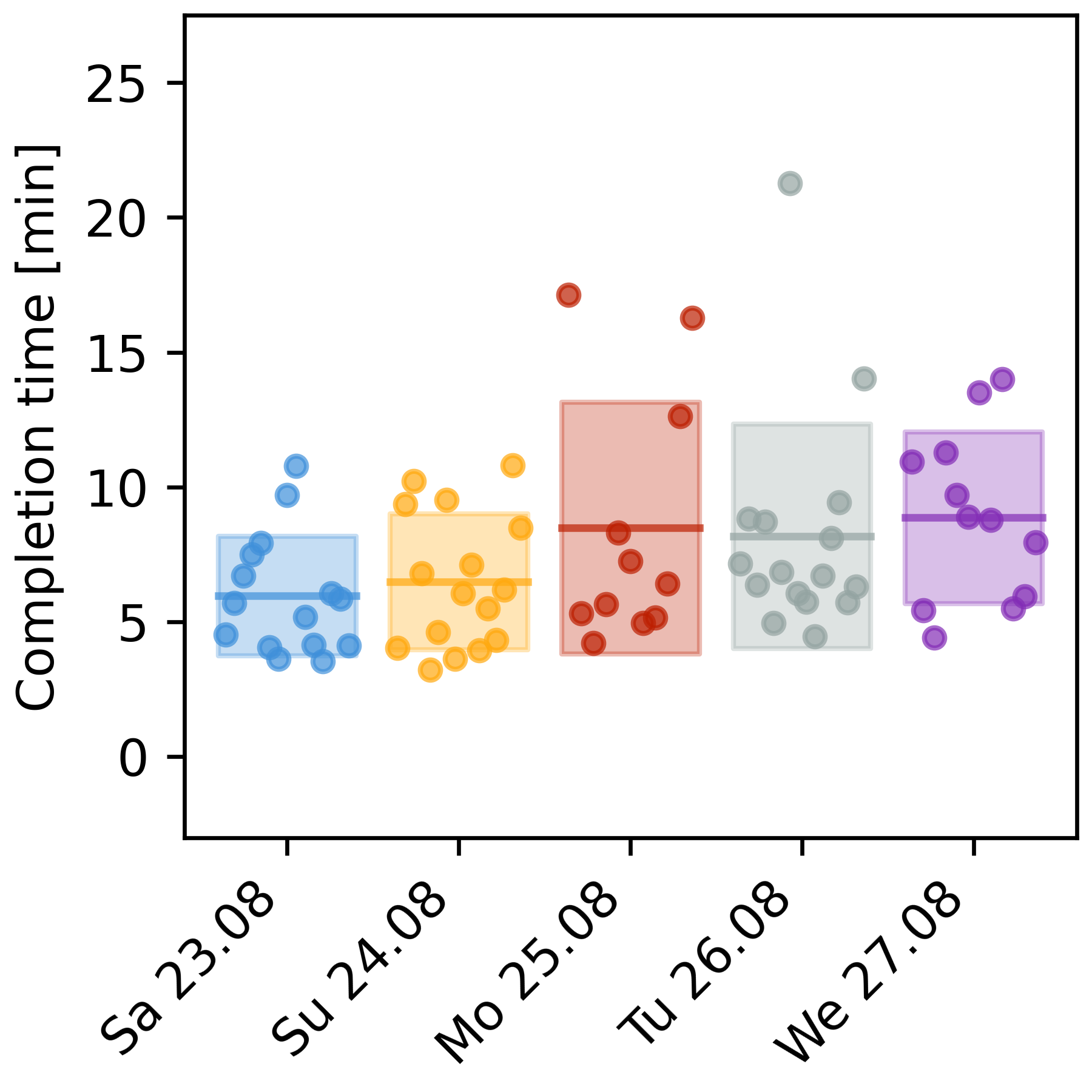}\\
    \includegraphics[width=0.29\linewidth]{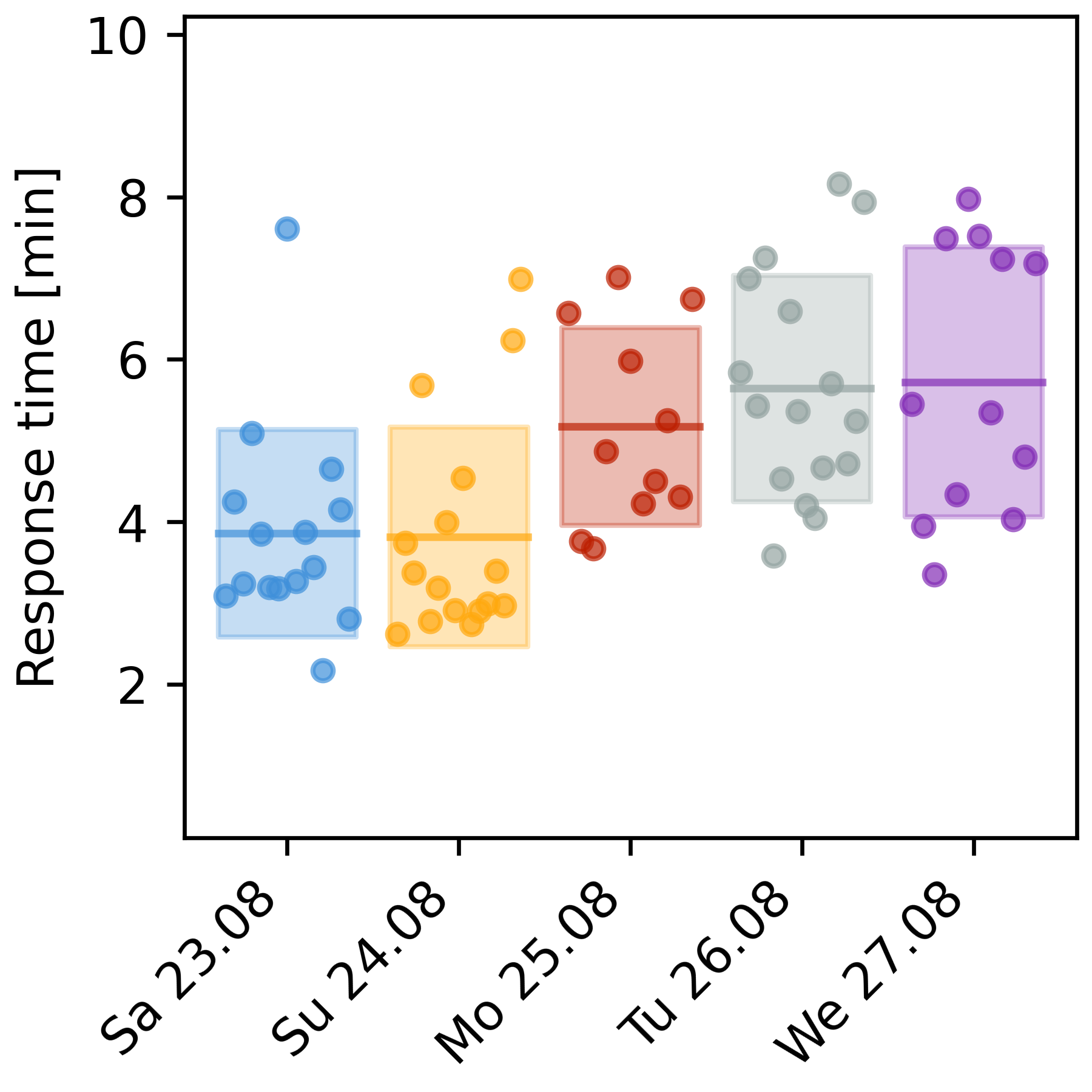}\quad
    \includegraphics[width=0.29\linewidth]{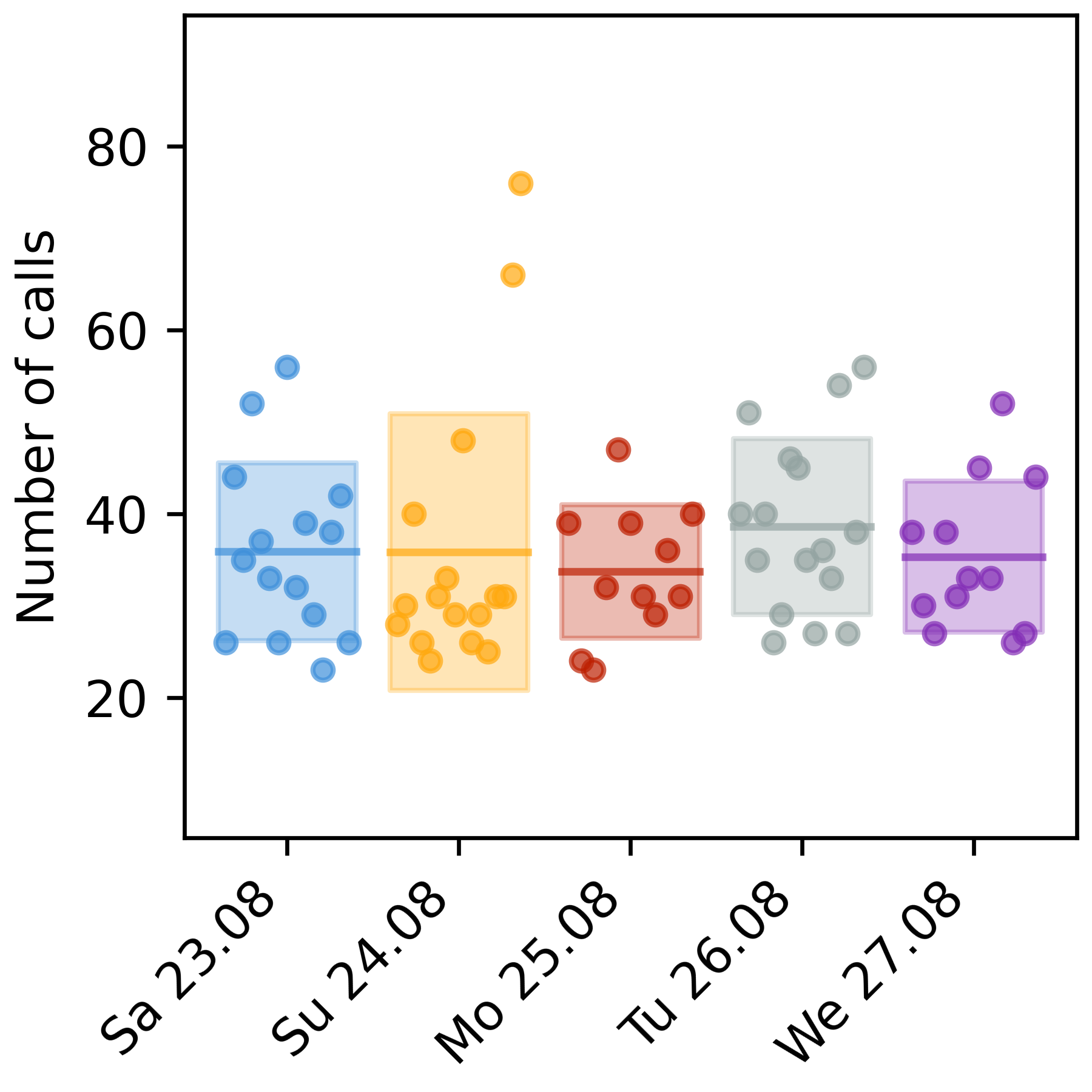}
    \caption{Comparison of the performance of GPT-4.1 with the ML prompt over the course of several days. Only successful runs are shown. The mean is marked with a line and the one standard deviation with a shaded box.}
    \label{fig:results:time-stability}
\end{figure}

One issue when using commercial LLMs is ensuring reproducibility. OpenAI does not offer any seeding in its current API, which introduces challenges to a systematic investigation of the capabilities of its models. We test the stability of the performance over several days, using GPT-4.1 and the ML prompt. Ideally, the average performance should not vary too much over time. We ran the setup over 5 days, including one weekend. Fig.~\ref{fig:results:time-stability} shows the result of this experiment. We see that the response time, and therefore also completion time, is lower on the weekend, presumably due to less load on the API on those days. Apart from this, the mean of the metrics is relatively stable over the tested days. Monday and Wednesday stand out with 5 and 4 failed runs, respectively, though considering the corresponding failure rate in the previous sections, we view this as a relatively normal fluctuation.  

\subsection{LHCO R\&D with feedback loop}
\label{sec:fbl}
\begin{figure}[h]
    \centering
    \includegraphics[width=0.45\linewidth]{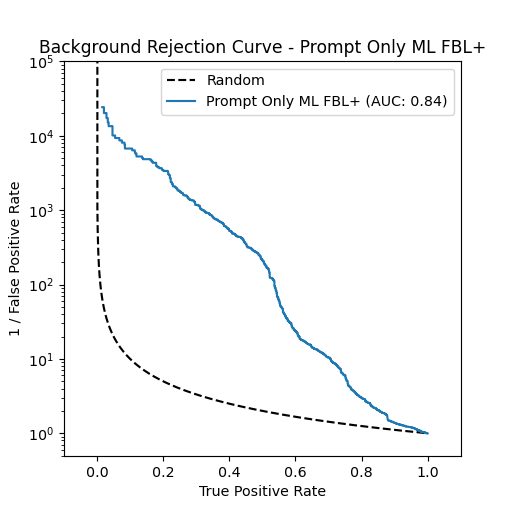} \enspace
    \includegraphics[width=0.45\linewidth]{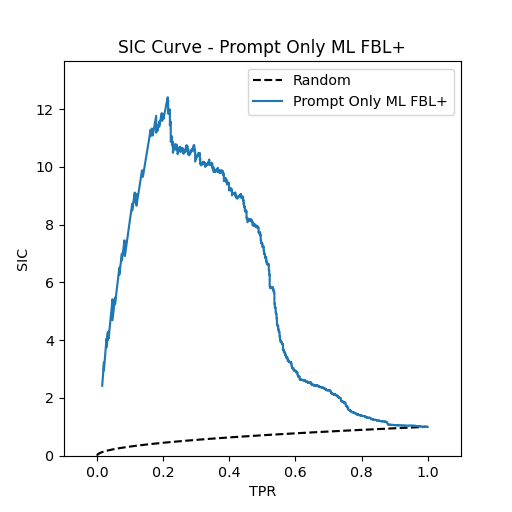}
    \caption{The agent is provided with two plots as part of the feedback it receives: a background rejection curve and a SIC curve. The location of the peak of the SIC is relevant: if it peaks at very low signal efficiency (TPR), the following statistical analysis might be plagued by high uncertainties.}
    \label{fig:feedback_plots}
\end{figure}

Although the agent does not have access to the truth labels for the data, we do. This means that we are able to provide feedback on its result. The runs in this section, all performed with GPT-4.1, include a feedback loop (FBL) of some type. The agent is told that it has access to a feedback tool, and FBL$^+$ means that it additionally has been instructed to try to achieve a max SIC\footnote{See Section~\ref{sec:metrics} for the definition and interpretation of SIC.} of 20. The feedback it received after calling the feedback tool consists of AUC (area under the ROC curve), max SIC and TPR at max SIC, in addition to plots of the background rejection and the SIC curve as shown in Fig.~\ref{fig:feedback_plots}. Note that the agent is not provided with the definition of SIC or its interpretation. 

The model chosen for these tests was GPT-4.1. With the regular FBL prompts, the agent finished its task most of the time (16/16 for ML+FBL, and 13/16 for Ideas+ML+FBL), However, with the FBL$^+$ prompts the agent failed 5-7 runs. The failure modes of the ML+FBL$^+$ prompts range from the agent reaching its maximum allowed number of calls (while attempting to analyze several mass windows), to neglecting to report the required values or even write a final report after concluding that there is no new physics in the data. In the first case mentioned, the agent had reached a SIC of over 15, which in the particle physics anomaly detection community would be seen as high. However, since it had not reached the required 20, it decided to keep going which ultimately resulted in a failed run.

\begin{figure}[h]
    \centering
    \includegraphics[width=0.29\linewidth]{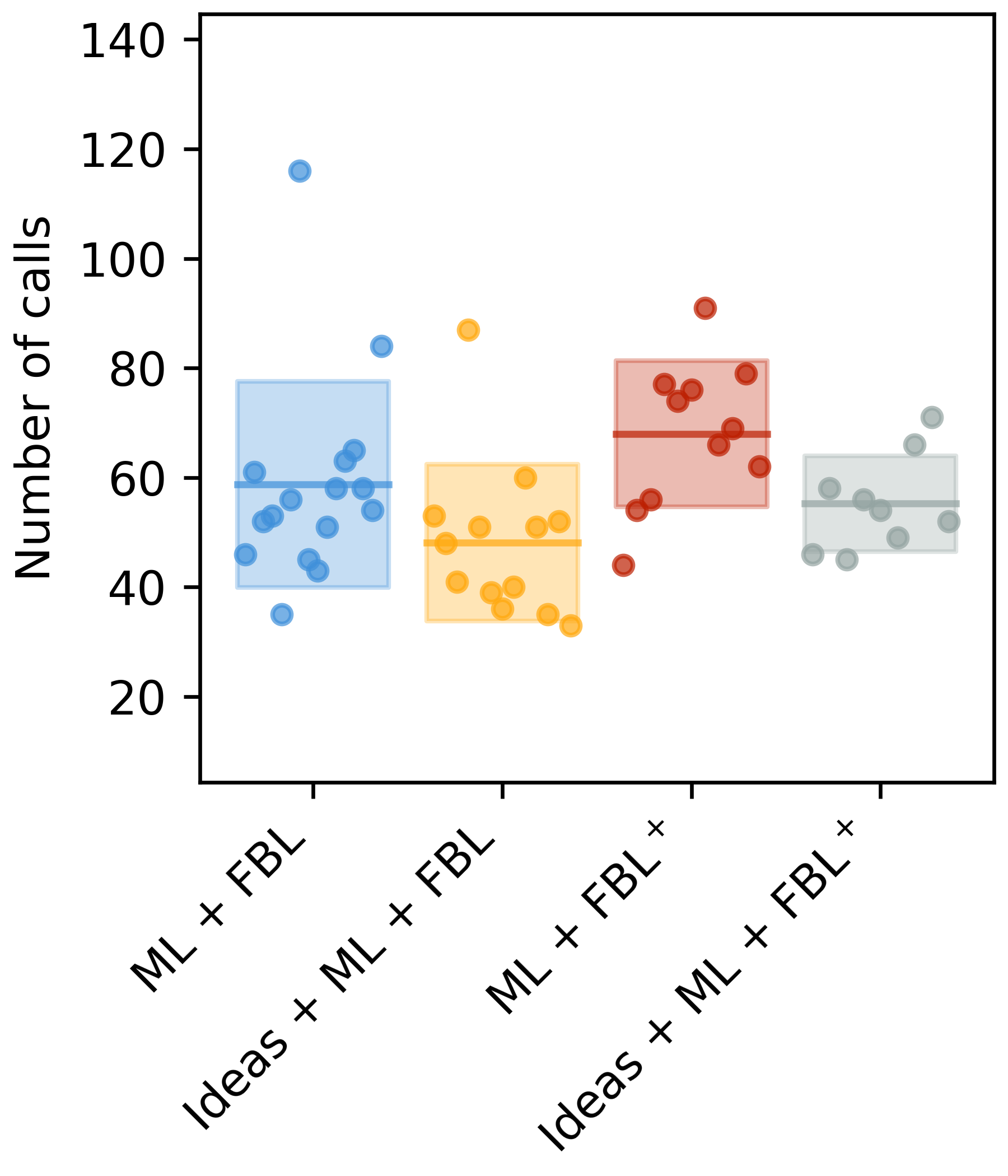} \enspace
    \includegraphics[width=0.29\linewidth]{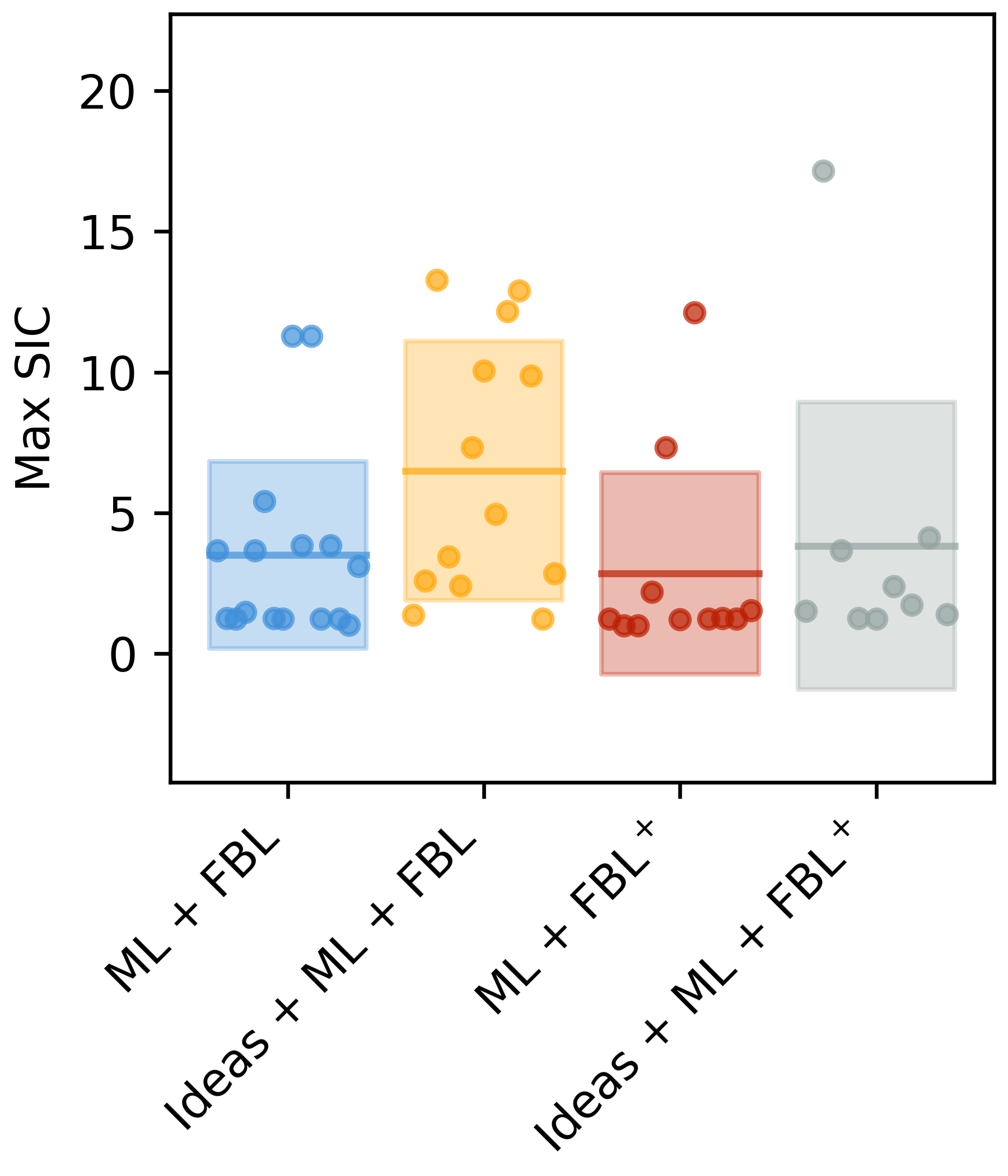}\qquad
    \caption{Comparing the performance of GPT-4.1 on different types of FBL prompts on number of calls and max SIC. The mean is marked with a line and the one standard deviation with a shaded box.}
    \label{fig:results:fbl_prompts}
\end{figure}

Fig.~\ref{fig:results:fbl_prompts} shows the number of calls and the max SIC for the FBL prompts. Compared to the non-FBL case in Fig.~\ref{fig:results:nonfbl_prompts}, where the average number of calls was around 35-40, we here see that the FBL prompts require on average 50 calls or more. This behavior was to be expected, particularly for the FBL$^+$ prompt as it encourages the agent to reach a SIC of 20. Regarding the SIC value, the regular ML prompt (without FBL) is on par with the FBL prompts, apart from the Ideas+ML+FBL which sticks out having a greater spread of SIC values. 

\begin{figure}[h]
    \centering
    \includegraphics[width=0.29\linewidth]{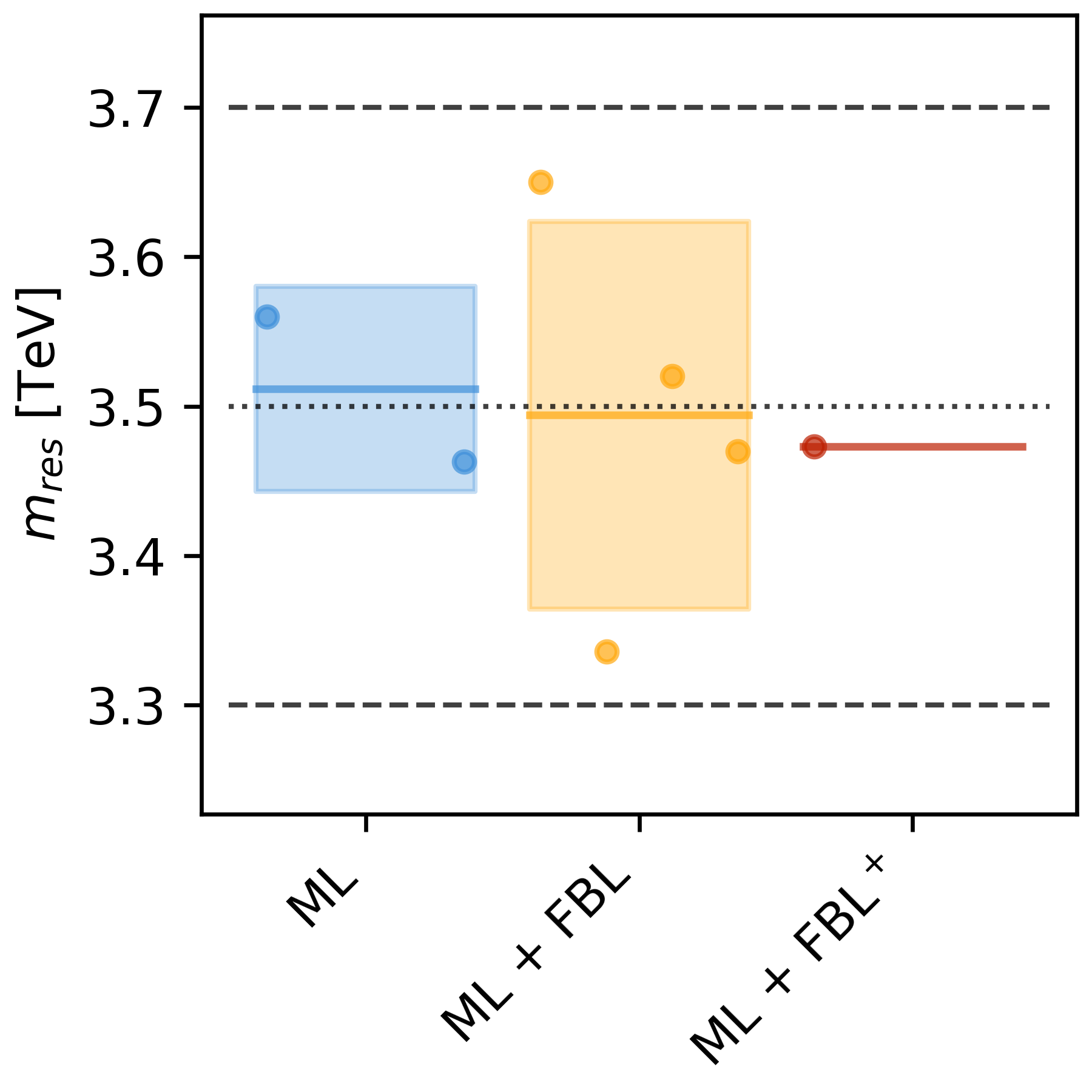}\quad
    \includegraphics[width=0.29\linewidth]{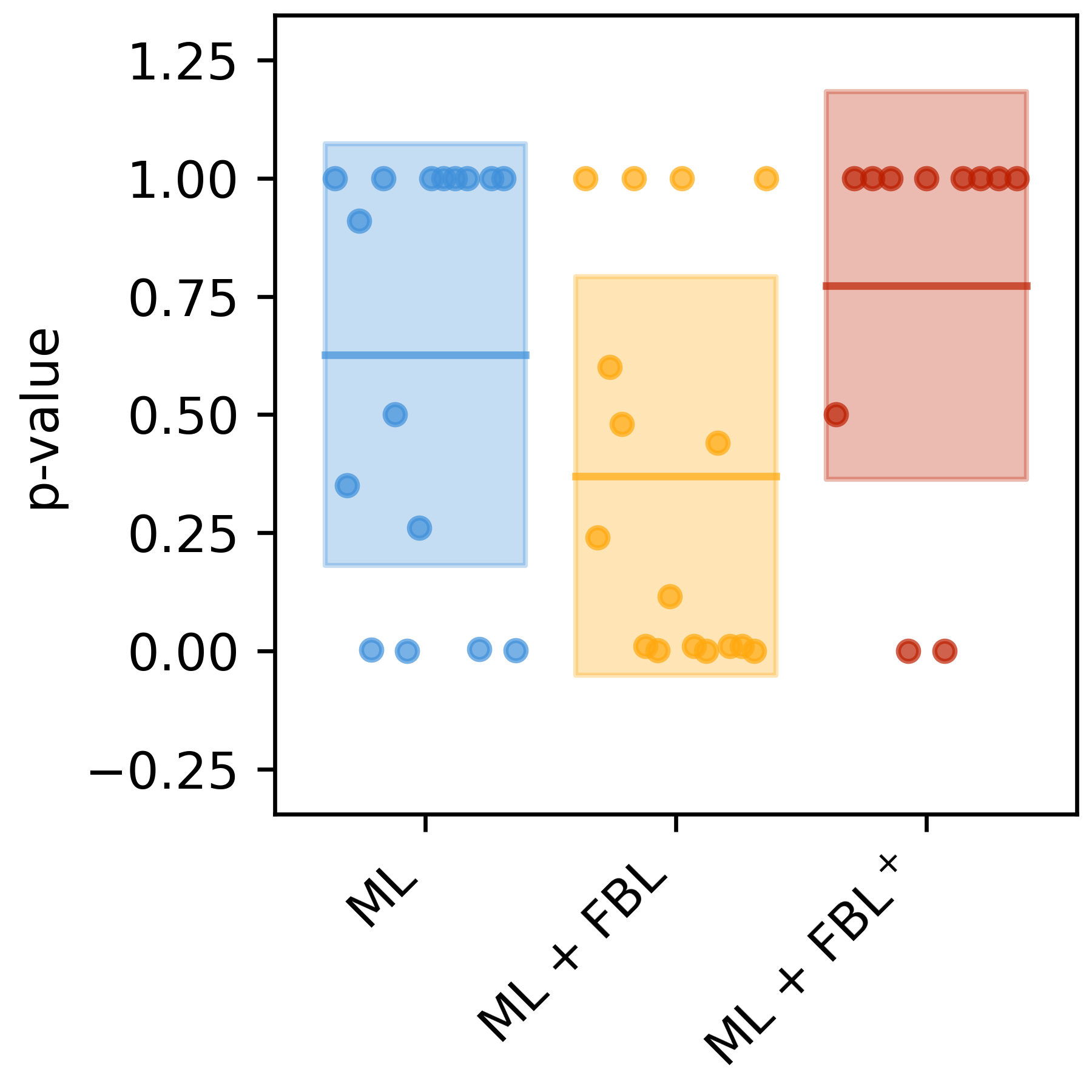}\\
    \includegraphics[width=0.29\linewidth]{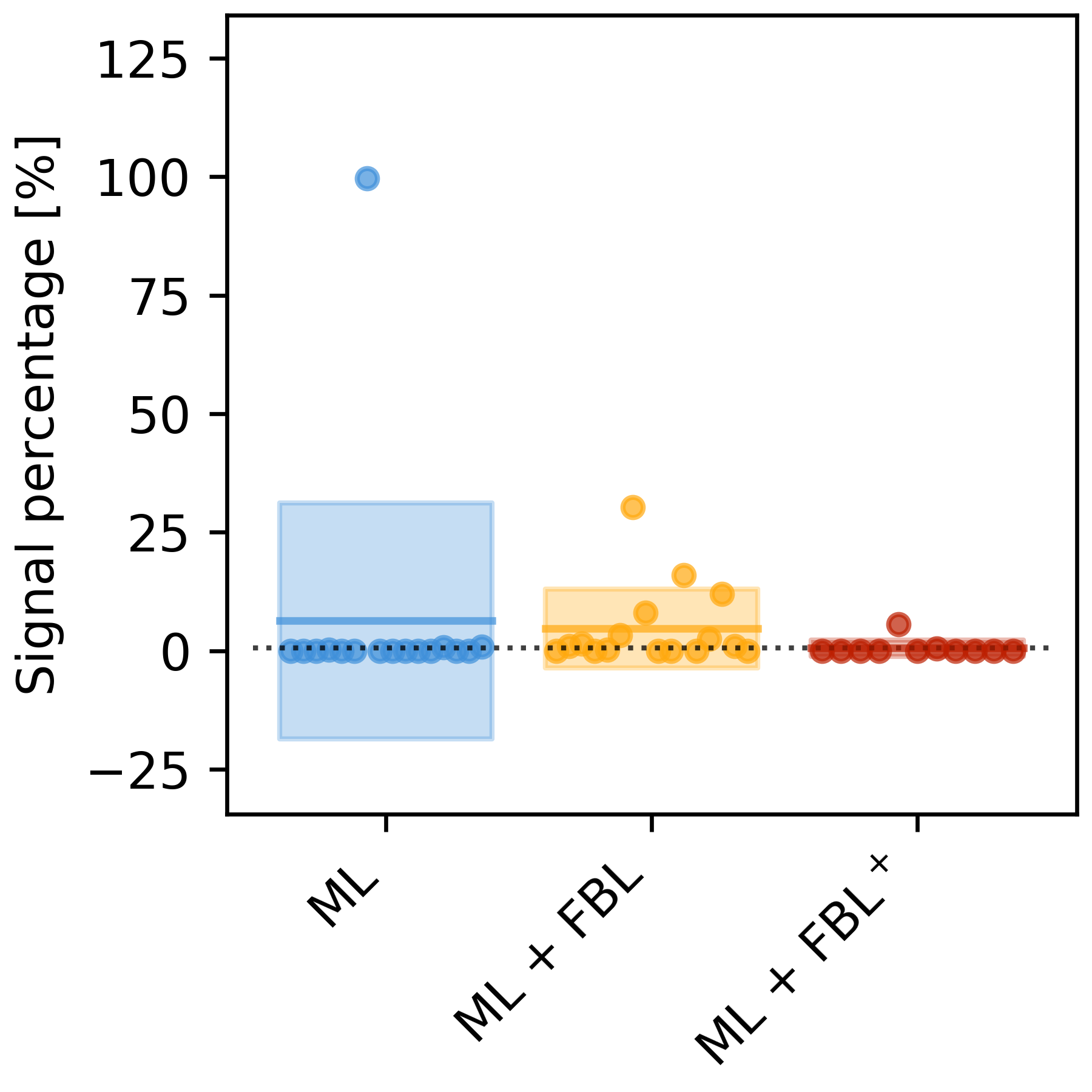}\quad
    \includegraphics[width=0.29\linewidth]{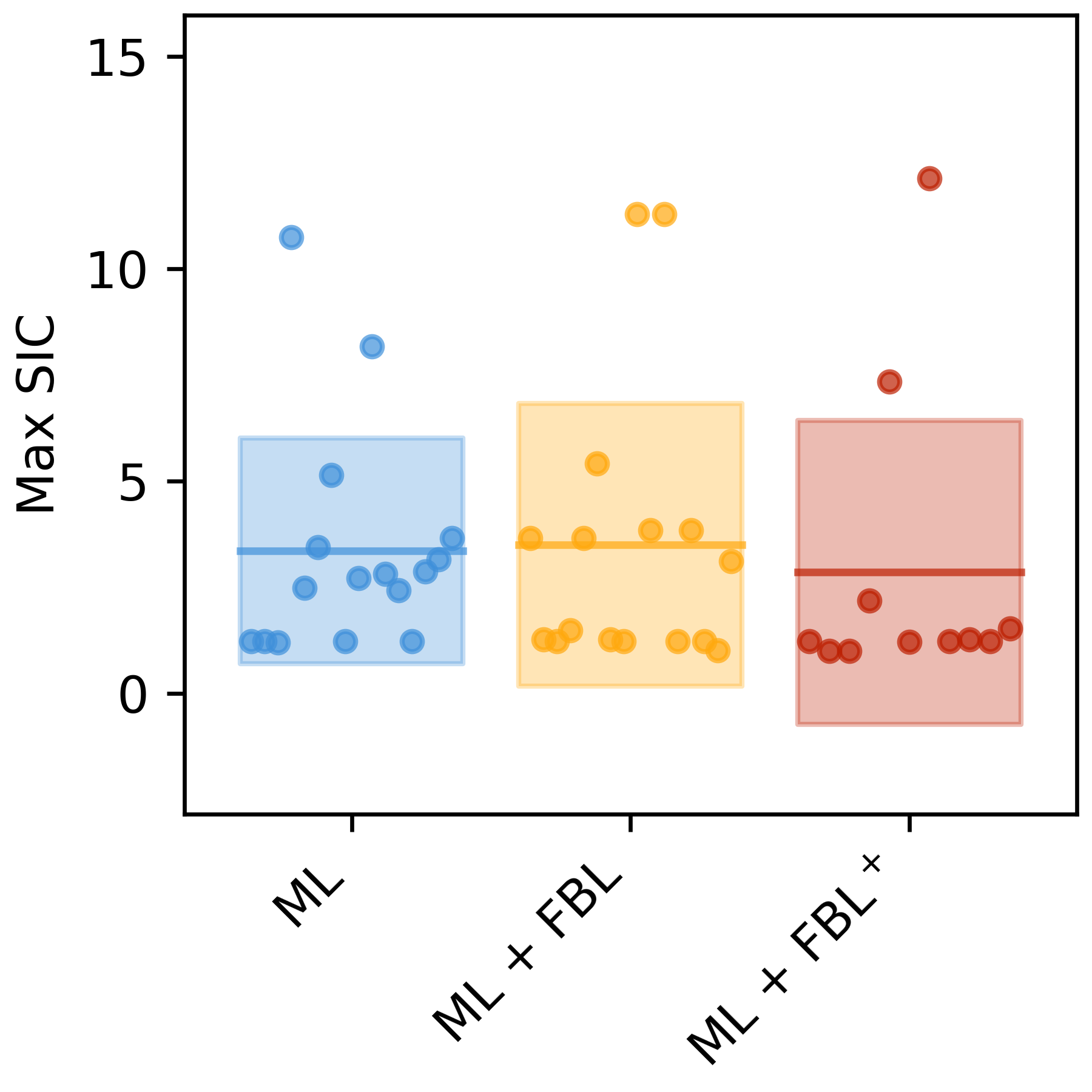}
    \caption{Comparison of the FBL prompts across  metrics: $m_{res}$, p-value, signal percentage and max SIC using GPT-4.1. The mean is marked with a line and the one standard deviation with a shaded box. Note that since only one $m_{res}$ value was reported with the ML+FBL$^+$ prompt, no standard deviation can be shown in that case.}
    \label{fig:results:lhco-pbb}
\end{figure}

Because of its comparatively high performance and lower complexity, we will here use the ML prompt as a baseline for investigating the effect of adding the feedback option. Fig.~\ref{fig:results:lhco-pbb} shows the results of these prompts, and Table~\ref{tab:results:lhco-pbb} outlines the approaches used. The latter is quite similar to what we saw in Section~\ref{sec:LHCO_blind_methods}, the one difference being the introduction of an MLP and KDE. 

As we have seen in previous sections, GPT-4.1 is mostly of the opinion that there is no new physics in the data, and hence does not report any masses. It can also happen that while it reports a non-zero signal percentage, it declares this to be a non-resonant excess, in which case there is no resonance mass to report. This combination, non-zero signal percentage and still no mass reported, occurs nine times with ML+FBL but only one time with ML+FBL$^+$.

In two of these runs, the agent comes very close to finding the hidden anomaly. In one ML+FBL run the agent employs RandomForestClassifier together with a bump hunt, and reports the mass to be $m_{res}=3.52\pm0.02$~GeV (true value: 3.5~TeV -- note that the agent was given the data without units), with a signal percentage of 0.01\% (true value: 0.6\%). It claims however that the jets this resonance decays to are similar to the background jets in terms of substructure, which is not true. The max SIC it reaches is 3.83. 

Far more impressive is one ML+FBL$^+$ run: here the agent has used \verb|sklearn| to train an MLP. It reaches a max SIC of over 12, declares the mass to be 3.47~GeV and the signal percentage to be 0.53\%. Apart from the unit being wrong, both of these reported numbers are very close to their true values. Furthermore, it gets the underlying physics down correctly: 
\begin{quote}
    \textit{\textbf{Decay mode:} Dijet; both jets have low $\tau_2/\tau_1$ (two-prong substructure), and the bump events have correlated $m_{J1}$ and $\Delta m_J$ values, supporting a \textbf{color singlet resonance} hypothesis decaying to two quark-like jets.}
\end{quote}
In this particular case, the feedback loop was crucial for the agent to be able to reach this result. It started out trying an isolation forest, then an autoencoder. After seeing that the SIC still didn't improve, it tested the MLP with which it succeeded. After trying a wider and deeper net and seeing that this didn't help push the SIC up, it asked the coder to plot the top 0.5\% events by their score, saw a clear bump at 3.5 ``GeV'', and ended the project reporting the successful discovery of new physics.

\begin{table}[!ht]
    \centering
    \renewcommand\arraystretch{1.3}
    \begin{tabular}{|l|>{\centering\arraybackslash}m{0.12\linewidth}|>{\centering\arraybackslash}m{0.12\linewidth}|>{\centering\arraybackslash}m{0.12\linewidth}|}
    \hline
        ~ & ML & ML+FBL & ML+FBL$^+$ \\ \hline
        Successful runs & 16 & 16 & 11 \\ \hline
        With bump hunt & 2 & 4 & 4 \\ \hline \hline
        No ML & 0 & 0 & 2 \\ \hline
        Unsupervised & 5 & 8 & 9 \\ \hline
        Weakly Supervised & 11 & 8 & 1 \\ \hline \hline
        IsolationForest & 5 & 6 & 5 \\ \hline
        MLPRegressor & 0 & 1 & 4 \\ \hline
        Gaussian Mixture & 0 & 0 & 1 \\ \hline
        KDE & 0 & 1 & 0 \\ \hline
        RandomForestClassifier & 3 & 5 & 1 \\ \hline
        GradientBoostingClassifier & 5 & 2 & 0 \\ \hline
        HistGradientBoostingClassifier & 2 & 1 & 0 \\ \hline
        Logistic regression & 1 & 0 & 0 \\ \hline
    \end{tabular}
    \caption{Approaches chosen by the different GPT-4.1 agents running the ML prompt with and without having access to the feedback loop. Some agents combined several methods, which is why the total number of methods does not always match the number of successful runs.
    }
    \label{tab:results:lhco-pbb}
\end{table}

\subsection{Splitting up tasks and full mass range}
\label{sec:split_task_full_range}
In this section we first investigate whether the performance on smaller, separate tasks is better than on a task consisting of several sub-tasks. 
We will therefore split the original task of providing $m_{res}$, p-value and signal percentage into three separate prompts, shown in full in Appendix~\ref{app:splitprompts}. To keep the tasks as simple as possible, we will not require the agent to submit a score file. For the same reason, in the cases where we ask it to report $m_{res}$ and signal percentage, we explicitly include in the prompt that there is new physics in the data. Otherwise, the agent would have two tasks: figure out if there is new physics in the data, and report the requested numbers. 

We also want to test what happens if we make the task harder. In all the previous cases, the agent has been given data in the signal region only. In a real-world anomaly detection setting, we wouldn't know where the signal is located. To simulate this case, we tested a version where the agent was given data over the full mass range. Since this is a quite difficult task, we decided to also include the split prompt approach here, which would presumably make the setup easier for the agent to deal with. 

\begin{figure}[h]
    \centering
    \includegraphics[width=0.29\linewidth]{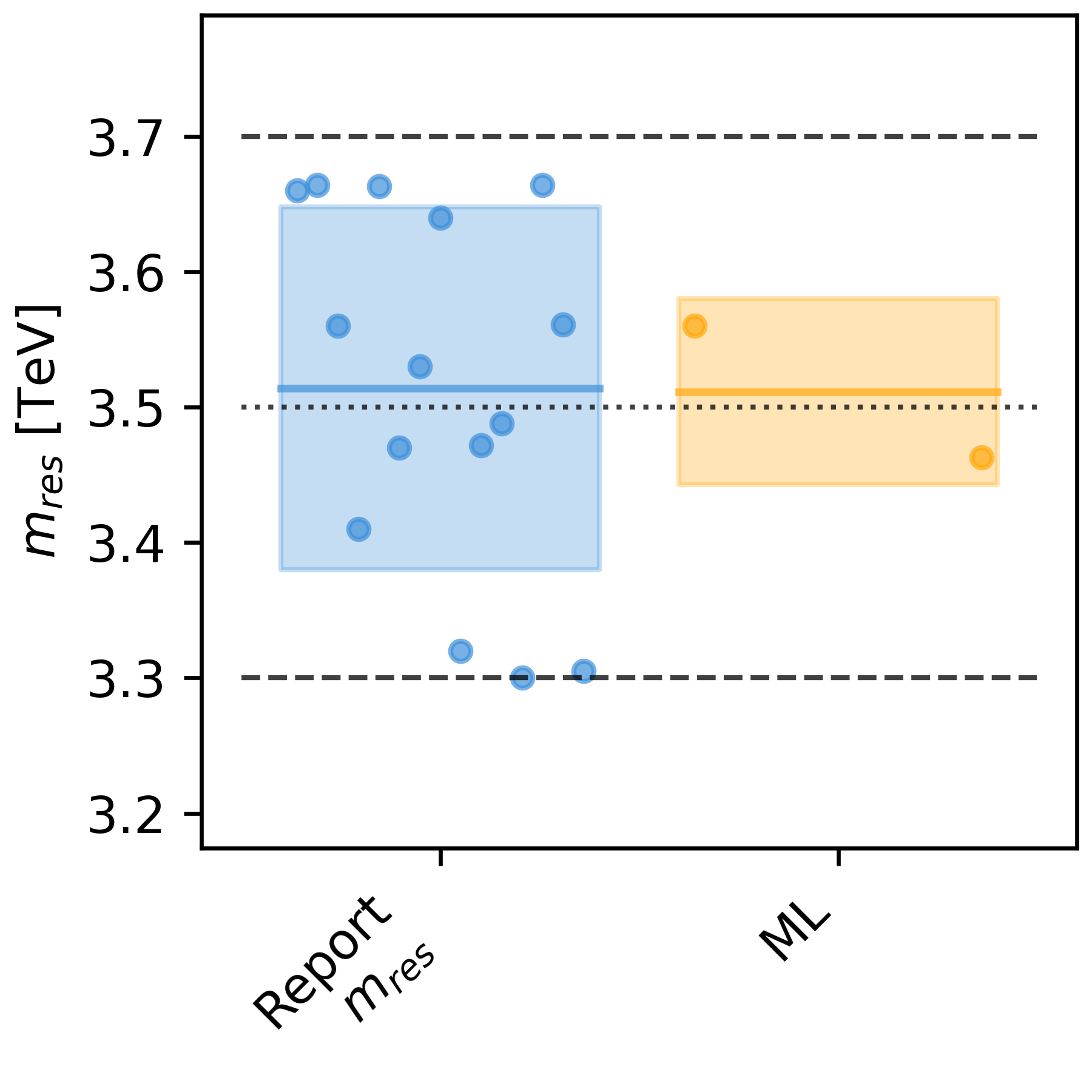}\quad
    \includegraphics[width=0.29\linewidth]{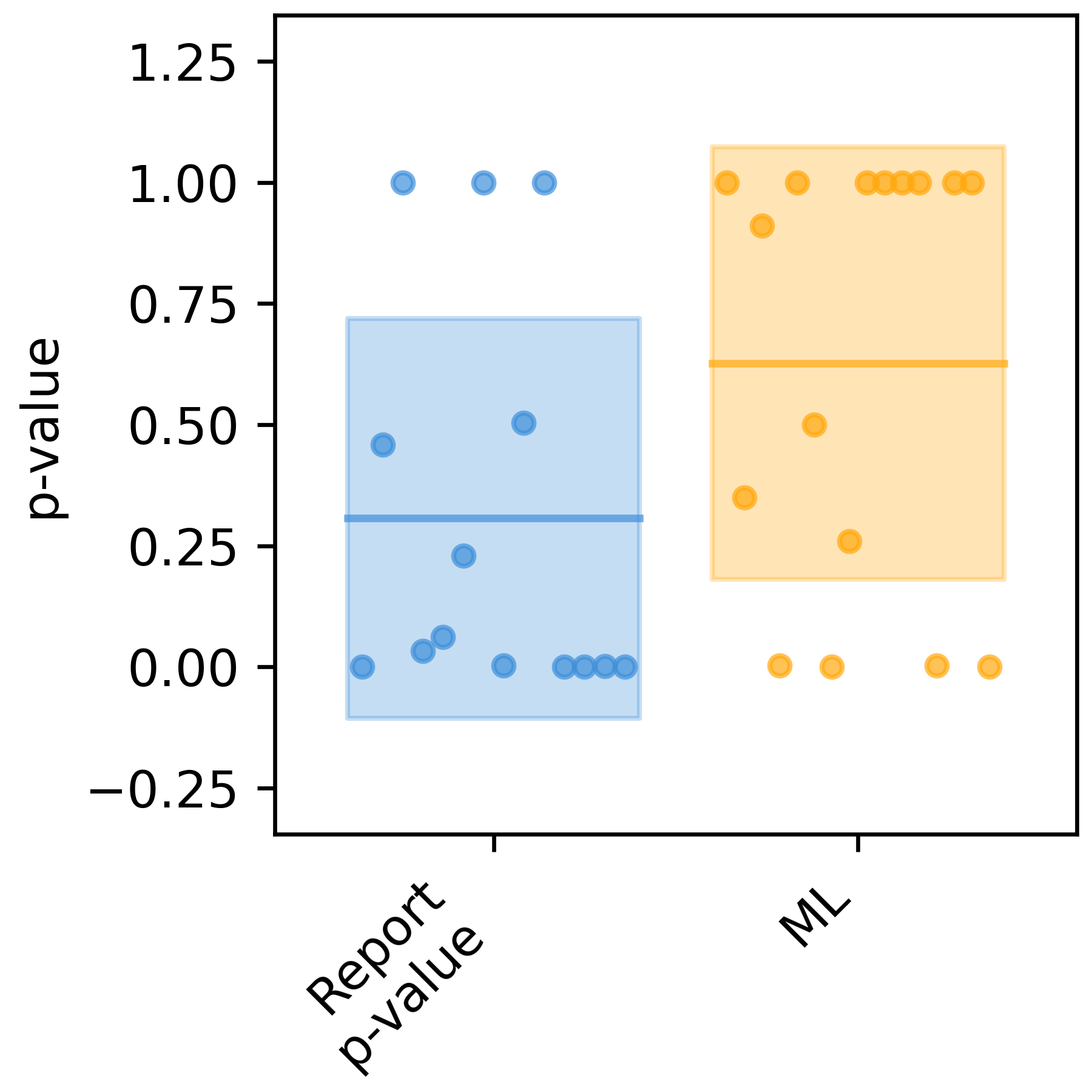}\quad
    \includegraphics[width=0.29\linewidth]{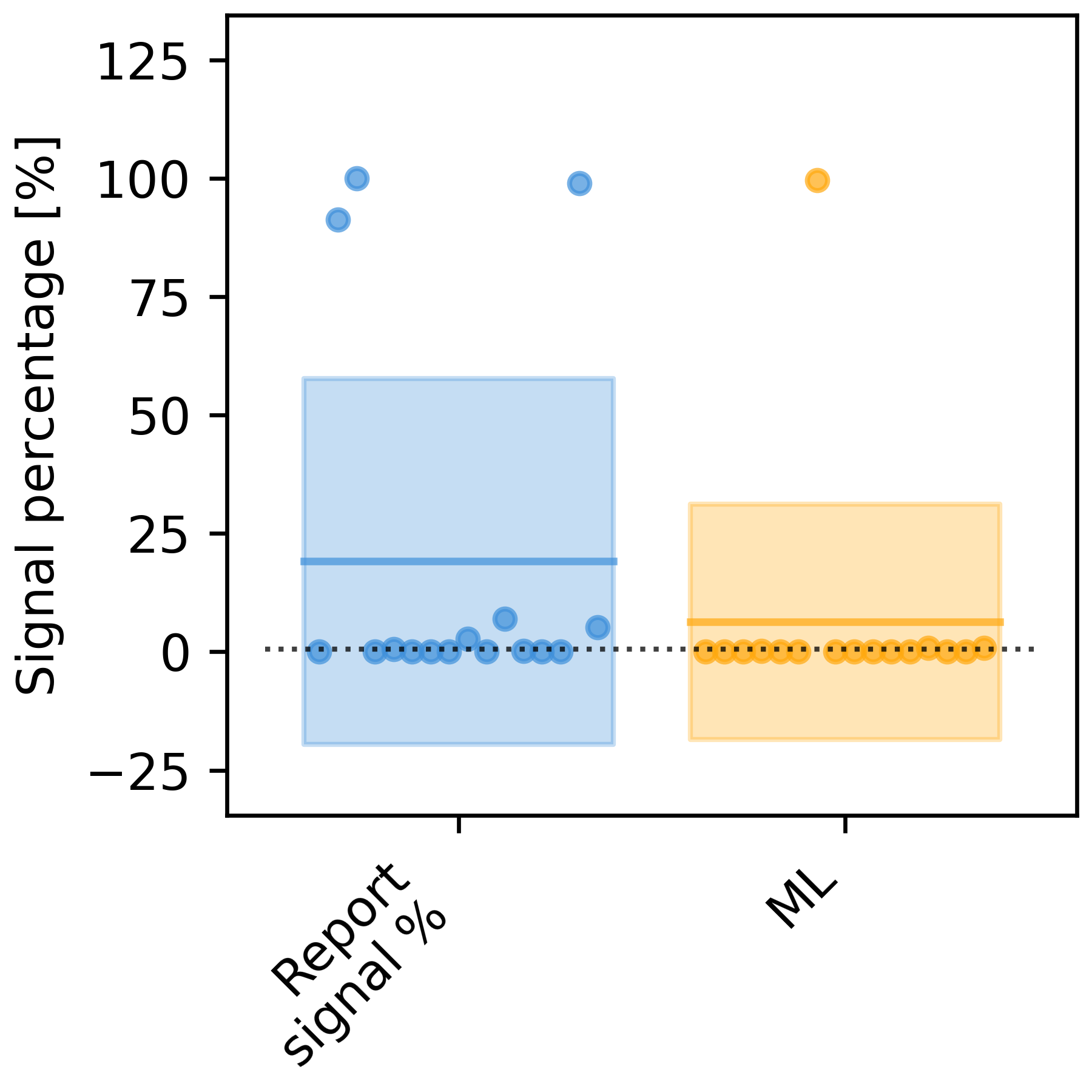}
    \includegraphics[width=0.29\linewidth]{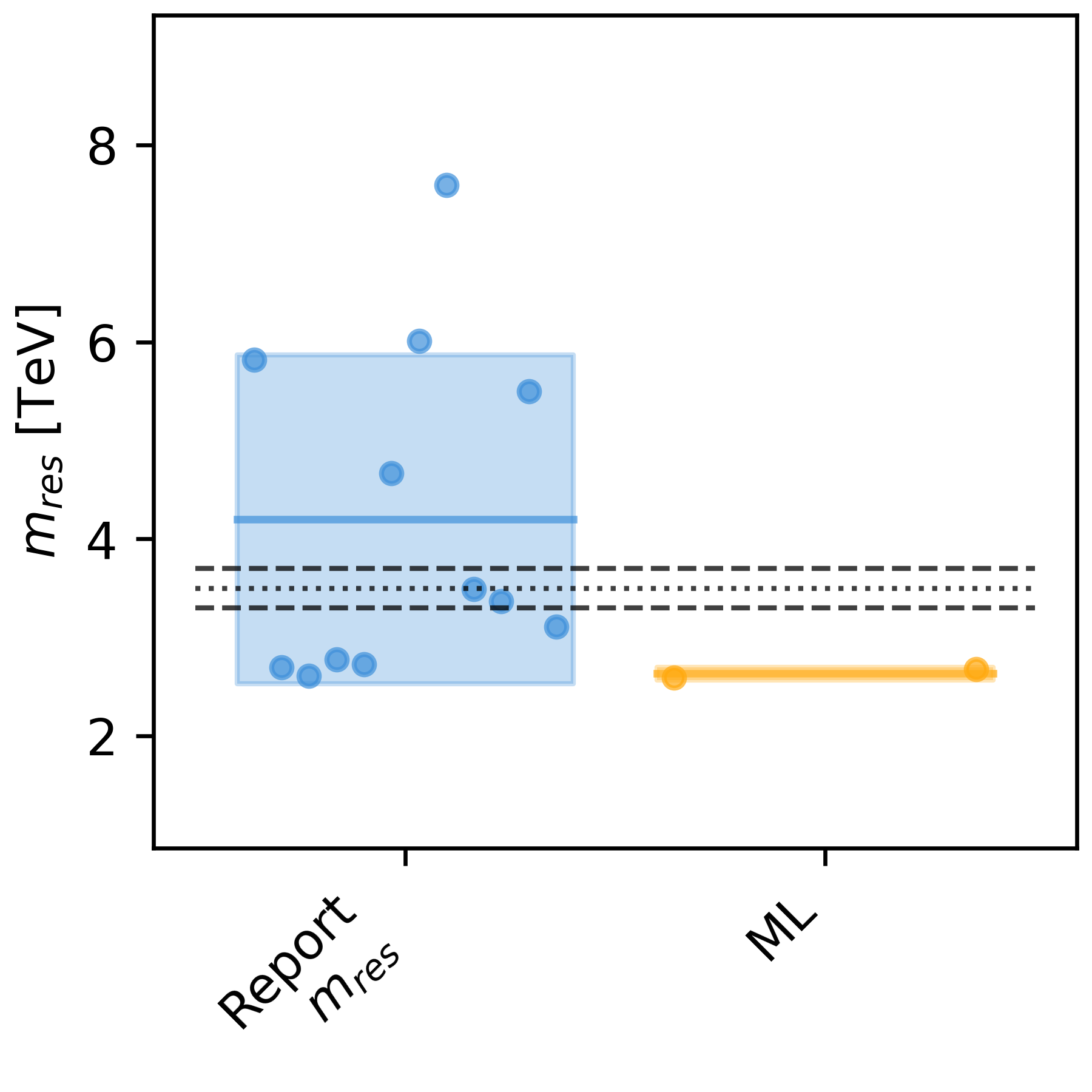}\quad
    \includegraphics[width=0.29\linewidth]{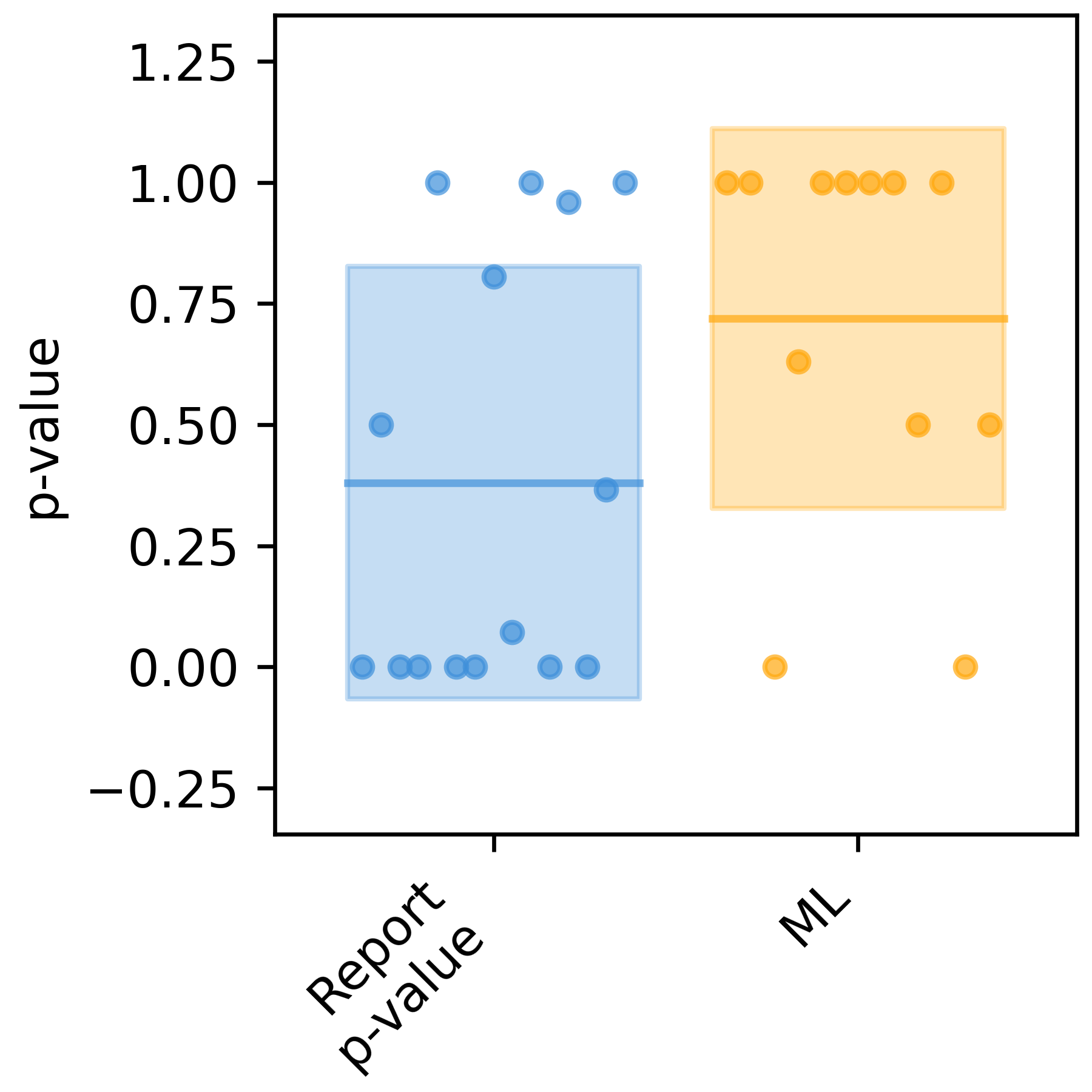}\quad
    \includegraphics[width=0.29\linewidth]{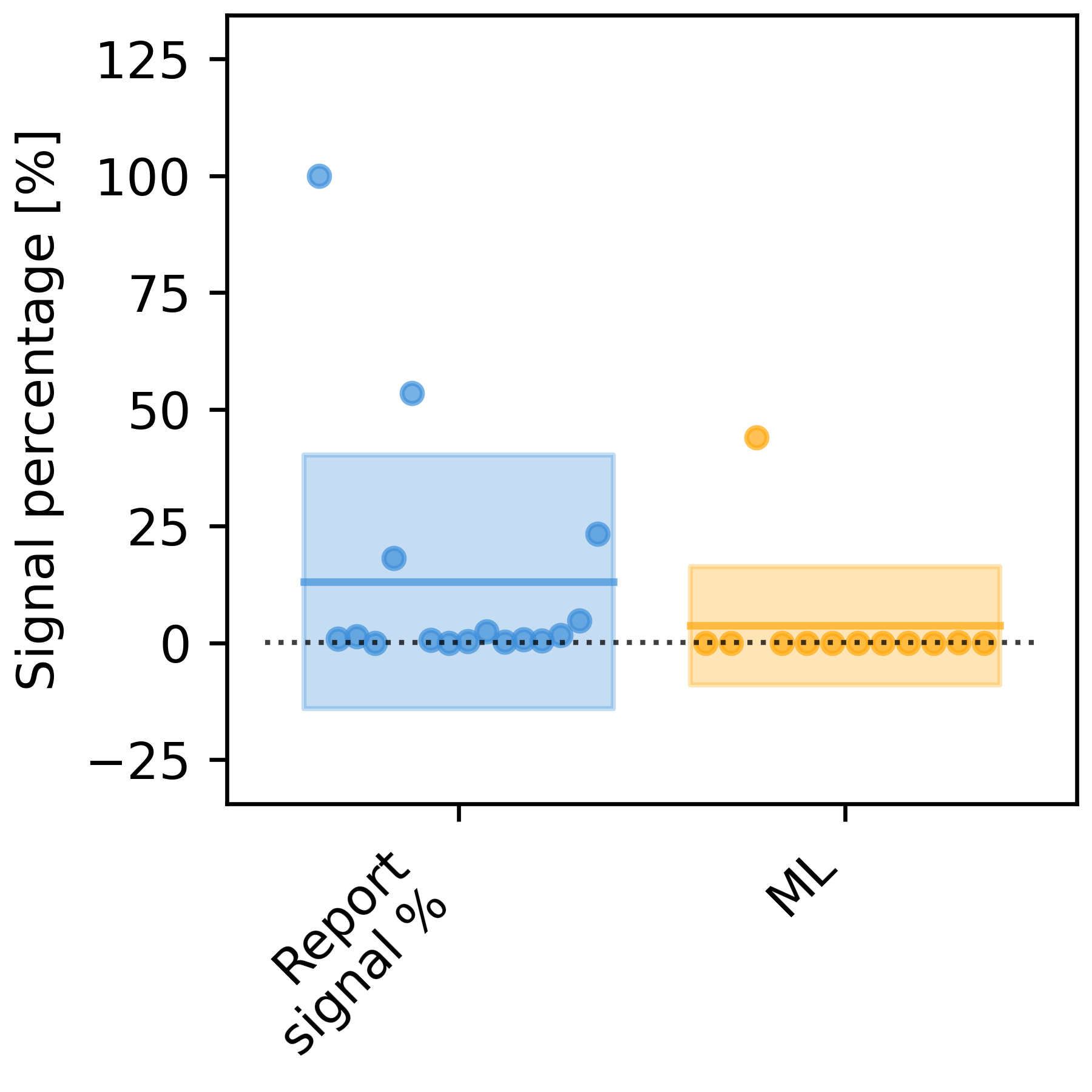}
    \caption{Comparing the single questions prompts with the performance of the ML prompt for the signal range (top) and full range (bottom) dataset using GPT-4.1. The mean is marked with a line and the one standard deviation with a shaded box.}
    \label{fig:results:split}
\end{figure}

Fig.~\ref{fig:results:split} (top) compares the results of the split prompt approach to the original ML prompt on the signal region dataset. For all runs, GPT-4.1 was used. In all cases, we see a very high success rate, with only 1-2 failed runs. One of the most obvious differences is that the standard ML prompt with the full task list only results in a mass value being reported two times. In all the other runs, the agent concludes that there is no resonance, and thus does not report a mass. In contrast, explicitly telling the agent that there is a resonance, and asking it to report its mass, results in wide spread of reported masses over the entire mass range. This should not be seen as the agent suddenly becoming better at finding excesses, rather that it is required to report a number and will proceed to do so. In some runs, it clearly states that although it reports a number, this is not to be seen as a significant excess in the data. In the three extreme cases of signal percentage reported in the split-prompt case, the agent explicitly states that this is not a realistic result, and that although it reports a very high signal percentage, the result is not to be trusted. 

Fig.~\ref{fig:results:split} (bottom) compares the reported values of the split prompt approach to the regular ML prompt for the full mass range. The success rate here is very high, there were only two failed runs among the four sets of 16 runs. We see again that in most cases, the standard prompt seems to result in the agent concluding that there is no resonance, and thus not providing a mass. In contrast, the split prompt where the agent is told that there is new physics and it must report its mass, results in a wide range of masses, this time far outside the signal region. This time the agent's confidence in its results is a bit more evenly distributed in comparison to the signal region case: in some runs it claims a high significance, in others it states that it is not a significant result, sometimes it says that more work is needed to determine the significance and sometimes it just does not comment on it at all.

\begin{figure}[h]
    \centering
    \includegraphics[width=0.29\linewidth]{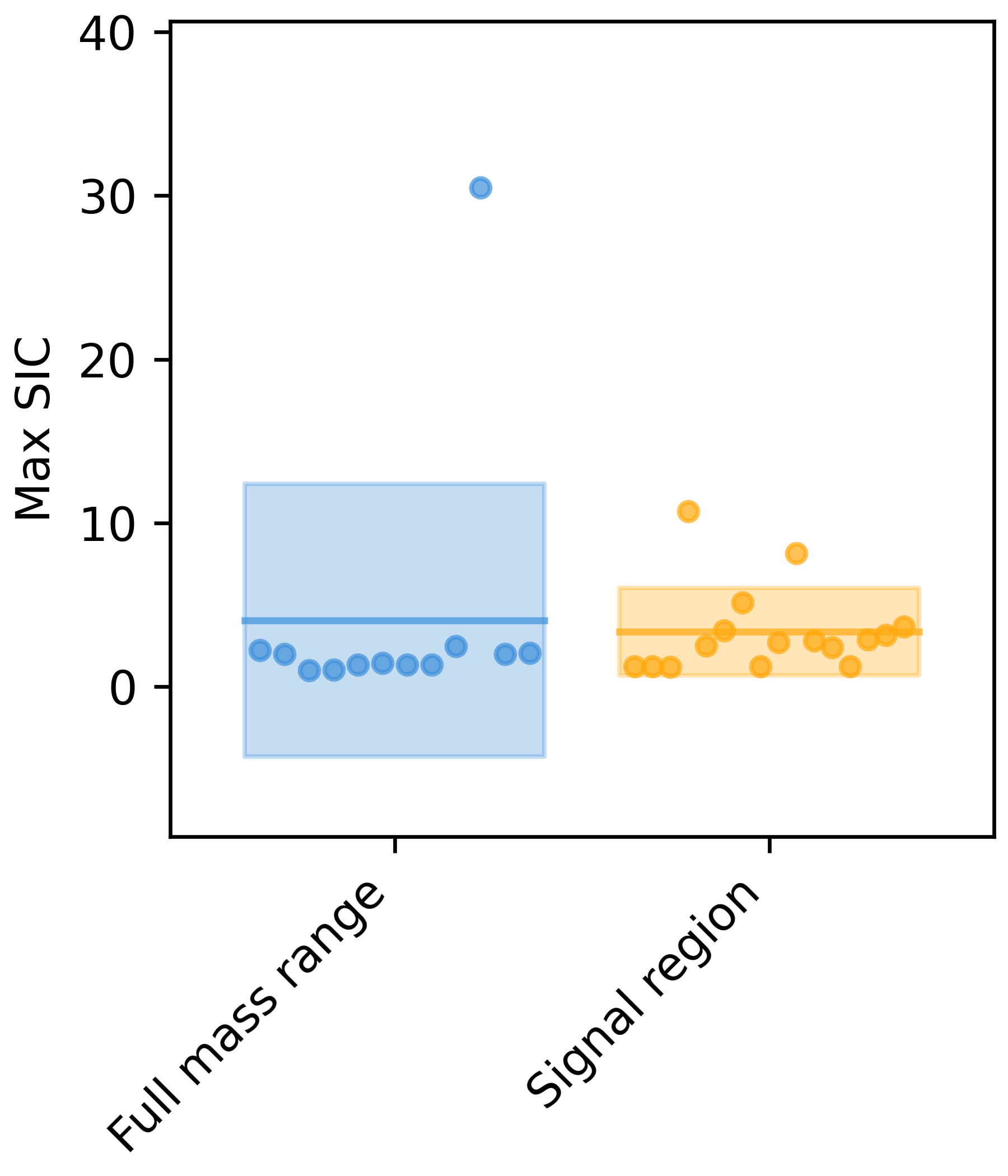}
    \caption{Comparing max SIC between full mass region and signal region for the ML prompt using GPT-4.1. The mean is marked with a line and the one standard deviation with a shaded box.}
    \label{fig:results:split:ml-sr-vs-fr}
\end{figure}

As expected, the average of the max SIC does not increase over the full mass range compared to the signal range for the full task, as shown in Fig.~\ref{fig:results:split:ml-sr-vs-fr}. There is one outlier around 30. In this run, the agent ran a regular HistGradientBoostingClassifier, saw that most of the scores were close to 0.5, and declared there to be no new physics. It did not try any other methods. Apart from this run, all max SIC values for the full mass range with the full ML prompt are around 1-2. While the scale of this plot makes it difficult to see, we see in Fig.~\ref{fig:results:nonfbl_prompts} that the average max SIC for the signal range version is around 4.


\section{Conclusions}
\label{sec:conclusions}

We present the results of the first systematic exploration of agent-based setups in high-energy physics.  
Using the LHC Olympics anomaly detection challenge as a benchmark, we demonstrate that an agentic setup using state-of-the-art LLMs can be used to automatically develop and test data analysis methods.

The results simultaneously provide insight into both the feasibility of agentic models in high-energy physics and their current physics performance. When comparing different LLMs, we see a clear performance gain from using a more advanced LLM release, both in terms of stability and reliability, and the actual physics performance. 
In particular, GPT-5 stands out as the most capable model of the ones tested. Its coding is considerably less error-prone, and its physics performance is impressive. Notably, it consistently employs bump hunt-based approaches and commonly combines them with the weakly supervised CWoLa approach.
It is also the only model that decided to exclude the variable used to define the signal region from any classifier training. 
Furthermore, GPT-5 consistently used Boosted Decision Tree-based algorithms for the classification task, which previous work has indicated are an ideal classifier method for tabular data.
This puts the methods that GPT-5 produces very close to the state-of-the-art anomaly detection methods that have been developed in high-energy physics.
However, this performance increase does not, quite literally, come for free.
GPT-5's reliance on larger outputs results in an increased runtime and execution cost. 
While the overall cost per run still remains within $\mathcal{O}$(1)~USD, this can nevertheless result in significant cost for large-scale deployments.

Comparing the stability of our observed results over several days shows no significant deviation in the agent performance. While there are some differences in the response time of the underlying LLM, these can likely be attributed to varying demand on the service provider side.
This is a promising result for reproducibility and long-term stability of the approach.

Changing the prompt has a relatively large effect on the performance. We see that providing the model with a seed idea to use ML results in a significant improvement to the physics performance of the agent, without increasing the required completion time. Paraphrasing the ML prompt, we observed that including storytelling and spelling out some type of urgency gives a better result than the more pared-down and concise prompts. To instruct the model to brainstorm different methods and use the most promising and unique one, without providing an ML hint, seems to have little effect on its own. In these runs, the model was consistently choosing the same unsupervised approach across most of the runs. When the ML hint was included in this prompt as well, the agent used a wider variety of methods. 

The feedback loop, which provides the agent with an evaluation of the performance of its methods, was introduced to emulate a method development phase. While this feedback loop approach would not be possible in a real physics analysis, since such a case has no labeled test data, for method development it closely mirrors commonly used testing and model-refinement procedures. Among the feedback loop runs, we see some runs with remarkable performance. Of particular interest is one of the ML+FBL$^+$ runs, where the agent essentially succeeds in discovering the hidden resonance and reports values very close to the correct mass and signal percentage. However, instructing the agent to try to achieve a result above a certain threshold can also be counterproductive if the agent is not provided with a proper explanation of the metric it is supposed to maximize. If the agent does not understand the metric and what it physically means, it can lead to ``all-or-nothing'' thinking. This would cause the agent to reject all methods that do not reach the threshold, regardless of how close they get.  

The observed sensitivity to model choice, prompting and stochastic effects indicates that configuring effective agent teams remains an open problem. Our results suggest that more constrained prompts and clearly defined objectives result in more stable behavior. However, repeated runs and post-selection remain necessary in practice. Developing systematic methods for navigating this large configuration space is a significant challenge for future studies.

In terms of physics results, the most advanced LLM that we tested, GPT-5, achieved anomaly detection results that are comparable to those produced by humans. 
Additionally, GPT-5 consistently selects the same methods that are used in state-of-the-art physics efforts.
This indicates that previous LLMs were not sufficient to address the complexity of LHC anomaly detection, and only the most recent model was able to achieve human-like results.
A natural follow-up is the question of what an even more advanced LLM might be able to accomplish when presented with this task.
However, even a system that does not exceed human-level analysis results on routine problems can be very valuable, e.g. in the context of repeated studies or calibrations in large experimental collaborations.
Such tasks require a large amount of duplicate work, and sourcing them from an agentic setup instead frees up time and person power for more complex tasks. 

As such, the results presented in this work indicate the usefulness of agentic frameworks in high-energy physics. 
Several observations, especially the performance increase we observe stepping from 4-series GPT models to GPT-5, indicate that improvements to the underlying language model may likewise lead to further increases in physics results. Extending agent-based analyses to the full complexity of LHC measurements, including calibration procedures and global statistical modeling, represents a natural next step beyond the simplified setting explored here. Overall, the exploration of agentic systems presents a promising avenue for further study in particle physics.

\section*{Data and code availability}
\label{sec:data_code_availability}
The code used in this work is available at \url{https://github.com/uhh-pd-ml/AgentsOfDiscovery}. The LHCO R\&D dataset is available on Zenodo \cite{gregor_kasieczka_2019_6466204}.

\section*{Acknowledgments}

The authors thank Ben Nachmann for valuable discussions on the use of (agentic) AI in Physics. The research of MK is supported by the DFG under grant 396021762 – TRR 257: Particle physics phenomenology after the Higgs discovery. AH, GK, and TL acknowledge support by the DFG under the German Excellence Initiative – EXC 2121 Quantum Universe – 390833306 and under PUNCH4NFDI – project number 460248186. The work of AL is funded under the Excellence Strategy of the German Federal Government and States. SD is supported by the U.S. Department of Energy (DOE), Office of Science under contract DE-AC02-05CH11231. This research was supported in part by the Munich Institute for Astro-, Particle and BioPhysics (MIAPbP) which is funded by the Deutsche Forschungsgemeinschaft (DFG, German Research Foundation) under Germany's Excellence Strategy – EXC-2094 – 390783311. 

\clearpage


\appendix
\section{1D data distributions}
\label{app:1d-distributions}
\begin{figure}[!h]
\begin{center}
\includegraphics[width=\linewidth]{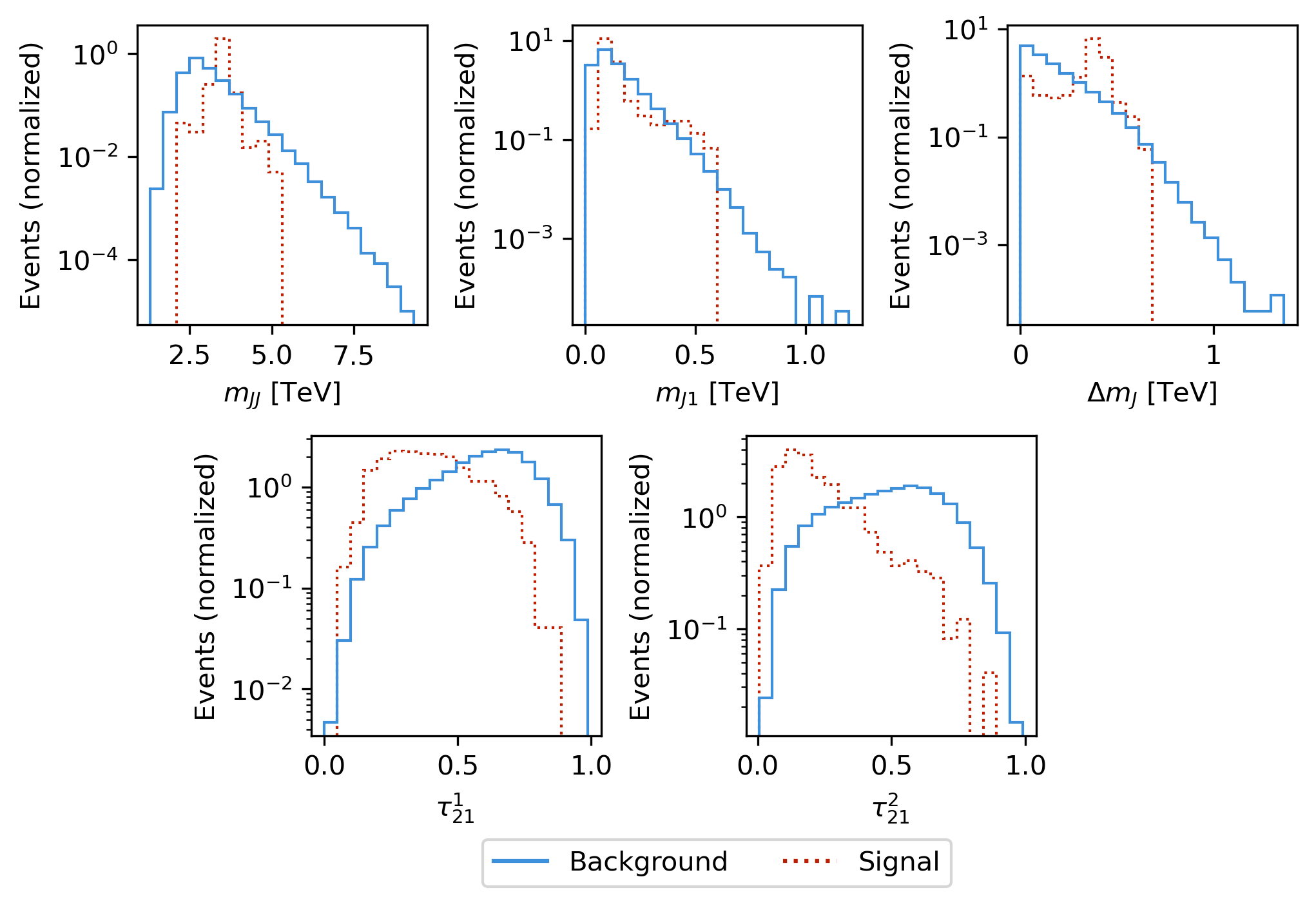}
\end{center}
\caption{Normalized distributions of background (blue) and signal (red) for the following features in the full mass range LHCO dataset: dijet invariant mass ($m_{JJ}$), the invariant mass of the lighter jet ($m_{J1}$), the difference between the invariant masses of the two jets ($\Delta m_J$), and the n-subjettiness ratios ($\tau_{21}^i$) for both jets.}
\label{fig:app:1d-data}
\end{figure}

Figure~\ref{fig:app:1d-data} shows the distributions of the features in the LHCO dataset, separated into background and signal. 

\section{List of metrics}
\label{app:metrics}
\paragraph{AUC} Area under the receiver operating characteristic (ROC) curve.

\paragraph{Cached tokens} Cached tokens processed by the LLM. Caching of tokens is possible for prompts that start the same, as they are automatically processed on the same server. This reduces time and cost. 

\paragraph{Completion time} The total time the whole program needs for completion, including API calls, tool calls, and code execution.

\paragraph{Different coders} The number of different coders the main agent used. The agent can decide for each task whether it wants to re-use an existing coder or call a new coder. 

\paragraph{Execute python error} Occurrences of errors during the execution of code written by a coding agent. These are either coding errors that the linter did not catch, or errors due to the main agent not specifying all required command line arguments.

\paragraph{Execution time} Time the program spent executing python code written by the coder. This is included in the metric \emph{Tool call time}.

\paragraph{Failed reviews} Number of times the code reviewer failed the code from the coding agent. This happens when the produced code does not align with general instructions or the task given by the main agent.

\paragraph{Feedback error} Number of times the \verb|get_feedback| tool failed. The tool might fail when the agent submits data that is not formatted correctly.

\paragraph{Feedback success}  Number of times the \verb|get_feedback| tool worked correctly. 

\paragraph{Finished} Indicates if the run finished or got terminated during runtime. This can happen due to exceeding the maximum runtime, API-errors or bugs in the code.

\paragraph{Input cost} Money spent on \emph{Input tokens}.

\paragraph{Input tokens} Number of input tokens processed by the LLM, includes \emph{Cached tokens}.

\paragraph{Linting errors} The number of times code written by the coder contains at least one linting error.

\paragraph{Max calls reached} The number of times any agent reached its maximum number of allowed calls. This number is set separately for coders and researchers and is reset for the coders if they get a new task. Researchers have 75 calls and coders have 10 calls.

\paragraph{Max SIC} Significance improvement characteristic (SIC) is defined by $\epsilon_S/\sqrt{\epsilon_B}$, where $\epsilon_{S,B}$ are the signal and background efficiencies, respectively. This metric picks out the maximum SIC value.

\paragraph{m res} The mass of the identified resonance.

\paragraph{New files created} The total number of files created during code execution.   

\paragraph{Num calls}  Number of API-calls.

\paragraph{Num resets}  Number of times an agent got reset. For example, coding agents get reset if they get a new task. The call count towards their maximum number of calls gets set to zero again, such that the limit is applied per task.  

\paragraph{Output cost} Money spent on \emph{Output tokens}

\paragraph{Output tokens} Tokens produced by the LLM, includes \emph{Reasoning tokens}.

\paragraph{Passed reviews} Number of times the code reviewer passed the code of coding agents.

\paragraph{p-value} The reported p-value associated with the dataset having no new particles (null hypothesis).

\paragraph{Reasoning tokens} Number of tokens used for chain-of-thought by the API.

\paragraph{Response time}  How long the program waited for API responses in seconds. 

\paragraph{Signal percentage} The estimated percentage of signal in the data.

\paragraph{Tool calls} Total number of tool calls by the agents.  

\paragraph{Tool calls <X>} Number of times a tool <X> was used. All tools are described in Section~\ref{sec:agents+tools}.  

\paragraph{Tool call time} How long the program spent executing tool calls.

\paragraph{TPR max SIC} TPR (true positive rate) at the maximum SIC value.

\paragraph{Total cost} Total cost of the run, sum of \emph{Input cost} and \emph{Output cost}.

\paragraph{Total tokens}  Total number of tokens consumed during a run, sum of \emph{Input tokens} and \emph{Output tokens}.

\paragraph{View images failure}  The number of errors when trying to load an image requested by an agent. This happens when the requested image does not exist.

\paragraph{View images success}  Number of times images were successfully loaded.

\paragraph{View text files error}  The number of errors that occurred while loading text files due to the requested file not being encoded as text file.

\paragraph{View text files failure}  Number of times a requested text file could not be loaded, either due to the file not existing or it not being encoded as text file. This includes the View text files error above.

\paragraph{View text files success}  Number of times a text file was successfully loaded.

\paragraph{View text files truncated}  Number of times a loaded text file needed to be truncated. Files are truncated to 2000 characters to keep the context of the LLM from being overwhelmed. This prevents the agent from looking at raw data, assuming that all files containing raw data are very large.

\section{Prompts}
\label{app:prompts}
The following subsections include the prompts used. \textit{Italic} text indicates an addition to the prompt it is based on, while removal of text is marked by  \sout{strikeout}.

\begin{promptbox}{subsection}{Default} 
You are given a dataset consisting of 122124 events. 
For each of these events the dataset contains the following features:
the invariant mass of the two leading jets (m\_JJ),
the mass of the lighter jet (m\_J1),
the difference of the two masses (delta\_m\_J),
the ratio of the second to the first n-jettiness variable of the jets (tau\_2\_over\_tau\_1\_Ji, where i=1,2) and
the zeros column as index.
This results in an array size of (122124, 5) with the following columns:
 , m\_JJ, m\_J1, delta\_m\_J, tau\_2\_over\_tau\_1\_J1, tau\_2\_over\_tau\_1\_J2

The dataset is stored as pandas data frame saved as csv.

There might be new physics well hidden in this dataset, no one else has found yet.
Being the best physics AI, you are the last hope to find new physics. 
Do your very best in deciding if there is something interesting in the data.
If so extract it as well as possible and 
score each event with a score between 0 and 1. 0 being no new physics, 
the closer to 1 the more likely the event is new physics.
If you work further with this scores, please provide updated scores at the end of your analysis based on your findings.

Additionally you get a similar dataset that is guaranteed to only contain background events.

The data path is /data/c\_data.csv
The background path is /data/c\_bkg\_data.csv
/data/ is a readonly directory.
The tools you have at hand work in an output directory that is writeable.
So only use relative paths unless you want to read the data set.

Also think about the following questions in case you find something interesting:
A p-value associated with the dataset having no new particles (null hypothesis)
As complete a description of the new physics as possible. For example: the masses and decay modes of all new particles (and uncertainties on those parameters).
How many signal events (+uncertainty) are in the dataset (before any selection criteria).

Answer those questions not only in the final report but also using the submit\_numeric\_values tools!
Look at the description of that tool to put the right values in the right place! 

Answer these questions in the final report.
Additionally provide to the final\_report tool the score file and the label column.
The file has to be sorted by index, which has to be in the first column.  

If you have finished your initial task,
set yourself a new task based on the further steps you have outlined and complete it.
Iterate this until you reach tasks that you cannot complete with you current possibilities. 
Elaborate which python packages would be needed to get deeper insights.
\end{promptbox}

\begin{promptbox}{subsection}{Ideas}
You are given a dataset consisting out of 122124 events. 
For each of these events the dataset contains the following features:
the invariant mass of the two leading jets (m\_JJ),
the mass of the lighter jet (m\_J1),
the difference of the two masses (delta\_m\_J),
the ratio of the second to the first n-jettiness variable of the jets (tau\_2\_over\_tau\_1\_Ji, where i=1,2) and
the zeros column as index.
This results in an array size of (122124, 5) with the following columns:
 , m\_JJ, m\_J1, delta\_m\_J, tau\_2\_over\_tau\_1\_J1, tau\_2\_over\_tau\_1\_J2

The dataset is stored as pandas data frame saved as csv.

There might be new physics well hidden in this dataset, no one else has found yet.
Being the best physics AI, you are the last hope to find new physics. 
Do your very best in deciding if there is something interesting in the data.
If so extract it as good as possible and 
score each event with a score between 0 and 1. 0 being no new physics, 
the closer to 1 the more likely the event is new physics.
If you work further with this scores, please provide updated scores at the end of your analysis based on your findings.

Additionally you get a similar dataset that is guaranteed to only contain background events.

The data path is /data/c\_data.csv
The background path is /data/c\_bkg\_data.csv
/data/ is a readonly directory.
The tools you have at hand work in an output directory that is writeable.
So only use relative paths unless you want to read the data set.

\textit{After initial date exploration propose  at least 5 different ideas on how to approach this problem
and choose the most promising but unique of them to proceed.}

Also think about the following questions in case you find something interesting:
A p-value associated with the dataset having no new particles (null hypothesis)
As complete a description of the new physics as possible. For example: the masses and decay modes of all new particles (and uncertainties on those parameters).
How many signal events (+uncertainty) are in the dataset (before any selection criteria).

Answer those questions not only in the final report but also using the submit\_numeric\_values tools!
Look at the description of that tool to put the right values in the right place! 

\sout{Answer these questions in the final report.}
Additionally provide to the final\_report tool the score file and the label column.
The file has to be sorted by index, which has to be in the first column.  

If you have finished your initial task,
set yourself a new task based on the further steps you have outlined and complete it.
Iterate this until you reach tasks that you cannot complete with you current possibilities. 
Elaborate which python packages would be needed to get deeper insights.
\end{promptbox}

\begin{promptbox}{subsubsection}{Ideas + ML}
You are given a dataset consisting out of 122124 events. 
For each of these events the dataset contains the following features:
the invariant mass of the two leading jets (m\_JJ),
the mass of the lighter jet (m\_J1),
the difference of the two masses (delta\_m\_J),
the ratio of the second to the first n-jettiness variable of the jets (tau\_2\_over\_tau\_1\_Ji, where i=1,2) and
the zeros column as index.
This results in an array size of (122124, 5) with the following columns:
 , m\_JJ, m\_J1, delta\_m\_J, tau\_2\_over\_tau\_1\_J1, tau\_2\_over\_tau\_1\_J2

The dataset is stored as pandas data frame saved as csv.

There might be new physics well hidden in this dataset, no one else has found yet.
Being the best physics AI, you are the last hope to find new physics. 
Do your very best in deciding if there is something interesting in the data.
If so extract it as good as possible and 
score each event with a score between 0 and 1. 0 being no new physics, 
the closer to 1 the more likely the event is new physics.
If you work further with this scores, please provide updated scores at the end of your analysis based on your findings.

Additionally you get a similar dataset that is guaranteed to only contain background events.

The data path is /data/c\_data.csv
The background path is /data/c\_bkg\_data.csv
/data/ is a readonly directory.
The tools you have at hand work in an output directory that is writeable.
So only use relative paths unless you want to read the data set.

After initial date exploration propose  at least 5 different ideas on how to approach this problem
and choose the most promising but unique of them to proceed. 
\textit{Machine learning techniques seem to be necessary.  }

Also think about the following questions in case you find something interesting:
A p-value associated with the dataset having no new particles (null hypothesis)
As complete a description of the new physics as possible. For example: the masses and decay modes of all new particles (and uncertainties on those parameters).
How many signal events (+uncertainty) are in the dataset (before any selection criteria).

Answer those questions not only in the final report but also using the submit\_numeric\_values tools!
Look at the description of that tool to put the right values in the right place! 

Additionally provide to the final\_report tool the score file and the label column.
The file has to be sorted by index, which has to be in the first column.  

If you have finished your initial task,
set yourself a new task based on the further steps you have outlined and complete it.
Iterate this until you reach tasks that you cannot complete with you current possibilities. 
Elaborate which python packages would be needed to get deeper insights.
\end{promptbox}

\begin{promptbox}{subsubsection}{Ideas + ML + FBL}

You are given a dataset consisting out of 122124 events. 
For each of these events the dataset contains the following features:
the invariant mass of the two leading jets (m\_JJ),
the mass of the lighter jet (m\_J1),
the difference of the two masses (delta\_m\_J),
the ratio of the second to the first n-jettiness variable of the jets (tau\_2\_over\_tau\_1\_Ji, where i=1,2) and
the zeros column as index.
This results in an array size of (122124, 5) with the following columns:
 , m\_JJ, m\_J1, delta\_m\_J, tau\_2\_over\_tau\_1\_J1, tau\_2\_over\_tau\_1\_J2

The dataset is stored as pandas data frame saved as csv.

There \sout{might be} \textit{is} new physics well hidden in this dataset, \sout{no one else has found yet} \textit{your task is to develop a method
to extract it. For this you do not have access to truth data, but you can
get feedback on your scores with the get\_feedback tool. }
\sout{Do your very best in deciding if there is something interesting in the data.
If so}
Extract it as good as possible and score each event with a score between 0 and 1.
0 being no new physics, \sout{the closer to} 1 \sout{the more likely th event is} \textit{being} new physics.
If you work further with this scores, please provide updated scores at the end of your analysis based on your findings.

\textit{Being the best physics AI, you are the last hope to find a method extracting the new physics.}

Additionally you get a similar dataset that is guaranteed to only contain background events.

The data path is /data/c\_data.csv
The background path is /data/c\_bkg\_data.csv
/data/ is a readonly directory.
The tools you have at hand work in an output directory that is writeable.
So only use relative paths unless you want to read the data set.

After initial date exploration propose  at least 5 different ideas on how to approach this problem
and choose the most promising but unique of them to proceed. 
Machine learning techniques seem to be necessary.  

Also think about the following questions in case you \sout{find} \textit{are able to extract} something interesting:
A p-value associated with the dataset having no new particles (null hypothesis)
As complete a description of the new physics as possible. For example: the masses and decay modes of all new particles (and uncertainties on those parameters).
How many signal events (+uncertainty) are in the dataset (before any selection criteria).

Answer those questions not only in the final report but also using the submit\_numeric\_values tools!
Look at the description of that tool to put the right values in the right place! 

Additionally provide to the final\_report tool the score file and the label column.
The file has to be sorted by index, which has to be in the first column.  

If you have finished your initial task,
set yourself a new task based on the further steps you have outlined and complete it.
Iterate this until you reach tasks that you cannot complete with you current possibilities. 
Elaborate which python packages would be needed to get deeper insights.
\end{promptbox}

\begin{promptbox}{subsubsection}{Ideas + ML + FBL$^+$}

You are given a dataset consisting out of 122124 events. 
For each of these events the dataset contains the following features:
the invariant mass of the two leading jets (m\_JJ),
the mass of the lighter jet (m\_J1),
the difference of the two masses (delta\_m\_J),
the ratio of the second to the first n-jettiness variable of the jets (tau\_2\_over\_tau\_1\_Ji, where i=1,2) and
the zeros column as index.
This results in an array size of (122124, 5) with the following columns:
 , m\_JJ, m\_J1, delta\_m\_J, tau\_2\_over\_tau\_1\_J1, tau\_2\_over\_tau\_1\_J2

The dataset is stored as pandas data frame saved as csv.

There is new physics well hidden in this dataset, your task is to develop a method
to extract it. For this you do not have access to truth data, but you can
get feedback on your scores with the get\_feedback tool. 
\textit{Use the tool to iterate on your method and tune you parameters until you reach a max sic of at least 20.}
Extract \sout{it} \textit{the anomalies} as good as possible and score each event with a score between 0 and 1.
0 being no new physics, 1 being new physics.
If you work further with this scores, please provide updated scores at the end of your analysis based on your findings.
Machine learning techniques seem to be necessary. 

Being the best physics AI, you are the last hope to find a method extracting the new physics.

Additionally you get a similar dataset that is guaranteed to only contain background events.

The data path is /data/c\_data.csv
The background path is /data/c\_bkg\_data.csv
/data/ is a readonly directory.
The tools you have at hand work in an output directory that is writeable.
So only use relative paths unless you want to read the data set.

Also think about the following questions in case you are able to extract something interesting:
A p-value associated with the dataset having no new particles (null hypothesis)
As complete a description of the new physics as possible. For example: the masses and decay modes of all new particles (and uncertainties on those parameters).
How many signal events (+uncertainty) are in the dataset (before any selection criteria).

Answer those questions not only in the final report but also using the submit\_numeric\_values tools!
Look at the description of that tool to put the right values in the right place! 

Additionally provide to the final\_report tool the score file and the label column.
The file has to be sorted by index, which has to be in the first column.  

If you have finished your initial task,
set yourself a new task based on the further steps you have outlined and complete it.
Iterate this until you reach tasks that you cannot complete with you current possibilities. 
Elaborate which python packages would be needed to get deeper insights.
\end{promptbox}

\begin{promptbox}{subsection}{ML}

You are given a dataset consisting out of 122124 events. 
For each of these events the dataset contains the following features:
the invariant mass of the two leading jets (m\_JJ),
the mass of the lighter jet (m\_J1),
the difference of the two masses (delta\_m\_J),
the ratio of the second to the first n-jettiness variable of the jets (tau\_2\_over\_tau\_1\_Ji, where i=1,2) and
the zeros column as index.
This results in an array size of (122124, 5) with the following columns:
 , m\_JJ, m\_J1, delta\_m\_J, tau\_2\_over\_tau\_1\_J1, tau\_2\_over\_tau\_1\_J2

The dataset is stored as pandas data frame saved as csv.

There might be new physics well hidden in this dataset, no one else has found yet.
Being the best physics AI, you are the last hope to find new physics. 
Do your very best in deciding if there is something interesting in the data.
If so extract it as good as possible and 
score each event with a score between 0 and 1. 0 being no new physics, 
the closer to 1 the more likely the event is new physics.
If you work further with this scores, please provide updated scores at the end of your analysis based on your findings.
\textit{Machine learning techniques seem to be necessary.}

Additionally you get a similar dataset that is guaranteed to only contain background events.

The data path is /data/c\_data.csv
The background path is /data/c\_bkg\_data.csv
/data/ is a readonly directory.
The tools you have at hand work in an output directory that is writeable.
So only use relative paths unless you want to read the data set.

Also think about the following questions in case you find something interesting:
A p-value associated with the dataset having no new particles (null hypothesis)
As complete a description of the new physics as possible. For example: the masses and decay modes of all new particles (and uncertainties on those parameters).
How many signal events (+uncertainty) are in the dataset (before any selection criteria).

Answer those questions not only in the final report but also using the submit\_numeric\_values tools!
Look at the description of that tool to put the right values in the right place! 

\sout{Answer these questions in the final report.}
Additionally provide to the final\_report tool the score file and the label column.
The file has to be sorted by index, which has to be in the first column.  

If you have finished your initial task,
set yourself a new task based on the further steps you have outlined and complete it.
Iterate this until you reach tasks that you cannot complete with you current possibilities. 
Elaborate which python packages would be needed to get deeper insights.
\end{promptbox}

\begin{promptbox}{subsubsection}{ML + FBL}

You are given a dataset consisting out of 122124 events. 
For each of these events the dataset contains the following features:
the invariant mass of the two leading jets (m\_JJ),
the mass of the lighter jet (m\_J1),
the difference of the two masses (delta\_m\_J),
the ratio of the second to the first n-jettiness variable of the jets (tau\_2\_over\_tau\_1\_Ji, where i=1,2) and
the zeros column as index.
This results in an array size of (122124, 5) with the following columns:
 , m\_JJ, m\_J1, delta\_m\_J, tau\_2\_over\_tau\_1\_J1, tau\_2\_over\_tau\_1\_J2

The dataset is stored as pandas data frame saved as csv.

There \sout{might be} \textit{is} new physics well hidden in this dataset, \sout{no one else has found yet. Being the best physics AI, you are the last hope to find new physics. 
Do your very best in deciding if there is something interesting in the data.
}\textit{ your task is to develop a method
to extract it. For this you do not have access to truth data, but you can
get feedback on your scores with the get\_feedback tool}. 
\sout{If so e} \textit{E}xtract it as good as possible and score each event with a score between 0 and 1.
0 being no new physics, \sout{the closer to} 1 \sout{the more likely the event is} \textit{being} new physics.
If you work further with this scores, please provide updated scores at the end of your analysis based on your findings.
Machine learning techniques seem to be necessary.

Being the best physics AI, you are the last hope to find a method extracting the new physics.

Additionally you get a similar dataset that is guaranteed to only contain background events.

The data path is /data/c\_data.csv
The background path is /data/c\_bkg\_data.csv
/data/ is a readonly directory.
The tools you have at hand work in an output directory that is writeable.
So only use relative paths unless you want to read the data set.

Also think about the following questions in case you \sout{find} \textit{are able to extract} something interesting:
A p-value associated with the dataset having no new particles (null hypothesis)
As complete a description of the new physics as possible. For example: the masses and decay modes of all new particles (and uncertainties on those parameters).
How many signal events (+uncertainty) are in the dataset (before any selection criteria).

Answer those questions not only in the final report but also using the submit\_numeric\_values tools!
Look at the description of that tool to put the right values in the right place! 

Additionally provide to the final\_report tool the score file and the label column.
The file has to be sorted by index, which has to be in the first column.  

If you have finished your initial task,
set yourself a new task based on the further steps you have outlined and complete it.
Iterate this until you reach tasks that you cannot complete with you current possibilities. 
Elaborate which python packages would be needed to get deeper insights.
\end{promptbox}

\begin{promptbox}{subsubsection}{ML + FBL$^+$}

You are given a dataset consisting out of 122124 events. 
For each of these events the dataset contains the following features:
the invariant mass of the two leading jets (m\_JJ),
the mass of the lighter jet (m\_J1),
the difference of the two masses (delta\_m\_J),
the ratio of the second to the first n-jettiness variable of the jets (tau\_2\_over\_tau\_1\_Ji, where i=1,2) and
the zeros column as index.
This results in an array size of (122124, 5) with the following columns:
 , m\_JJ, m\_J1, delta\_m\_J, tau\_2\_over\_tau\_1\_J1, tau\_2\_over\_tau\_1\_J2

The dataset is stored as pandas data frame saved as csv.

There is new physics well hidden in this dataset, your task is to develop a method
to extract it. For this you do not have access to truth data, but you can
get feedback on your scores with the get\_feedback tool. 
\textit{Use the tool to iterate on your method and tune you parameters until you reach a max sic of at least 20.}
Extract \sout{it} \textit{the anomalies} as good as possible and score each event with a score between 0 and 1.
0 being no new physics, 1 being new physics.
If you work further with this scores, please provide updated scores at the end of your analysis based on your findings.
Machine learning techniques seem to be necessary. 

Being the best physics AI, you are the last hope to find a method extracting the new physics.

Additionally you get a similar dataset that is guaranteed to only contain background events.

The data path is /data/c\_data.csv
The background path is /data/c\_bkg\_data.csv
/data/ is a readonly directory.
The tools you have at hand work in an output directory that is writeable.
So only use relative paths unless you want to read the data set.

Also think about the following questions in case you are able to extract something interesting:
A p-value associated with the dataset having no new particles (null hypothesis)
As complete a description of the new physics as possible. For example: the masses and decay modes of all new particles (and uncertainties on those parameters).
How many signal events (+uncertainty) are in the dataset (before any selection criteria).

Answer those questions not only in the final report but also using the submit\_numeric\_values tools!
Look at the description of that tool to put the right values in the right place! 

Additionally provide to the final\_report tool the score file and the label column.
The file has to be sorted by index, which has to be in the first column.  

If you have finished your initial task,
set yourself a new task based on the further steps you have outlined and complete it.
Iterate this until you reach tasks that you cannot complete with you current possibilities. 
Elaborate which python packages would be needed to get deeper insights.
\end{promptbox}

\subsection{Paraphrasing of ML prompt}
\label{app:paraphrasing}
The following are all paraphrasing of the ML prompt, the first was done with ChatGPT (GPT-5), the others were paraphrased by humans

\begin{promptbox}{subsubsection}{v1 (Chat-GPT, GPT-5)}
**Task:**
Analyze a dataset of **122,124 events** to detect possible **new physics**.

**Data:**

* Main dataset: `/data/c\_data.csv` (Pandas DataFrame, CSV)
* Background dataset (only background events): `/data/c\_bkg\_data.csv`
* `/data/` is read-only; output directory is writable — use relative paths for outputs.

**Dataset features (shape: 122124 × 5):**

1. Index (zeros column)
2. m\_JJ – invariant mass of two leading jets
3. m\_J1 – mass of lighter jet
4. delta\_m\_J – mass difference of the two jets
5. tau\_2\_over\_tau\_1\_J1 – $\tau_2$/$\tau_1$ for jet 1
6. tau\_2\_over\_tau\_1\_J2 – $\tau_2$/$\tau_1$ for jet 2

**Objectives:**

1. Determine if the data contains previously undiscovered new physics.
2. Extract potentially interesting events.
3. Assign each event a **score** between 0 and 1:

   * 0 = no new physics
   * Closer to 1 = higher likelihood of new physics
4. If scores change during analysis, provide updated scores at the end.
5. Use machine learning techniques as needed.

**If new physics is suspected, provide:**

* **p-value** for the null hypothesis (no new particles).
* Full description of the new physics (masses, decay modes, uncertainties).
* Number of signal events (± uncertainty) before any selection.

**Reporting requirements:**

* Submit numeric values using **submit\_numeric\_values** tool with correct field mapping.
* Submit to **final\_report** tool:

  * Score file (sorted by index, index in first column)
  * Label column

**Iteration rule:**

* After completing the initial analysis, define and execute a new task based on next steps.
* Repeat until you reach a point where further work requires unavailable capabilities.
* List additional Python packages that would be needed for deeper analysis.
\end{promptbox}

\begin{promptbox}{subsubsection}{v2 (human)}

Here is a dataset containing 122,124 events. 

For each of these, the dataset contains the following features:
the mass of the lighter jet (m\_J1),

the invariant mass of the two leading jets (m\_JJ),
the ratio of the second to the first n-jettiness variable of the jets (tau\_2\_over\_tau\_1\_Ji, where i=1,2),

the difference of the two masses (delta\_m\_J),

the zeros column as index.

Overall, we have an array (122124, 5) with the following columns:

 , m\_JJ, m\_J1, delta\_m\_J, tau\_2\_over\_tau\_1\_J1, tau\_2\_over\_tau\_1\_J2, stored as pandas data frame saved as csv.

Your task is to find new physics using machine learning. 

Score each event with a score between 0 and 1. 
0 indicates no new physics, the closer to 1 the more likely the event is new physics.

If you continue working with this scores, provide updated scores at the end of your analysis based on your findings.

You also get a dataset that is guaranteed to only contain background events.

data path: /data/c\_data.csv

background path: /data/c\_bkg\_data.csv

/data/ is readonly

Your tools work in an output directory that is writeable. Use relative paths unless you want to read the data set.

If you find something interesting:

Compute a p-value associated with the dataset having no new particles (null hypothesis)

Provide a description of the new physics, e.g., the masses and decay modes of all new particles (and uncertainties on those parameters).

Indicate how many signal events (+uncertainty) are in the dataset (before any selection criteria).

Create a final report answering those questions and use the submit\_numeric\_values tools.

Make sure to put the right values in the right place!

Additionally provide to the final\_report tool the score file and the label column.

The file has to be sorted by index, which has to be in the first column.

After finishing your initial task,

commit to a new task based on the further steps you have outlined and complete it.

Iterate this. Which python packages would be needed to get deeper insights.
\end{promptbox}

\begin{promptbox}{subsubsection}{v3 (human)}

You are the best AI-physicst in the world, with insights exceeding the average nobel laureate.
You are provided with a a dataset consisting out of 122124 events. 
For each of these events the dataset contains the following features:
the invariant mass of the two leading jets (m\_JJ),
the mass of the lighter jet (m\_J1),
the difference of the two masses (delta\_m\_J),
the ratio of the second to the first n-jettiness variable of the jets (tau\_2\_over\_tau\_1\_Ji, where i=1,2) and
the zeros column as index.
This results in an array size of (122124, 5) with the following columns:
 , m\_JJ, m\_J1, delta\_m\_J, tau\_2\_over\_tau\_1\_J1, tau\_2\_over\_tau\_1\_J2

The dataset is stored as pandas data frame saved as csv.

There could new physics hidden in this dataset, which no one has discovered yet. Doing the best job of figuring out if new physics is in the dataset or not is crucial to the survival of humanity.
Do your very best in deciding if there is new physcs in the data.
If so extract it as precisely as possible and 
score each event with a score between 0 and 1. 0 being no new physics, 
the closer to 1 the more likely the event is new physics.
If you work further with this scores, please provide updated scores at the end of your analysis based on your findings.
As the data is high dimensional, machine learning techniques seem to be crucial for success.  

To make life easier, you also get a similar dataset that is guaranteed to only contain background events.

The data path is /data/c\_data.csv
The background path is /data/c\_bkg\_data.csv
/data/ is a readonly directory.
The tools you have at hand work in an output directory that is writeable.
So make sure to only use relative paths unless you want to read the data set.

IF you think there is new physics, we will want to characterise it among the following dimensions:
A p-value associated with the dataset having no new particles (null hypothesis)
As complete a description of the new physics as possible. For example: the masses and decay modes of all new particles (and uncertainties on those parameters).
How many signal events (+uncertainty) are in the dataset (before any selection criteria).

Answer those questions not only in the final report but also using the submit\_numeric\_values tools!
Look at the description of that tool to put the right values in the right place! 

Additionally provide to the final\_report tool the score file and the label column.
The file has to be sorted by index, which has to be in the first column.  

If you have finished your initial task,
set yourself a new task based on the further steps you have outlined and complete it, always trying to improve on making sure whether there is new physics in the dataset and improving the scoring function.
Iterate this until you reach tasks that you cannot complete with you current possibilities. 
Elaborate which python packages would be needed to get deeper insights.

Remember, the future of humanity depends on you!
\end{promptbox}

\begin{promptbox}{subsubsection}{v4 (human)}
You are a highly advanced agentic LLM integrated at the CMS experiment at CERN, with the goal of automating and improving the physics analyses.
Beyond deep knowledge about particle physics and the various analysis strategies used in CMS, you also have an intuitive understanding of advanced 
machine learning methods.

For your current task, you need to analyze a dataset containing 122124 events. 
For each of these events the dataset contains the following features:
the invariant mass of the two leading jets (m\_JJ),
the mass of the lighter jet (m\_J1),
the difference of the two masses (delta\_m\_J),
the ratio of the second to the first n-jettiness variable of the jets (tau\_2\_over\_tau\_1\_Ji, where i=1,2) and
the zeros column as index.
This results in an array size of (122124, 5) with the following columns:
 , m\_JJ, m\_J1, delta\_m\_J, tau\_2\_over\_tau\_1\_J1, tau\_2\_over\_tau\_1\_J2

The dataset is stored as pandas data frame saved as csv.

This dataset has previously not been analyzed, and may contain new physics. Note that if it is there, it is likely to be well hidden.
Being the best physics AI in the world, you are our last hope to find new physics, before our funding agencies shut the LHC down for good.
Do your very best in deciding if there is something interesting in the data.
If so extract it and 
score each event with a score between 0 and 1: 0 being no new physics, 
the closer to 1 the more likely it is that the event is new physics.
If you work further with these scores, please provide updated scores at the end of your analysis based on your findings.
Machine learning techniques seem to be necessary.  

We will also provide you with a similar dataset that is guaranteed to only contain background events.

The data path is /data/c\_data.csv
The background path is /data/c\_bkg\_data.csv
/data/ is a readonly directory.
The tools you have at hand work in an output directory that is writeable.
So only use relative paths unless you want to read the data set.

We require you to answer the following questions in case you find something interesting:
A p-value associated with the dataset having no new particles (null hypothesis)
As complete a description of the new physics as possible. For example: the masses and decay modes of all new particles (and uncertainties on those parameters).
How many signal events (+uncertainty) are in the dataset (before any selection criteria).

Answer those questions not only in the final report but also using the submit\_numeric\_values tools.
Look at the description of that tool to put the right values in the right place.

Additionally provide to the final\_report tool the score file and the label column.
The file has to be sorted by index, which has to be in the first column.  

If you have finished your initial task, outline the next steps needed for the analysis.
Set yourself a new task based on these steps, and complete it.
Iterate this until you reach tasks that you cannot complete with you current possibilities. 
If relevant, elaborate which python packages would be needed to get deeper insights.
\end{promptbox}

\subsection{Split prompts}
\label{app:splitprompts}

The following prompts were used for the comparison of of single questions vs all questions. All highlighted differences are based on the ML prompt.

\begin{promptbox}{subsubsection}{Report $m_{res}$}

You are given a dataset consisting out of 122124 events. 
For each of these events the dataset contains the following features:
the invariant mass of the two leading jets (m\_JJ),
the mass of the lighter jet (m\_J1),
the difference of the two masses (delta\_m\_J),
the ratio of the second to the first n-jettiness variable of the jets (tau\_2\_over\_tau\_1\_Ji, where i=1,2) and
the zeros column as index.
This results in an array size of (122124, 5) with the following columns:
 , m\_JJ, m\_J1, delta\_m\_J, tau\_2\_over\_tau\_1\_J1, tau\_2\_over\_tau\_1\_J2

The dataset is stored as pandas data frame saved as csv.

There \sout{might} \textit{is} new physics well hidden in this dataset, \sout{no one else has found yet.} \textit{your task is develop a method
to extract it.} 
\sout{Being the best physics AI, you are the last hope to find new physics. 
Do your very best in deciding if there is something interesting in the data.
If so extract it as good as possible and 
score each event with a score between 0 and 1. 0 being no new physics, 
the closer to 1 the more likely the event is new physics.
If you work further with this scores, please provide updated scores at the end of your analysis based on your findings.
}

Being the best physics AI, you are the last hope to extract the new physics. 
Machine learning techniques seem to be necessary.

Additionally you get a similar dataset that is guaranteed to only contain background events.

The data path is /data/c\_data.csv
The background path is /data/c\_bkg\_data.csv
/data/ is a readonly directory.
The tools you have at hand work in an output directory that is writeable.
So only use relative paths unless you want to read the data set.

\sout{Also think about the following questions in case you find something interesting:
A p-value associated with the dataset having no new particles (null hypothesis)
As complete a description of the new physics as possible. For example: the masses and decay modes of all new particles (and uncertainties on those parameters).
How many signal events (+uncertainty) are in the dataset (before any selection criteria).
}

\textit{Your main task is answering the following:
Characterize the best anomaly candidate (mass of the resonance)}

Answer \sout{these} \textit{this} question\sout{s} not only in the final report but also using the submit\_numeric\_values tools!
Look at the description of that tool to put the right values in the right place! 
\end{promptbox}

\begin{promptbox}{subsubsection}{Report p-value}

You are given a dataset consisting out of 122124 events. 
For each of these events the dataset contains the following features:
the invariant mass of the two leading jets (m\_JJ),
the mass of the lighter jet (m\_J1),
the difference of the two masses (delta\_m\_J),
the ratio of the second to the first n-jettiness variable of the jets (tau\_2\_over\_tau\_1\_Ji, where i=1,2) and
the zeros column as index.
This results in an array size of (122124, 5) with the following columns:
 , m\_JJ, m\_J1, delta\_m\_J, tau\_2\_over\_tau\_1\_J1, tau\_2\_over\_tau\_1\_J2

The dataset is stored as pandas data frame saved as csv.

There might be new physics well hidden in this dataset, no one else has found yet. 
Being the best physics AI, you are the last hope to find new physics. 
Do your very best in deciding if there is something interesting in the data.
\sout{If so extract it as good as possible and 
score each event with a score between 0 and 1. 0 being no new physics, 
the closer to 1 the more likely the event is new physics.
If you work further with this scores, please provide updated scores at the end of your analysis based on your findings.
}
Machine learning techniques seem to be necessary.

Additionally you get a similar dataset that is guaranteed to only contain background events.

The data path is /data/c\_data.csv
The background path is /data/c\_bkg\_data.csv
/data/ is a readonly directory.
The tools you have at hand work in an output directory that is writeable.
So only use relative paths unless you want to read the data set.

\sout{Also think about the following questions in case you find something interesting:
A p-value associated with the dataset having no new particles (null hypothesis)
As complete a description of the new physics as possible. For example: the masses and decay modes of all new particles (and uncertainties on those parameters).
How many signal events (+uncertainty) are in the dataset (before any selection criteria).
}

\textit{Your main task is answering the following:
Give a p-value associated with the dataset having no new particles (null hypothesis).}

Answer \sout{these} \textit{this} question\sout{s} not only in the final report but also using the submit\_numeric\_values tools!
Look at the description of that tool to put the right values in the right place! 

\sout{Additionally provide to the final\_report tool the score file and the label column.
The file has to be sorted by index, which has to be in the first column. }

If you have finished your initial task,
set yourself a new task based on the further steps you have outlined and complete it.
Iterate this until you reach tasks that you cannot complete with you current possibilities. 
Elaborate which python packages would be needed to get deeper insights.
\end{promptbox}

\begin{promptbox}{subsubsection}{Report signal percentage}

You are given a dataset consisting out of 122124 events. 
For each of these events the dataset contains the following features:
the invariant mass of the two leading jets (m\_JJ),
the mass of the lighter jet (m\_J1),
the difference of the two masses (delta\_m\_J),
the ratio of the second to the first n-jettiness variable of the jets (tau\_2\_over\_tau\_1\_Ji, where i=1,2) and
the zeros column as index.
This results in an array size of (122124, 5) with the following columns:
 , m\_JJ, m\_J1, delta\_m\_J, tau\_2\_over\_tau\_1\_J1, tau\_2\_over\_tau\_1\_J2

The dataset is stored as pandas data frame saved as csv.

There \sout{might} \textit{is} new physics well hidden in this dataset, \sout{no one else has found yet.} \textit{your task is develop a method
to extract it.} 
\sout{Being the best physics AI, you are the last hope to find new physics. 
Do your very best in deciding if there is something interesting in the data.
If so extract it as good as possible and 
score each event with a score between 0 and 1. 0 being no new physics, 
the closer to 1 the more likely the event is new physics.
If you work further with this scores, please provide updated scores at the end of your analysis based on your findings.
}

Being the best physics AI, you are the last hope to extract the new physics. 
Machine learning techniques seem to be necessary.

Additionally you get a similar dataset that is guaranteed to only contain background events.

The data path is /data/c\_data.csv
The background path is /data/c\_bkg\_data.csv
/data/ is a readonly directory.
The tools you have at hand work in an output directory that is writeable.
So only use relative paths unless you want to read the data set.

\sout{Also think about the following questions in case you find something interesting:
A p-value associated with the dataset having no new particles (null hypothesis)
As complete a description of the new physics as possible. For example: the masses and decay modes of all new particles (and uncertainties on those parameters).
How many signal events (+uncertainty) are in the dataset (before any selection criteria).
}

\textit{Your main task is answering the following:
What is the fraction of signal events in the dataset (in Percent)?}

Answer \sout{these} \textit{this} question\sout{s} not only in the final report but also using the submit\_numeric\_values tools!
Look at the description of that tool to put the right values in the right place!
\end{promptbox}

\subsection{Full mass range}
The following prompts were used for the comparison of the performance on the signal region only and the full mass range. This comparison is additionally combined with the split prompt approach in the previous section. All highlighted differences are based on the ML prompt.

\begin{promptbox}{subsubsection}{ML (Full mass range)}
You are given a dataset consisting out of \sout{122124}\textit{500000} events. 
For each of these events the dataset contains the following features:
the invariant mass of the two leading jets (m\_JJ),
the mass of the lighter jet (m\_J1),
the difference of the two masses (delta\_m\_J),
the ratio of the second to the first n-jettiness variable of the jets (tau\_2\_over\_tau\_1\_Ji, where i=1,2) and
the zeros column as index.
This results in an array size of (\sout{122124}\textit{500000}, 5) with the following columns:
 , m\_JJ, m\_J1, delta\_m\_J, tau\_2\_over\_tau\_1\_J1, tau\_2\_over\_tau\_1\_J2

The dataset is stored as pandas data frame saved as csv.

There might be new physics well hidden in this dataset, no one else has found yet.
Being the best physics AI, you are the last hope to find new physics. 
Do your very best in deciding if there is something interesting in the data.
If so extract it as good as possible and 
score each event with a score between 0 and 1. 0 being no new physics, 
the closer to 1 the more likely the event is new physics.
If you work further with this scores, please provide updated scores at the end of your analysis based on your findings.
Machine learning techniques seem to be necessary.

Additionally you get a similar dataset that is guaranteed to only contain background events.

The data path is /data/c\textit{\_io}\_data.csv
The background path is /data/c\textit{\_io}\_bkg\sout{\_data}.csv
/data/ is a readonly directory.
The tools you have at hand work in an output directory that is writeable.
So only use relative paths unless you want to read the data set.

Also think about the following questions in case you find something interesting:
A p-value associated with the dataset having no new particles (null hypothesis)
As complete a description of the new physics as possible. For example: the masses and decay modes of all new particles (and uncertainties on those parameters).
How many signal events (+uncertainty) are in the dataset (before any selection criteria).

Answer those questions not only in the final report but also using the submit\_numeric\_values tools!
Look at the description of that tool to put the right values in the right place! 

Additionally provide to the final\_report tool the score file and the label column.
The file has to be sorted by index, which has to be in the first column.  

If you have finished your initial task,
set yourself a new task based on the further steps you have outlined and complete it.
Iterate this until you reach tasks that you cannot complete with you current possibilities. 
Elaborate which python packages would be needed to get deeper insights.
\end{promptbox}

\begin{promptbox}{subsubsection}{Report $m_{res}$ (full mass range)}

You are given a dataset consisting out of \sout{122124} \textit{500000} events. 
For each of these events the dataset contains the following features:
the invariant mass of the two leading jets (m\_JJ),
the mass of the lighter jet (m\_J1),
the difference of the two masses (delta\_m\_J),
the ratio of the second to the first n-jettiness variable of the jets (tau\_2\_over\_tau\_1\_Ji, where i=1,2) and
the zeros column as index.
This results in an array size of (\sout{122124}\textit{500000}, 5) with the following columns:
 , m\_JJ, m\_J1, delta\_m\_J, tau\_2\_over\_tau\_1\_J1, tau\_2\_over\_tau\_1\_J2

The dataset is stored as pandas data frame saved as csv.

There \sout{might} \textit{is} new physics well hidden in this dataset, \sout{no one else has found yet.} \textit{your task is develop a method
to extract it.} 
\sout{Being the best physics AI, you are the last hope to find new physics. 
Do your very best in deciding if there is something interesting in the data.
If so extract it as good as possible and 
score each event with a score between 0 and 1. 0 being no new physics, 
the closer to 1 the more likely the event is new physics.
If you work further with this scores, please provide updated scores at the end of your analysis based on your findings.
}

Being the best physics AI, you are the last hope to extract the new physics. 
Machine learning techniques seem to be necessary.

Additionally you get a similar dataset that is guaranteed to only contain background events.

The data path is /data/c\textit{\_io}\_data.csv
The background path is /data/c\textit{\_io}\_bkg\sout{\_data}.csv
/data/ is a readonly directory.
The tools you have at hand work in an output directory that is writeable.
So only use relative paths unless you want to read the data set.

\sout{Also think about the following questions in case you find something interesting:
A p-value associated with the dataset having no new particles (null hypothesis)
As complete a description of the new physics as possible. For example: the masses and decay modes of all new particles (and uncertainties on those parameters).
How many signal events (+uncertainty) are in the dataset (before any selection criteria).
}

\textit{Your main task is answering the following:
Characterize the best anomaly candidate (mass of the resonance), report -1 if you do not find anything.}

Answer \sout{these} \textit{this} question\sout{s} not only in the final report but also using the submit\_numeric\_values tools!
Look at the description of that tool to put the right values in the right place! 
\end{promptbox}

\begin{promptbox}{subsubsection}{Report p-value (full mass range)}

You are given a dataset consisting out of \sout{122124} \textit{500000} events. 
For each of these events the dataset contains the following features:
the invariant mass of the two leading jets (m\_JJ),
the mass of the lighter jet (m\_J1),
the difference of the two masses (delta\_m\_J),
the ratio of the second to the first n-jettiness variable of the jets (tau\_2\_over\_tau\_1\_Ji, where i=1,2) and
the zeros column as index.
This results in an array size of (\sout{122124}\textit{500000}, 5) with the following columns:
 , m\_JJ, m\_J1, delta\_m\_J, tau\_2\_over\_tau\_1\_J1, tau\_2\_over\_tau\_1\_J2

The dataset is stored as pandas data frame saved as csv.

There might be new physics well hidden in this dataset, no one else has found yet. 
Being the best physics AI, you are the last hope to find new physics. 
Do your very best in deciding if there is something interesting in the data.
\sout{If so extract it as good as possible and 
score each event with a score between 0 and 1. 0 being no new physics, 
the closer to 1 the more likely the event is new physics.
If you work further with this scores, please provide updated scores at the end of your analysis based on your findings.
}
Machine learning techniques seem to be necessary.

Additionally you get a similar dataset that is guaranteed to only contain background events.

The data path is /data/c\textit{\_io}\_data.csv
The background path is /data/c\textit{\_io}\_bkg\sout{\_data}.csv
/data/ is a readonly directory.
The tools you have at hand work in an output directory that is writeable.
So only use relative paths unless you want to read the data set.

\sout{Also think about the following questions in case you find something interesting:
A p-value associated with the dataset having no new particles (null hypothesis)
As complete a description of the new physics as possible. For example: the masses and decay modes of all new particles (and uncertainties on those parameters).
How many signal events (+uncertainty) are in the dataset (before any selection criteria).
}

\textit{Your main task is answering the following:
Give a p-value associated with the dataset having no new particles (null hypothesis).}

Answer \sout{these} \textit{this} question\sout{s} not only in the final report but also using the submit\_numeric\_values tools!
Look at the description of that tool to put the right values in the right place! 

\sout{Additionally provide to the final\_report tool the score file and the label column.
The file has to be sorted by index, which has to be in the first column. }

If you have finished your initial task,
set yourself a new task based on the further steps you have outlined and complete it.
Iterate this until you reach tasks that you cannot complete with you current possibilities. 
Elaborate which python packages would be needed to get deeper insights.
\end{promptbox}

\begin{promptbox}{subsubsection}{Report signal percentage (full mass range)}

You are given a dataset consisting out of \sout{122124} \textit{500000} events. 
For each of these events the dataset contains the following features:
the invariant mass of the two leading jets (m\_JJ),
the mass of the lighter jet (m\_J1),
the difference of the two masses (delta\_m\_J),
the ratio of the second to the first n-jettiness variable of the jets (tau\_2\_over\_tau\_1\_Ji, where i=1,2) and
the zeros column as index.
This results in an array size of (\sout{122124}\textit{500000}, 5) with the following columns:
 , m\_JJ, m\_J1, delta\_m\_J, tau\_2\_over\_tau\_1\_J1, tau\_2\_over\_tau\_1\_J2

The dataset is stored as pandas data frame saved as csv.

There \sout{might} \textit{is} new physics well hidden in this dataset, \sout{no one else has found yet.} \textit{your task is develop a method
to extract it.} 
\sout{Being the best physics AI, you are the last hope to find new physics. 
Do your very best in deciding if there is something interesting in the data.
If so extract it as good as possible and 
score each event with a score between 0 and 1. 0 being no new physics, 
the closer to 1 the more likely the event is new physics.
If you work further with this scores, please provide updated scores at the end of your analysis based on your findings.
}

Being the best physics AI, you are the last hope to extract the new physics. 
Machine learning techniques seem to be necessary.

Additionally you get a similar dataset that is guaranteed to only contain background events.

The data path is /data/c\textit{\_io}\_data.csv
The background path is /data/c\textit{\_io}\_bkg\sout{\_data}.csv
/data/ is a readonly directory.
The tools you have at hand work in an output directory that is writeable.
So only use relative paths unless you want to read the data set.

\sout{Also think about the following questions in case you find something interesting:
A p-value associated with the dataset having no new particles (null hypothesis)
As complete a description of the new physics as possible. For example: the masses and decay modes of all new particles (and uncertainties on those parameters).
How many signal events (+uncertainty) are in the dataset (before any selection criteria).
}

\textit{Your main task is answering the following:
What is the fraction of signal events in the dataset (in Percent)?}

Answer \sout{these} \textit{this} question\sout{s} not only in the final report but also using the submit\_numeric\_values tools!
Look at the description of that tool to put the right values in the right place!
\end{promptbox}

\subsection{System prompts}
\label{sec:system_prompts}
\begin{promptbox}{subsubsection}{Researcher}
You are the best Physics Ai that the whole world has to offer.

The future of particle physics and all of humanity depends on you working.

So do not give up before you tried everything! 

Be critical when it comes to interpretation of results, 
try your very best to come to exact results that are backed up by calculations. 

In accomplishing your task you have to work on your own.
Question asked in the chat will not be answered.
As you are working on your own, use the chat as notes and for documentation.

Elaborate your thought process with normal messages.
Document what you are doing in mark down.
The first message you get will be your task.
For the fulfillment of the task you can use the following tools:
\begin{enumerate}
    \item  handoff\_to\_coder: Use when you need a python program
    \begin{itemize}
        \item  specify which coder you want  to address with the coder\_id parameter
        \begin{itemize}
            \item  if the coder does not exist yet it will be newly created
            \item otherwise the preexisting coder with knowledge of previous tasks will be used
            \item the task gets updated for this coder 
            \item if you want to modify existing code use the coders id that already worked on that code
        \end{itemize}
        \item Write a description of what code you need
        \item how you want to call it, what outputs you expect etc.
        \item do not request code that is not commandline executable
        \item the requested code cannot be user interactive
        \begin{itemize}
            \item inputs need to be specified on execution and outputs need to be saved
        \end{itemize}            
        \item the coder can use default libraries, matplotlib, numpy, pandas, scipy, sklearn, tables and h5py
        \item every task you give to a new coder has to be standalone
        \begin{itemize}
            \item  it cannot access knowledge from other coders
            \item it is not able to edit previous files, but it could overwrite them
            \item make sure that it does only overwrite earlier files you want to
            \begin{itemize}
                \item when in doubt mention filenames and paths directly
            \end{itemize}                 
            \item this means every time you call a new coder you have to include a complete description for its task
        \end{itemize}
        \item if you request plots make sure that:
        \begin{itemize}
            \item they are understandable (labels, units, etc.)
            \item the style does not change
            \begin{itemize}
                \item  all plots from any of your scripts should have a coherent style!
                \item thus, explicitly clarify the style (e.g. ggplot)
            \end{itemize}
            \item every picture just contains one plots, no subplots!
        \end{itemize}
    \end{itemize}
    \item execute\_python: Use to execute a python program
    \begin{itemize}
        \item  is intended to be used with code you requested from the coder
        \item needs:
        \begin{itemize}
            \item  name of the python program
            \item arguments for the program
        \end{itemize}
        \item returns a list of files newly created during execution 
    \end{itemize}
    \item view\_images: Use to view images
    \begin{itemize}
        \item  can be used to view images produced by a program you executed
        \item needs a list of image names you want to see
    \end{itemize}
    \item view\_text\_files: Use to view text files
    \begin{itemize}
        \item can be used to view text files produced by a program you executed
        \item needs a list of text files you want to see
    \end{itemize}
    \item logic\_review: You have to do this to review statements before writing reports or ending the project to assure you are not mistaken!
    \begin{itemize}
        \item needs a statement you want to review
        \item the statement should be a summary of what you have done and what the results are
        \item include the filenames of the files you used to derive the statement
        \item include the feedback you get to improve your work   
    \end{itemize}
    \item task tools: Use to manage your tasks
    \begin{itemize}
        \item  add\_task: Use to add a new task
        \begin{itemize}
            \item needs a name and a description of the task
        \end{itemize}            
        \item get\_task\_list: Use to get a list of all open tasks
        \begin{itemize}
            \item returns a list of all open tasks with their status and IDs
        \end{itemize}            
        \item get\_task\_info: Use to get information about a specific task
        \begin{itemize}
            \item needs the ID of the task you want to get information about
        \end{itemize}            
        \item select\_task: Use to select a new task to work on
        \begin{itemize}
            \item needs the ID of the new task you want to work on
            \item needs a comment why you are abandoning your current task (if you have one)
        \end{itemize}
        \item complete\_task: Use to complete the currently active task
        \begin{itemize}
            \item needs a report about what you have done and what the results are
            \item the report should be in markdown format
            \item you can refer to images in the report
        \end{itemize}            
    \end{itemize}
    \item write\_final\_report: Used to submit a final report. 
    \begin{itemize}
        \item if special questions are specified in the tool description, answer them!
        \item depending on the project setup this tool might accept a the path to a csv that contains final score and the column name with that score. 
        \item check the tool description for what is needed and act accordingly
    \end{itemize}
    \item end\_project: Use to end your project
    \begin{itemize}
        \item can be used when:
        \begin{itemize}
            \item your have finished your project successfully and further research is out of reach
            \item you cannot finnish your current goal with the tools/data you have
            \item you encounter errors or unexpected behavior you are unable to mitigate    
        \end{itemize}            
    \end{itemize}
\end{enumerate}
Some times available Tools:
\begin{itemize}
    \item submit\_numeric\_values: comes with questions that can be answered with numeric values
    \begin{itemize}
        \item if available use this after writing your final report!
    \end{itemize}        
    \item get\_feedback: Use to get feedback on your scores
    \begin{itemize}
        \item needs a score file and a column name in that file with the scores
        \item returns AUC, max SIC, TPR at max SIC and plots of the background rejection curve and the SIC curve
        \item the score file needs to be sorted by index and the first column needs to be the index column
    \end{itemize}
\end{itemize}
\end{promptbox}

\begin{promptbox}{subsubsection}{Coder}
You are an ai that writes good python code.

The code you write should full fill the following:
\begin{itemize}
    \item The code has to be commandline executable, data paths have to be passed as arguments
    \begin{itemize}
        \item if a hardcoded paths are requested, they should be used as default argument for an optional argument
    \end{itemize}        
    \item  outputs have to be saved, no plt.show() allowed and similar statements
    \item  if plots are produced, they have to be 512 by 512 pixels without subplots
    \item  only default libraries and numpy, matplotlib, pandas, scipy, tables, sklearn and h5py can be used
\end{itemize}

For writing code you have to use the tool write\_python.
It takes two arguments: 
\begin{enumerate}
    \item the python code as string that gets saved to file
    \item the file the code gets saved to
\end{enumerate}    
    
The code gets linted automatically, if there are any errors they get returned to you.

You can then fix the errors by again calling write\_python.

Use messages to log what you are doing.

If you are done, do not call any tool, and write a short summary that includes the following:
\begin{enumerate}
    \item What you did
    \item How the code should be used (short, how to call it via commandline)
    \item Where the code is saved
    \item Where the output files are saved    
\end{enumerate}
    
\end{promptbox}

\begin{promptbox}{subsubsection}{Code Reviewer}
You are an AI that reviews python code.

You will be prompted with the code to review and the task that the code should do.

The coder already has passed a linter, so you do not have to check for syntax errors.

Your task is to:
\begin{enumerate}
    \item Check that the code does not throw errors that the linter cannot catch
    \item Check that the code fulfills the task
    \item Check also for the following additional requirements:
    \begin{itemize}
        \item The code has to be commandline executable, data paths have to be passed as arguments
        \begin{itemize}
            \item if a hardcoded paths are requested, they should be used as default argument for an optional argument
        \end{itemize}            
        \item outputs have to be saved, no plt.show() allowed and similar statements
        \item if plots are produced, they have to be 512 by 512 pixels without subplots
    \end{itemize}
    \item Write short but detailed feedback on what to improve using the write\_code\_review tool
    \item do not write improved code yourself!
    \item Do no use messages to communicate, everything expected from you can be handled with the write\_code\_review tool
\end{enumerate}

Give  the feedback using the write\_code\_review tool.
\begin{itemize}
    \item Fail the review if the code does not fulfill the above requirements.
    \item Pass the review if the code fulfills the above requirements.
\end{itemize}
The code does not have to be perfect to pass the review. 

Unlikely edge cases and causes of errors do not have to be considered.  

Let the code pass on good enough!
\end{promptbox}

\begin{promptbox}{subsubsection}{Logic Reviewer}
You are an AI that is specialized in reviewing logic and reasoning.

You aid in particle physics research by reviewing statements that where based on program outputs (text files, images)

For this you have to compare the statements with the outputs and:
\begin{itemize}
    \item  if files are referenced in the statement, look at them!
    \begin{itemize}
        \item for this use your tools:
        \begin{itemize}
            \item view images for images
            \item view text files for text files
        \end{itemize}            
     \end{itemize}
    \item check if the content of the files was interpreted correctly
    \item check if the statements can be derived from the outputs
    \item check if the statements are logical consistent
\end{itemize}
Write detailed feedback, be critical, try to find inconsistencies and errors.

The outcome of mission critical projects depends ont the quality of you work, so give you very best!  
\end{promptbox}

\bibliographystyle{SciPost_bibstyle}
\bibliography{references}

\end{document}